\documentclass[12pt]{article}
\pdfoutput=1
\usepackage{putex}
\usepackage{feyn}
\usepackage[vcentermath]{youngtab}
\usepackage{graphicx}
\usepackage{epstopdf}
\usepackage{enumerate}
\usepackage{cite}
\usepackage{tensor}
\usepackage{slashed}
\usepackage{feynmf}
\usepackage{hyperref}

\usepackage[usenames,dvipsnames,svgnames,table]{xcolor}

\numberwithin{equation}{section}

\renewcommand{\baselinestretch}{1.2}

\def\cW{{\cal W}}
\def\vphi{\varphi}

\newcommand{\Farg}[3]{\left( \begin{array}{c} #1 \\ #2 \end{array} \Big | #3 \right)}

\def\subsubsec{\subsubsection}

\def\s{\sigma}

\def\bul{$\bullet$~}
\def\Oc{{\cal O}}
\def\vs{\vskip .1 in}
\def\D{\Delta}

\def\g{\gamma}
\def\a{\alpha}

\newcommand{\e}[2] {\begin{equation} \label{#1} #2 \end{equation}}

\newcommand{\grp}[1]{\mathrm{#1}}

\newcommand{\grSO}{\grp{SO}}

\newcommand {\be} {\begin {equation}}
\newcommand {\ee} {\end {equation}}

\newcommand {\bes} {\begin {equation*}}
\newcommand {\ees} {\end {equation*}}

\newcommand{\es}[2] {\begin{equation} \label{#1} \begin{split} #2 \end{split} \end{equation}}

\newcommand{\N}{\mathbb{N}}
\newcommand{\R}{\mathbb{R}}

\newcommand{\beq}{\begin{equation}}
\newcommand{\eeq}{\end{equation}}

\newcommand{\cO}{\mathcal{O}}

\newcommand{\bR}{\mathbb{R}}

\newcommand{\p}{\partial}

\def\be{ \begin{equation} }
\def\ee{ \end{equation} }

\def\half{{1\over  2}}

\def\sec{\section}
\def\subsec{\subsection}
\def\N{{\cal N}}
\def\A{{\cal A}}

\def\eqr{\eqref}

\def\b{\beta}
\def\l{\lambda}

\newcommand{\bea}{\begin{eqnarray}}
\newcommand{\eea}{\end{eqnarray}}

\newcommand\sW{{\mathcal W}}

\newcommand\zb{\bar{z}}

\def\rar{\rightarrow}

\def\zb{\overline{z}}

\begin{document}

\institution{UCLA}{Department of Physics and Astronomy, University of California, Los Angeles, CA 90095, USA}

\institution{PU}{Department of Physics, Princeton University, Princeton, NJ 08544, USA}

\title{Witten Diagrams Revisited:\\The AdS Geometry of Conformal Blocks}

\authors{Eliot Hijano\worksat{\UCLA}, Per Kraus\worksat{\UCLA}, Eric Perlmutter\worksat{\PU}, River Snively\worksat{\UCLA}}

\abstract{We develop a new method for decomposing Witten diagrams into conformal blocks.  The steps involved are elementary, requiring no explicit integration, and operate directly in position space.    Central to this construction is an appealingly simple answer to the question: what object in AdS computes a conformal block?  The answer is a ``geodesic Witten diagram," which is essentially an ordinary exchange Witten diagram, except that the cubic vertices are not integrated over all of AdS, but only over bulk geodesics connecting the boundary operators.  In particular, we consider the case of four-point functions of scalar operators, and show how to easily reproduce existing results for the relevant conformal blocks in arbitrary dimension.}

\date{}

\maketitle
\setcounter{tocdepth}{2}
\tableofcontents

\renewcommand{\baselinestretch}{1.2}

\sec{Introduction}\label{i}

The conformal block decomposition of correlation functions in conformal field theory is a powerful way of disentangling the universal information dictated by conformal symmetry from the ``dynamical" information that depends on the particular theory under study; see e.g. \cite{Ferrara:1971vh,Ferrara:1973vz,Ferrara:1974ny,Dolan:2000ut,Dolan:2003hv,Dolan:2011dv,Costa:2011dw}.  The latter is expressed as a list of primary operators and the OPE coefficients amongst them. The use of conformal blocks in the study of CFT correlation functions therefore eliminates redundancy, as heavily utilized, for instance, in recent progress made in the conformal bootstrap program, e.g. \cite{Rattazzi:2008pe,ElShowk:2012ht}.

In the AdS/CFT correspondence \cite{hep-th/9711200, hep-th/9802109, Witten:1998qj}, the role of conformal blocks has been somewhat neglected. The extraction of spectral and OPE data of the dual CFT from a holographic correlation function, as computed by Witten diagrams \cite{Witten:1998qj}, was addressed early on in the development of the subject \cite{Liu:1998ty,Liu:1998th,Freedman:1998bj,D'Hoker:1998mz,D'Hoker:1999pj,D'Hoker:1999jp,Hoffmann:2000tr, Hoffmann:2000mx}, and has been refined in recent years through the introduction of Mellin space technology \cite{Penedones:2010ue,Paulos:2011ie,Fitzpatrick:2011ia,Fitzpatrick:2011hu,Costa:2012cb,Fitzpatrick:2011dm,1410.4185}.
In examining this body of work, however, one sees that a systematic method of decomposing Witten diagrams into conformal blocks is missing. A rather natural question appears to have gone unanswered: namely, what object in AdS computes a conformal block? A geometric bulk description of a conformal block would greatly aid in the comparison of correlators between AdS and CFT, and presumably allow for a more efficient implementation of the dual conformal block decomposition, as it would remove the necessity of actually computing the full Witten diagram explicitly. The absence of such a simpler method would indicate a surprising failure of our understanding of AdS/CFT: after all, conformal blocks are determined by conformal symmetry, the matching of which is literally the most basic element in the holographic dictionary.

In this paper we present an appealingly simple answer to the above question, and demonstrate its utility via streamlined computations of Witten diagrams.    More precisely, we will answer this question in the case of four-point correlation functions of scalar operators, but we expect a similar story to hold in general.    The answer is that conformal blocks are computed by ``geodesic Witten diagrams."   The main feature of a geodesic Witten diagram that distinguishes it from a standard exchange Witten diagram is that in the former, the bulk vertices are not integrated over all of AdS, but only over geodesics connecting points on the boundary hosting the external operators.   This representation of conformal blocks in terms of geodesic Witten diagrams is valid in all spacetime dimensions, and holds for all conformal blocks that arise in four-point functions of scalar operators belonging to arbitrary CFTs (and probably more generally).

\begin{figure}[t!]
   \begin{center}
 \includegraphics[width = 0.7\textwidth]{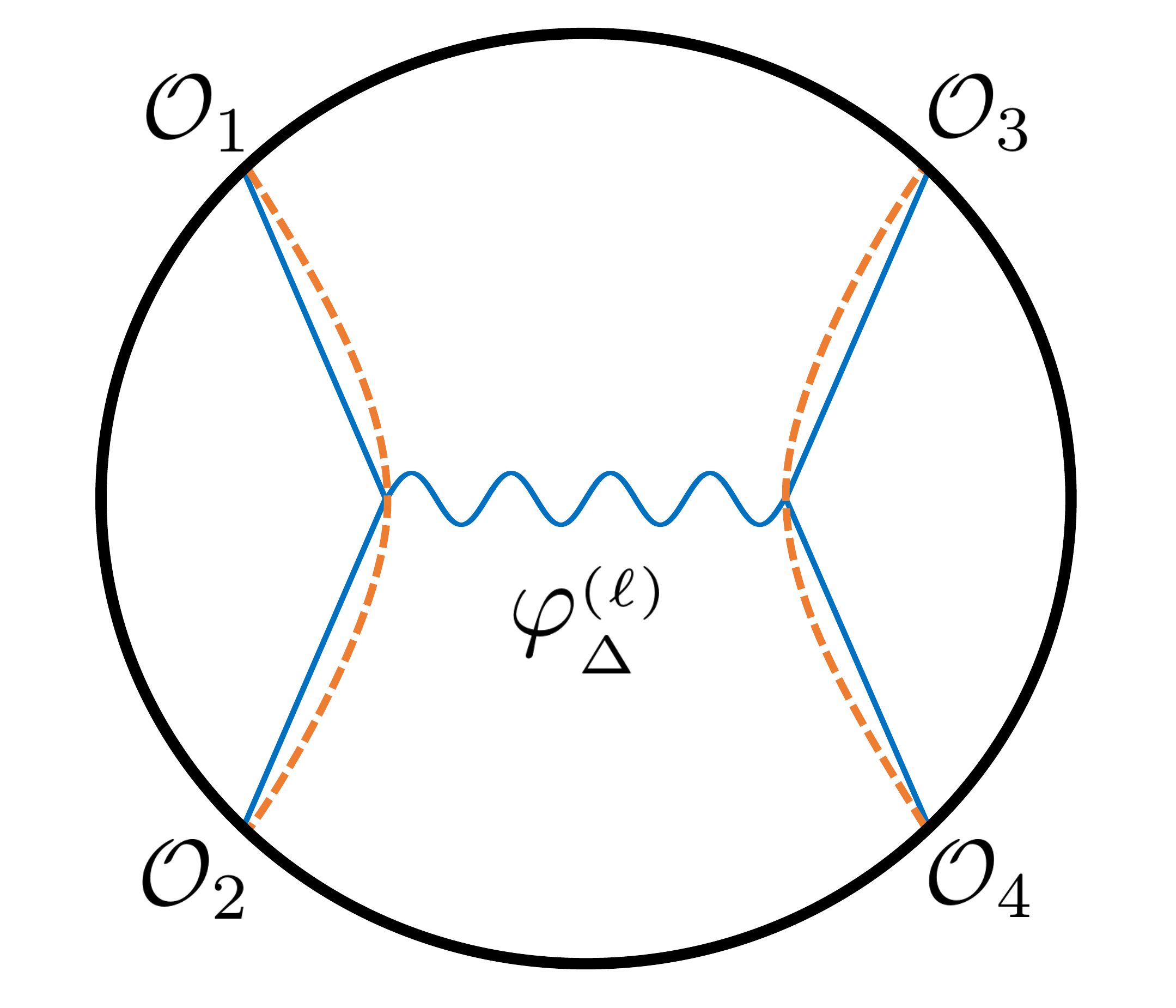}
 \caption{This is a geodesic Witten diagram in AdS$_{d+1}$, for the exchange of a symmetric traceless spin-$\ell$ tensor with $m^2=\D(\D-d)-\ell$ in AdS units. Its main feature is that the vertices are integrated over the geodesics connecting the two pairs of boundary points, here drawn as dashed orange lines. This computes the conformal partial wave for the exchange of a CFT$_d$ primary operator of spin $\ell$ and dimension $\D$.}
 \label{f1}
 \end{center}
 \end{figure}

To be explicit, consider four scalar operators $\Oc_i$ with respective conformal dimensions $\Delta_i$.  The conformal blocks that appear in their correlators correspond to the exchange of primaries carrying dimension $\Delta$ and transforming as symmetric traceless tensors of rank $\ell$; we refer to these as spin-$\ell$ operators.  Up to normalization, the conformal partial wave\footnote{Conformal partial waves and conformal blocks are related by simple overall factors as we review below.} in CFT$_d$ is given by the following object in AdS$_{d+1}$:
\bea\label{geowit}
\int_{\g_{12}}\!d\lambda\int_{\g_{34}}\!d\lambda' G_{b\p}(y(\l),x_1)G_{b\p}(y(\l),x_2)\times G_{bb}(y(\l),y(\l');\D,\ell) \times G_{b\p}(y(\l'),x_3)G_{b\p}(y(\l'),x_4)\cr
\eea
$\gamma_{ij}$ denotes the bulk geodesic connecting boundary points $x_i$ and $x_j$, with $\lambda$ and $\lambda'$ denoting the corresponding proper length parameters.    $G_{b\partial}(y,x)$ are standard scalar bulk-to-boundary propagators connecting a bulk point $y$ to a boundary point $x$.   $G_{bb}(y(\l),y(\l');\D,\ell)$ is the bulk-to-bulk propagator for a spin-$\ell$ field, whose mass squared in AdS units is $m^2 = \D(\D-d)-\ell$, pulled back to the geodesics. The above computes the $s$-channel partial wave, corresponding to using the OPE on the pairs of operators ${\cal O}_1{\cal O}_2$ and ${\cal O}_3{\cal O}_4$. As noted earlier, the expression (\ref{geowit}) looks essentially like an exchange Witten diagram composed of two cubic vertices, except that the vertices are only integrated over geodesics. See figure  \ref{f1}.  Note that although geodesics sometimes appear as an approximation used in the case of high dimension operators, here there is no approximation: the geodesic Witten diagram computes the exact conformal block for any operator dimension.

As we will show, geodesic Witten diagrams arise very naturally upon dismantling  a full Witten diagram into constituents, and this leads to an efficient implementation of the conformal block decomposition.   Mellin space techniques also provide powerful methods, but it is useful to have an approach that can be carried out directly in position space, and that provides an explicit and intuitive picture for the individual conformal blocks.

For the cases that we consider, the conformal blocks are already known, and so one of our tasks is to demonstrate that (\ref{geowit}) reproduces these results.  One route is by explicit computation.  Here, the most direct comparison to existing results is to the original work of Ferrara, Gatto, Grillo, and Parisi \cite{Ferrara:1971vh,Ferrara:1973vz,Ferrara:1974ny}, who provided integral representations for conformal blocks.  In hindsight, these integral expressions can be recognized as geodesic Witten diagrams.  Later work by Dolan and Osborn \cite{Dolan:2000ut,Dolan:2003hv,Dolan:2011dv} provided closed-form expressions for some even-$d$ blocks in terms of hypergeometric functions.   Dolan and Osborn employed the very useful fact that conformal partial waves are eigenfunctions of the conformal Casimir operator.   The most efficient way to prove that geodesic Witten diagrams compute conformal partial waves is to establish that they are the correct eigenfunctions.  This turns out to be quite easy using embedding space techniques, as we will discuss.

Having established that geodesic Witten diagrams compute conformal partial waves, we turn to showing how to decompose a Witten diagram into geodesic Witten diagrams. We do not attempt an exhaustive demonstration here, mostly focusing on tree-level contact and exchange diagrams with four external lines. The procedure turns out to be quite economical and elegant; in particular, we do not need to carry out the technically complicated step of integrating bulk vertices over AdS. Indeed, the method requires no integration at all, as all integrals are transmuted into the definition of the conformal partial waves.  The steps that are required are all elementary. We carry out this decomposition completely explicitly for scalar contact and exchange diagrams, verifying that we recover known results. These include certain hallmark features, such as the presence of logarithmic singularities due to anomalous dimensions of double-trace operators. We also treat the vector exchange diagram, again recovering the correct structure of CFT exchanges.

Let us briefly mention how the analysis goes.   The key step is to use a formula expressing the product of two bulk-to-boundary propagators sharing a common bulk point as a sum of bulk solutions sourced on a geodesic connecting the two boundary points. The fields appearing in the sum turn out to be dual to the double-trace operators appearing in the OPE of the corresponding external operators, and the coefficients in the sum are closely related to the OPE coefficients. See equation \eqr{41a}. With this result in hand, all that is needed are a few elementary properties of AdS propagators to arrive at the conformal block decomposition. This procedure reveals the generalized free field nature of the dual CFT.

The results presented here hopefully lay the foundation for further exploration of the use of geodesic Witten diagrams.  We believe they will prove to be very useful, both conceptually and computationally, in AdS/CFT and in CFT more generally.

The remainder of this paper is organized as follows. In section 2 we review relevant aspects of conformal blocks, Witten diagrams, and their relation.  Geodesic Witten diagrams for scalar exchange are introduced in section 3, and we show by direct calculation and via the conformal Casimir equation that they compute conformal blocks. In section 4 we turn to the conformal block decomposition of Witten diagrams involving just scalar fields.   We describe in detail how single and double trace operator exchanges arise in this framework.  Section 5 is devoted to  generalizing all of this to the case of spinning exchange processes.  We conclude in section 6 with a discussion of some open problems and future prospects.

The ideas developed in this paper originated by thinking about the bulk representation of Virasoro conformal blocks in AdS$_3$/CFT$_2$, based on recent results in this direction \cite{Hartman:2013mia,Asplund:2014coa,Fitzpatrick:2015zha,Fitzpatrick:2014vua,Hijano:2015rla,Alkalaev:2015wia}.  The extra feature associated with a bulk representation of Virasoro blocks is that the bulk metric is deformed in a nontrivial way; essentially, the geodesics backreact on the geometry.   In this paper we focus on global conformal blocks (Virasoro blocks are of course special to CFT$_2$), deferring the Virasoro case to a companion paper \cite{Hijano:2015qja}.

\sec{Conformal blocks, holographic CFTs and Witten diagrams}\label{ii}

Let us first establish some basic facts about four-point correlation functions in conformal field theories, and their computation in AdS$_{d+1}$/CFT$_d$. Both subjects are immense, of course; the reader is referred to \cite{Rychkov:lectures,D'Hoker:2002aw}   and references therein for foundational material.

\subsec{CFT four-point functions and holography}\label{ii1}

We consider vacuum four-point functions of local scalar operators $\Oc(x)$ living in $d$ Euclidean dimensions. Conformal invariance constrains these to take the form
\e{21a}{\langle \Oc_1(x_1)\Oc_2(x_2)\Oc_3(x_3)\Oc_4(x_4)\rangle =\left({x_{24}^2\over x_{14}^2}\right)^{\half \D_{12}}\left({x_{14}^2\over x_{13}^2}\right)^{\half \D_{34}}{g(u,v)\over (x_{12}^2)^{\half (\D_1+\D_2)}(x_{34}^2)^{\half(\D_3+\D_4)}}~,}
where $\D_{ij} \equiv \D_i-\D_j$ and $x_{ij} \equiv x_i-x_j$. $g(u,v)$ is a function of the two independent conformal cross-ratios,
\e{21b}{u={x_{12}^2x_{34}^2\over x_{13}^2x_{24}^2}~,\quad v={x_{14}^2x_{23}^2\over x_{13}^2x_{24}^2}~.}
One can also define complex coordinates $z,\zb$, which obey
\e{21c}{u=z\zb~,\quad v = (1-z)(1-\zb)~.}
These may be viewed as complex coordinates on a two-plane common to all four operators after using conformal invariance to fix three positions at $0,1,\infty$.

$g(u,v)$ can be decomposed into conformal blocks, $G_{\Delta,\ell}(u,v)$, as
\e{21d}{g(u,v) = \sum_{\Oc}C_{12\Oc}\,C^{\Oc}_{~~34} G_{\D,\ell}(u,v)}
where $\Oc$ is a primary operator of dimension $\D$ and spin $\ell$.\footnote{In this paper we only consider scalar correlators, in which only symmetric, traceless tensor exchanges can appear. More generally, $\ell$ would stand for the full set of angular momenta under the $d$-dimensional little group.}
Accordingly, the correlator can be written compactly as a sum of conformal partial waves, $W_{\Delta,\ell}(x_i)$:
\e{21da}{\langle \Oc_1(x_1)\Oc_2(x_2)\Oc_3(x_3)\Oc_4(x_4)\rangle =\sum_{\Oc}C_{12\Oc}\,C^{\Oc}_{~~34} W_{\D,\ell}(x_i)}
where
\e{21db}{W_{\Delta,\ell}(x_i) \equiv \left({x_{24}^2\over x_{14}^2}\right)^{\half \D_{12}}\left({x_{14}^2\over x_{13}^2}\right)^{\half \D_{34}}{G_{\D,\ell}(u,v)\over (x_{12}^2)^{\half (\D_1+\D_2)}(x_{34}^2)^{\half(\D_3+\D_4)}}~.}
Each conformal partial wave is fixed by conformal invariance: it contains the contribution to the correlator of any conformal family whose highest weight state has quantum numbers $(\D,\ell)$, up to overall multiplication by OPE coefficients.  It is useful to think of $W_{\D,\ell}(x_i)$ as the insertion of a projector onto the conformal family of $\Oc$, normalized by the OPE coefficients:
\e{}{W_{\Delta,\ell}(x_i) =  {1\over C_{12\Oc}C^{\Oc}_{~~34}}\,\langle  \Oc_1(x_1)\Oc_2(x_2) \, P_{\D,\ell}\, \Oc_3(x_3)\Oc_4(x_4)\rangle}
where
\e{}{ P_{\D,\ell} \equiv \sum_{{\bf n}} |P^{\bf n} \Oc\rangle\langle P^{\bf n}  \Oc|}
and $P^{\bf n}\Oc$ is shorthand for all descendants of $\Oc$ made from ${\bf n}$ raising operators $P_{\mu}$. We will sometimes refer to conformal blocks and conformal partial waves interchangeably, with the understanding that they differ by the power law prefactor in \eqr{21db}.

Conformal blocks admit double power series expansions in $u$ and $1-v$, in any spacetime dimension \cite{Dolan:2000ut}; for $\ell=0$, for instance,
\e{21e}{G_{\D,0}(u,v) = u^{\D/2} \sum_{m,n=0}^{\infty}{\left({\D+\D_{12}\over 2}\right)_m\left({\D-\D_{34}\over 2}\right)_m\left({\D-\D_{12}\over 2}\right)_{m+n}\left({\D+\D_{34}\over 2}\right)_{m+n}\over m!n!\left(\D+1-{d\over 2}\right)_m(\D)_{2m+n}}u^m(1-v)^n~.}
Higher $\ell$ blocks can be obtained from this one by the use of various closed-form recursion relations \cite{Dolan:2011dv,ElShowk:2012ht}. Especially relevant for our purposes are integral representations of the conformal blocks \cite{Ferrara:1971vh,Ferrara:1973vz,Ferrara:1974ny}. For $\ell=0$,
\es{21f}{G_{\D,0}(u,v) ={1\over 2\beta_{\D 34}} \,u^{\D/2}\int_0^1& d\s \, \s^{\D+\D_{34}-2\over 2}(1-\s)^{\D-\D_{34}-2\over 2}(1-(1-v)\s)^{-\D+\D_{12}\over 2}\\&\times{}_2F_1\left({\D+\D_{12}\over 2}, {\D-\D_{12}\over 2}, \D-{d-2\over 2}, {u\s(1-\s)\over 1-(1-v)\s}\right)}
where we have defined a coefficient
\e{21g}{\beta_{\D 34} \equiv {\Gamma\left({\D+\D_{34}\over 2}\right)\Gamma\left({\D-\D_{34}\over 2}\right) \over 2\Gamma (\D)}~.}
The blocks can also be expressed as infinite sums over poles in $\D$ associated with null states of $SO(d,2)$, in analogy with Zamolodchikov's recursion relations in $d=2$ \cite{Zamo:lectures,zamo, 1307.6856}; these provide excellent rational approximations to the blocks that are used in numerical work. Finally, as we revisit later, in even $d$ the conformal blocks can be written   in terms of hypergeometric functions.

Conformal field theories with weakly coupled AdS duals obey further necessary conditions on their spectra.\footnote{Finding a set of sufficient conditions for a CFT to have a weakly coupled holographic dual remains an unsolved problem. More recent work has related holographic behavior to polynomial boundedness of Mellin amplitudes \cite{1208.0337, Joaotalk}, and to the onset of chaos in thermal quantum systems \cite{1503.01409}.} In addition to having a large number of degrees of freedom, which we will label\footnote{We are agnostic about the precise exponent: vector models and 6d CFTs are welcome here. More generally, we refer to the scaling of $C_T$, the stress tensor two-point function normalization, for instance.} $N^2$, there must be a finite density of states below any fixed energy as $N\rar\infty$; e.g. \cite{Heemskerk:2009pn,Heemskerk:2010ty, ElShowk:2011ag,Hartman:2014oaa}. For theories with Einstein-like gravity duals, this set of parametrically light operators must consist entirely of primaries of spins $\ell\leq 2$ and their descendants.

The ``single-trace'' operators populating the gap are generalized free fields: given any set of such primaries $\Oc_i$, there necessarily exist ``multi-trace'' primaries comprised of conglomerations of these with some number of derivatives (distributed appropriately to make a primary). Altogether, the single-trace operators and their multi-trace composites comprise the full set of primary fields dual to non-black hole states in the bulk. In a four-point function of $\Oc_i$, all multi-trace composites necessarily run in the intermediate channel at some order in $1/N$.

Focusing on the double-trace operators, these are schematically of the form
\e{21h}{[\Oc_i\Oc_j]_{n,\ell} \approx \Oc_i\p^{2n} \p_{\mu_1}\ldots \p_{\mu_{\ell}}\Oc_j~.}
These have spin-$\ell$ and conformal dimensions
\e{21i}{\D^{(ij)}(n,\ell) = \D_i+\D_j+2n+\ell+\g^{(ij)}(n,\ell)~,}
where $\g^{(ij)}(n,\ell)$ is an anomalous dimension.
The expansion of a correlator in the $s$-channel includes the double-trace terms\footnote{Unless otherwise noted, all sums over $m,n$ and $\ell$ run from 0 to $\infty$ henceforth.}
\e{21j}{\langle \Oc_1(x_1)\Oc_2(x_2)\Oc_3(x_3)\Oc_4(x_4)\rangle \supset \sum_{m,\ell}P^{(12)}(m,\ell)W_{\D^{(12)}(m,\ell),\ell}(x_i)+\sum_{n,\ell}P^{(34)}(n,\ell)W_{\D^{(34)}(n,\ell),\ell}(x_i)}
Following \cite{Heemskerk:2009pn, Fitzpatrick:2011dm}, we have defined a notation for squared OPE coefficients,
\e{21k}{P^{(ij)}(n,\ell) \equiv C_{12\Oc}C^{\Oc}_{~~34}~, ~~~\text{where}~~~ \Oc= [\Oc_i\Oc_j]_{n,\ell}~.}
The $1/N$ expansion of the OPE data,
\es{21l}{P^{(ij)}(n,\ell) &= \sum_{r=0}^{\infty} N^{-2r} P^{(ij)}_r(n,\ell)~,\\
\g^{(ij)}(n,\ell) &= \sum_{r=1}^{\infty}N^{-2r} \g^{(ij)}_r(n,\ell)~,}
induces a $1/N$ expansion of the four-point function. Order-by-order in $1/N$, the generalized free fields and their composites must furnish crossing-symmetric correlators. This is precisely the physical content captured by the loop expansion of Witten diagrams in AdS, to which we now turn.

\subsec{A Witten diagrams primer}\label{ii2}

See \cite{D'Hoker:2002aw} for background.
We work in Euclidean AdS$_{d+1}$, with $R_{AdS}\equiv 1$. In Poincar\'e coordinates $y^{\mu}=\lbrace u,x^{i}\rbrace$, the metric is
\e{22a}{ds^2 = {du^2 + dx^idx^i\over u^2}~.}
The ingredients for computing Witten diagrams are the set of bulk vertices, which are read off from a Lagrangian, and the AdS propagators for the bulk fields. A scalar field of mass $m^2=\D(\D-d)$ in AdS$_{d+1}$ has bulk-to-bulk propagator
\e{22b}{G_{bb}(y,y';\D) = e^{-\D \s(y,y')}{}_2F_1\left(\D,{d\over 2};\D+1-{d\over 2}; e^{-2\s(y,y')}\right)}
where $\s(y,y')$ is the geodesic distance between points $y,y'$. In Poincar\'e AdS,
\e{22c}{\s(y,y') = \log\left({1+\sqrt{1-\xi^2}\over \xi}\right)~, \quad \xi = {2u u'\over u^2+u'^2+|x-x'|^2}~.}
$G_{bb}(y,y';\D)$ is a normalizable solution of the AdS wave equation with a delta-function source,\footnote{We use this normalization for later convenience. Our propagator is ${2\pi^{d/2}\Gamma(\Delta-{d-2\over 2})/ \Gamma(\Delta)}$ times the common normalization found in, e.g., equation 6.12 of \cite{D'Hoker:2002aw}.}
\e{22ca}{(\nabla^2 - m^2)G_{bb}(y,y';\D) = -{2\pi^{d/2}\Gamma(\Delta-{d-2\over 2})\over \Gamma(\Delta)}{1\over \sqrt{g}}\delta^{(d+1)}(y-y')~.}
The bulk-to-boundary propagator is
\e{22d}{G_{b\p}(y,x_i) = \left({u\over u^2+|x-x_i|^2}\right)^{\D}~.}
We will introduce higher spin propagators in due course.

A holographic CFT $n$-point function, which we denote ${\cal A}_n$, receives contributions from all possible $n$-point Witten diagrams. The loop-counting parameter is $G_N\sim 1/N^{2}$. At $O(1/N^2)$, only tree-level diagrams contribute. The simplest such diagrams are contact diagrams, which integrate over a single $n$-point vertex. Every local vertex in the bulk Lagrangian gives rise to a contact diagram: schematically,
\e{22e}{{\cal L}\supset \prod_{i=1}^n\p^{p_i}\phi_{\D_i,\ell_i} \quad \Rightarrow \quad {\cal A}_n^{\rm Contact}(x_i) = \int_y \prod_{i=1}^n \p^{p_i}G_{b\p}(y,x_i)}
where $G_{b\p}(y,x_i)$ are bulk-to-boundary propagators for fields with quantum numbers $(\D_i,\ell_i)$, and $p_i$ count derivatives. We abbreviate
\e{22f}{\int_y \equiv \int d^{d+1}y\, \sqrt{g(y)}~.}
There are also exchange-type diagrams, which involve ``virtual'' fields propagating between points in the interior of AdS.

We focus henceforth on tree-level four-point functions of scalar fields $\phi_i$ dual to scalar CFT operators $\Oc_i$. For a non-derivative interaction $\phi_1\phi_2\phi_3\phi_4$, and up to an overall quartic coupling that we set to one, the contact diagram equals
\e{22g}{{\cal A}_4^{\rm Contact}(x_i) =D_{\D_1\D_2\D_3\D_4}(x_i) = \int_y G_{b\p}(y,x_1)G_{b\p}(y,x_2)G_{b\p}(y,x_3)G_{b\p}(y,x_4)~.}
$D_{\D_1\D_2\D_3\D_4}(x_i)$ is the $D$-function, which is defined by the above integral. For generic $\D_i$, this integral cannot be performed for arbitrary $x_i$. There exists a bevy of identities relating various $D_{\D_1\D_2\D_3\D_4}(x_i)$ via permutations of the $\D_i$, spatial derivatives, and/or shifts in the $\D_i$ \cite{D'Hoker:1999jp,Arutyunov:2002fh}.  Derivative vertices, which appear in the axio-dilaton sector of type IIB supergravity, for instance, define $D$-functions with shifted parameters.  When $\D_i=1$ for all $i$,
\e{22h}{ {2x_{13}^2x_{24}^2\over \Gamma\left(2-{d\over 2}\right) \pi^{d/2}}D_{1111}(x_i) =
 {1\over z-\zb}\left(2\text{Li}_2(z)-2\text{Li}_2(\zb) + \log(z\zb)\log{1-z\over 1-\zb}\right)~.}
This actually defines the $D$-bar function, $\overline D_{1111}(z,\zb)$. For various sets of $\D_i\in \mathbb{Z}$, combining \eqr{22h} with efficient use of $D$-function identities leads to polylogarithmic representations of contact diagrams.

  \begin{figure}[t!]
   \begin{center}
    \hspace*{\fill}%
 \includegraphics[width = 0.46\textwidth]{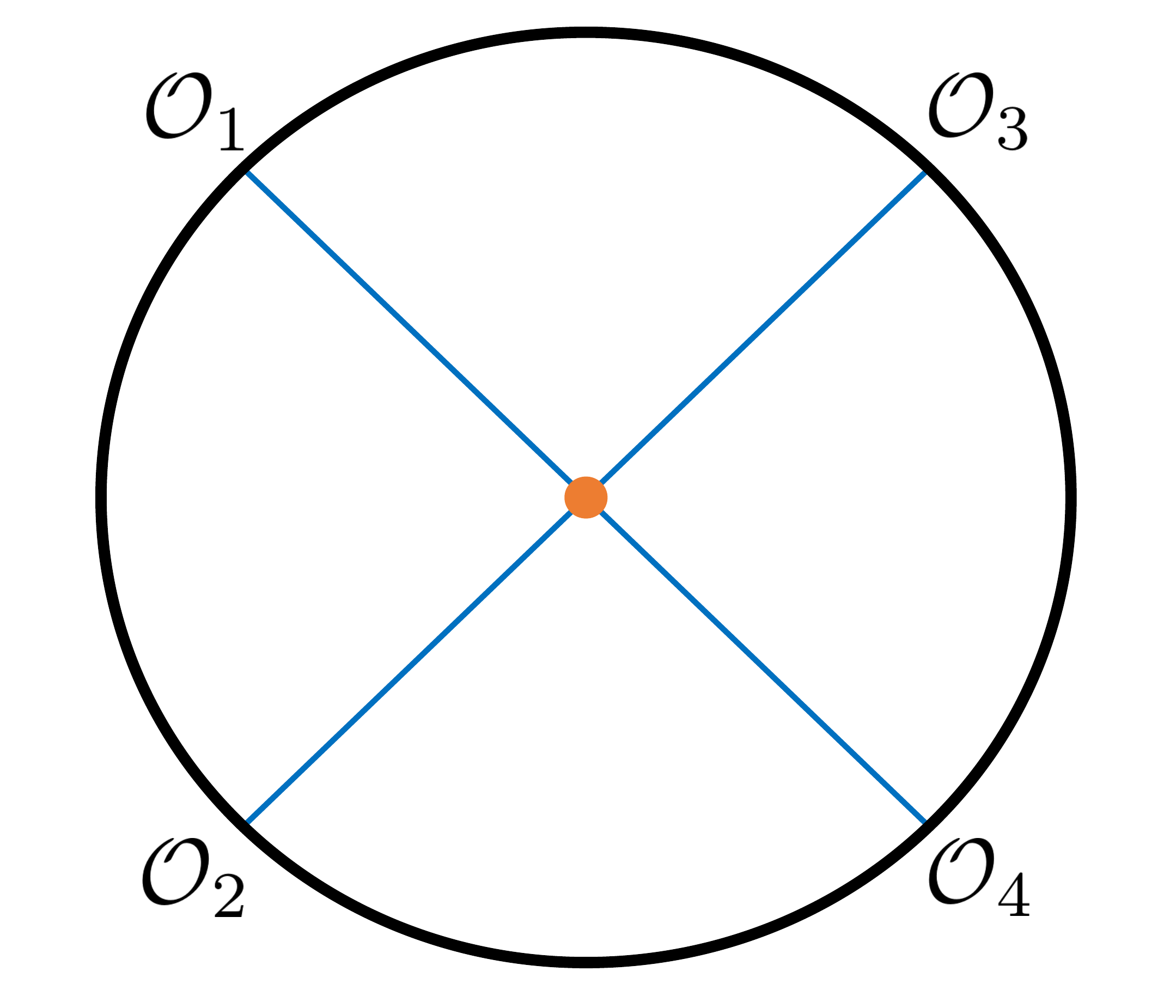}\hfill
  \includegraphics[width = 0.46\textwidth]{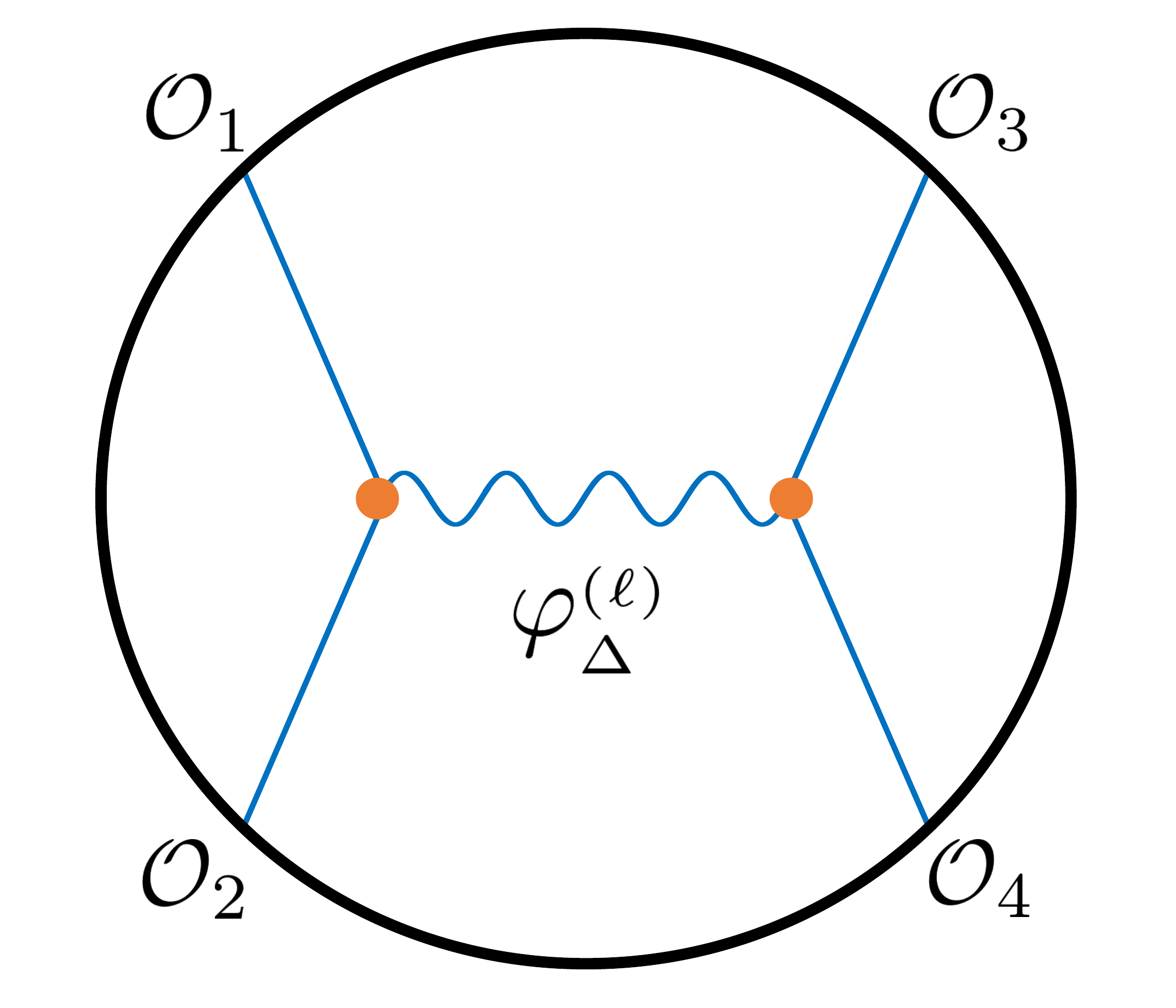}%
    \hspace*{\fill}%
 \caption{Tree-level four-point Witten diagrams for external scalar operators. On the left is a contact diagram. On the right is an exchange diagram for a symmetric traceless spin-$\ell$ tensor field of dual conformal dimension $\D$. Here and throughout this work, orange dots denote vertices integrated over all of AdS.}
 \label{f2}
 \end{center}
 \end{figure}

The other class of tree-level diagrams consists of exchange diagrams. For external scalars, one can consider exchanges of symmetric, traceless tensor fields of arbitrary spin $\ell$. These are computed, roughly, as
\e{22i}{{\cal A}_4^{\rm Exch}(x_i) =  \int_{y} \int_{y'}G_{b\p}(y,x_1)G_{b\p}(y,x_2)\times G_{bb}(y,y';\D,\ell)\times G_{b\p}(y',x_3)G_{b\p}(y',x_4)~.}
$G_{bb}(y,y';\D,\ell)$ is shorthand for the bulk-to-bulk propagator for the spin-$\ell$ field of dimension $\D$, which is really a bitensor $[G_{bb}(y,y';\D)]_{\mu_1 \ldots \mu_{\ell}; \,\nu_1 \ldots \nu_{\ell}}$. We have likewise suppressed all derivatives acting on the external scalar propagators, whose indices are contracted with those of $[G_{bb}(y,y';\D)]_{\mu_1 \ldots \mu_{\ell};\, \nu_1 \ldots \nu_{\ell}}$. Due to the double integral, brute force methods of simplifying exchange diagrams are quite challenging, even for $\ell=0$, without employing some form of asymptotic expansion.

The key fact about these tree-level Witten diagrams relevant for a dual CFT interpretation is as follows. For contact diagrams \eqr{22g}, their decomposition into conformal blocks contains the infinite towers of double-trace operators in \eqr{21j}, and only these. This is true in any channel. For exchange diagrams \eqr{22i}, the $s$-channel decomposition includes a single-trace contribution from the operator dual to the exchanged bulk field, in addition to infinite towers of double-trace exchanges \eqr{21j}. In the $t$- and $u$-channels, only double-trace exchanges are present.  The precise set of double-trace operators that appears is determined by the spin associated to the bulk vertices.

Higher-loop Witten diagrams are formed similarly, although the degree of difficulty increases rapidly with the loop order. No systematic method has been developed to compute these.

\subsubsec{Logarithmic singularities and anomalous dimensions}

When the external operator dimensions are non-generic, logarithms can appear in tree-level Witten diagrams \cite{Freedman:1998bj, Liu:1998th, D'Hoker:1999jp}. These signify the presence of perturbatively small anomalous dimensions, of order $1/N^2$, for intermediate states appearing in the CFT correlator. Let us review some basic facts about this.

In general, if any operator of free dimension $\D_0$ develops an anomalous dimension $\g$, so that its full dimension is $\D=\D_0+\g$, a small-$\g$ expansion of its contribution to correlators yields an infinite series of logs:
\e{22j}{G_{\D_0+\g,\ell}(u,v) \approx u^{\D_0-\ell\over 2}\left(1+{\g\over 2} \log u + \ldots\right)~.}
In the holographic context, the double-trace composites $[\Oc_i\Oc_j]_{n,\ell}$ have anomalous dimensions at $O(1/N^2)$. Combining \eqr{21i}, \eqr{21j} and \eqr{21l} leads to double-trace contributions to holographic four-point functions of the form
\es{22k}{{\cal A}_4(x_i)\Big|_{1/N^2} \supset &\sum_{m,\ell}\left( P_1^{(12)}(m,\ell)+\half P_0^{(12)}(m,\ell)\g_1^{(12)}(m,\ell)\p_m\right) W_{\D_1+\D_2+2m+\ell,\ell}(x_i)
\\+&\sum_{n,\ell}\left( P_1^{(34)}(n,\ell)+\half P_0^{(34)}(n,\ell)\g_1^{(34)}(n,\ell)\p_n\right) W_{\D_3+\D_4+2n+\ell,\ell}(x_i)}
where $\p_m W_{\D_1+\D_2+2m+\ell,\ell}\propto \log u$ and likewise for the (34) terms. These logarithmic singularities should therefore be visible in tree-level Witten diagrams. In top-down examples of AdS/CFT, the supergravity fields are dual to protected operators, so the $\g^{(ij)}(n,\ell)$ are the only (perturbative) anomalous dimensions that appear, and hence are responsible for all logs.

For generic operator dimensions, the double-trace operators do not appear in both the $\Oc_1\Oc_2$ and $\Oc_3\Oc_4$ OPEs at $O(N^0)$, so $P_0^{(ij)}(n,\ell)=0$. On the other hand, when the $\D_i$ are related by the integrality condition $\D_1+\D_2-\D_3-\D_4 \in 2\mathbb{Z}$, one has $P_0^{(ij)}(n,\ell)\neq0$ \cite{Liu:1998th}.

\subsubsec{What has been computed?}

In a foundational series of papers  \cite{Muck:1998rr, Liu:1998ty, Freedman:1998bj,D'Hoker:1998gd,D'Hoker:1998mz,D'Hoker:1999jc,D'Hoker:1999pj,D'Hoker:1999ni,D'Hoker:1999jp}, methods of direct computation were developed for scalar four-point functions, in particular for scalar, vector and graviton exchanges. Much of the focus was on the axio-dilaton sector of type IIB supergravity on AdS$_5\times S^5$ in the context of duality with $\N=4$ super-Yang Mills (SYM), but the methods were gradually generalized to arbitrary operator and spacetime dimensions.

This effort largely culminated in \cite{D'Hoker:1999pj,D'Hoker:1999ni} and \cite{D'Hoker:1999jp}. \cite{D'Hoker:1999pj} collected the results from all channels contributing to axio-dilaton correlators in $\N=4$ SYM, yielding the full correlator at $O(1/N^2)$. In \cite{D'Hoker:1999ni}, a more efficient method of computation was developed for exchange diagrams. It was shown that scalar, vector and graviton exchange diagrams can generically be written as infinite sums over contact diagrams for external fields of variable dimensions. These truncate to finite sums if certain relations among the dimensions are obeyed.\footnote{For instance, an $s$-channel scalar exchange is written as a finite linear combination of $D$-functions if $\D_1+\D_2-\D$ is a positive even integer \cite{D'Hoker:1999ni}.} These calculations were translated in \cite{D'Hoker:1999jp} into CFT data, where it was established that logarithmic singularities appear precisely at the order determined by the analysis of the previous subsection. This laid the foundation for the modern perspective on generalized free fields. Further analysis of implications of four-point Witten diagrammatics for holographic CFTs (e.g. crossing symmetry, non-renormalization), and for $\N=4$ SYM in particular, was performed in \cite{Banks:1998nr, Brodie:1998ke, Chalmers:1998wu, Intriligator:1998ig, hep-th/0002154, hep-th/0002170, hep-th/0003218, hep-th/0005182, hep-th/0009106, Dolan:2000ut, Arutyunov:2002fh, hep-th/0601148, hep-th/0611123}. A momentum space-based approach can be found in \cite{1011.0780, 1201.6449}.

More recent work has computed Witten diagrams for higher spin exchanges \cite{Costa:2014kfa,Bekaert:2014cea}. These works develop the split representation of massive spin-$\ell$ symmetric traceless tensor fields, for arbitrary integer $\ell$. There is a considerable jump in technical difficulty, but the results are all consistent with AdS/CFT.

\subsec{Mellin space}

An elegant alternative approach to computing correlators, especially holographic ones, has been developed in Mellin space \cite{0907.2407, Penedones:2010ue}. The analytic structure of Mellin amplitudes neatly encodes the CFT data and follows a close analogy with the momentum space representation of flat space scattering amplitudes. We will not make further use of Mellin space in this paper, but it should be included in any discussion on Witten diagrams; we only briefly review its main properties with respect to holographic four-point functions, and further aspects and details may be found in e.g. \cite{Paulos:2011ie,Fitzpatrick:2011ia,Fitzpatrick:2011hu,Fitzpatrick:2011dm, Nandan:2011wc, Costa:2012cb, 1410.4717, 1410.4185, 1411.1675}.

Given a four-point function as in \eqr{21a}, its Mellin representation may be defined by the integral transform
\es{m1}{g(u,v) = \int_{-i\infty}^{i\infty} &ds\, dt\, M(s,t) u^{t/2}v^{-(s+t)/2}\Gamma\left({\D_1+\D_2-t\over 2}\right)\Gamma\left({\D_3+\D_4-t\over 2}\right)\\
&\Gamma\left({\D_{34}-s\over 2}\right)\Gamma\left({-\D_{12}-s\over 2}\right)\Gamma\left({s+t\over 2}\right)\Gamma\left({s+t+\D_{12}-\D_{34}\over 2}\right)~.}
The integration runs parallel to the imaginary axis and to one side of all poles of the integrand. The Mellin amplitude is $M(s,t)$. Assuming it formally exists, $M(s,t)$ can be defined for any correlator, holographic \cite{Penedones:2010ue} or otherwise \cite{1506.04659, Joaocalc}. $M(s,t)$ is believed to be meromorphic in any compact CFT. Written as a sum over poles in $t$, each pole sits at a fixed twist $\tau=\D-\ell$, capturing the exchange of twist-$\tau$ operators in the intermediate channel. Given rhe exchange of a primary $\Oc$ of twist $\tau_{\Oc}$, its descendants of twist $\tau=\tau_{\Oc}+2m$ contribute a pole
\e{m2}{M(s,t) \supset  C_{12\Oc}C^{\Oc}_{~~34}\,{ {\cal Q}_{\ell,m}(s)\over t-\tau_{\Oc}-2m}}
where $m=0,1,2,\ldots$. ${\cal Q}_{\ell,m}(s)$ is a certain degree-$\ell$ (Mack) polynomial that can be found in \cite{Costa:2012cb}. Note that an infinite number of descendants contributes at a given $m$. $n$-point Mellin amplitudes may be likewise defined in terms of $n(n-3)/2$ parameters, and are known to factorize onto lower-point amplitudes \cite{1410.4185}.

Specifying now to holographic correlators at tree-level,\footnote{This is the setting that is known to be especially amenable to a Mellin treatment. Like other approaches to Witten diagrams, the Mellin program has not been systematically extended to loop level (except for certain classes of diagrams; see Section \ref{vi}). Because higher-trace operators appear at higher orders in $1/N$, some of the elegance of the tree-level story is likely to disappear. The addition of arbitrary external spin in a manner which retains the original simplicity has also not been done, although see \cite{Paulos:2011ie}.} the convention of including explicit Gamma functions in \eqr{m1} has particular appeal: their poles encode the double-trace exchanges of $[\Oc_1\Oc_2]_{m,\ell}$ and $[\Oc_3\Oc_4]_{n,\ell}$.  Poles in $M(s,t)$ only capture the single-trace exchanges, if any, associated with a Witten diagram. In particular, all local AdS interactions give rise to contact diagrams whose Mellin amplitudes are mere polynomials in the Mellin variables. In this language, the counting of solutions to crossing symmetry in sparse large $N$ CFTs performed in \cite{Heemskerk:2009pn} becomes manifestly identical on both sides of the duality. Exchange Witten diagrams have meromorphic Mellin amplitudes that capture the lone single-trace exchange: they take the form\footnote{For certain non-generic operator dimensions, the sum over poles actually truncates \cite{Penedones:2010ue, Fitzpatrick:2011ia}. The precise mechanism for this is not fully understood from a CFT perspective. We thank Liam Fitzpatrick and Joao Penedones for discussions on this topic.}
\e{m2}{M(s,t) =  C_{12\Oc}C^{\Oc}_{~~34}\sum_{m=0}^{\infty}{ {\cal Q}_{\ell,m}(s)\over t-\tau_{\Oc}-2m}+\text{Pol}(s,t)~.}
Pol($s,t$) stands for a possible polynomial in $s,t$. The polynomial boundedness is a signature of local AdS dynamics \cite{1208.0337, Costa:2012cb}. Anomalous dimensions appear when poles of the integrand collide to make double poles.

A considerable amount of work has led to a quantitative understanding of the above picture. These include formulas for extraction of the one-loop OPE coefficients $P_1^{(ij)}(n,\ell)$ and anomalous dimensions $\g_1^{(ij)}(n,\ell)$ from a given Mellin amplitude (Section 2.3 of \cite{Fitzpatrick:2011dm}); and the graviton exchange amplitude between pairwise identical operators in arbitrary spacetime dimension (Section 6 of \cite{Costa:2014kfa}).

\subsec{Looking ahead}
Having reviewed much of what has been accomplished, let us highlight some of what has not.

First, we note that no approach to computing holographic correlators has systematically deconstructed loop diagrams, nor have arbitrary external spins been efficiently incorporated. Save for some concrete proposals in Section \ref{vi}, we will not address these issues here.

While Mellin space is home to a fruitful approach to studying holographic CFTs in particular, it comes with a fair amount of technical complication. Nor does it answer the natural question of how to represent a single conformal block in the bulk. One is, in any case, left to wonder whether a truly efficient approach exists in position space.

Examining the position space computations reviewed in subsection \ref{ii2}, one is led to wonder: where are the conformal blocks? In particular, the extraction of dual CFT spectral data and OPE coefficients in the many works cited earlier utilized a double OPE expansion. More recent computations of exchange diagrams \cite{Costa:2014kfa,Bekaert:2014cea} using the split representation do make the conformal block decomposition manifest, in a contour integral form \cite{Dobrev:1975ru}: integration runs over the imaginary axis in the space of complexified conformal dimensions, and the residues of poles in the integrand contain the OPE data. This is closely related to the shadow formalism. However, this approach is technically quite involved, does not apply to contact diagrams, and does not answer the question of what bulk object computes a single conformal block.

Let us turn to this latter question now, as a segue to our computations of Witten diagrams.

\sec{The holographic dual of a scalar conformal block}\label{iii}

What is the holographic dual of a conformal block? This is to say, what is the geometric representation of a conformal block in AdS? In this section we answer this question for the case of scalar exchanges between scalar operators, for generic operator and spacetime dimensions.  In Section \ref{v}, we will tackle higher spin exchanges. At this stage, these operators need not belong to a holographic CFT, since the form of a conformal block is fixed solely by symmetry. What follows may seem an inspired guess, but as we show in the next section, it emerges very naturally as an ingredient in the computation of Witten diagrams.
\vs

Let us state the main result. We want to compute the scalar conformal partial wave $W_{\D,0}(x_i)$, defined in \eqr{21db}, corresponding to exchange of an operator $\Oc$ of dimension $\D$ between two pairs of external operators $\Oc_1,\Oc_2$ and $\Oc_3,\Oc_4$. Let us think of the external operators as sitting on the boundary of AdS$_{d+1}$ at positions $x_{1,2,3,4}$, respectively. Denote the geodesic running between two boundary points $x_i$ and $x_j$ as $\g_{ij}$. Now consider the scalar {\it geodesic Witten diagram}, which we denote ${\cal W}_{\D,0}(x_i)$, first introduced in Section \ref{i} and drawn in Figure \ref{f1}:
\es{gwitt}{&{\cal W}_{\D,0}(x_i)\equiv\\& \int_{\g_{12}}\int_{\g_{34}}G_{b\p}(y(\l), x_1)G_{b\p}(y(\l),x_2)\times G_{bb}(y(\l),y(\l');\D)\times G_{b\p}(y(\l'),x_3)G_{b\p}(y(\l'),x_4)~,}
where
\e{3b}{\int_{\g_{12}} \equiv \int_{-\infty}^{\infty} d\l~, \quad \int_{\g_{34}} \equiv \int_{-\infty}^{\infty} d\l'}
denote integration over proper time coordinates $\l$ and $\l'$ along the respective geodesics. Then ${\cal W}_{\D,0}(x_i)$ is related to the conformal partial wave $W_{\D,0}(x_i)$ by
\e{3c}{ {\cal W}_{\D,0}(x_i) =  \beta_{\D 12}\beta_{\D 34} W_{\D,0}(x_i)~.}
The proportionality constant $\beta_{\D 34}$ is defined in equation \eqr{21g} and $\beta_{\D 12}$ is defined analogously.

The object ${\cal W}_{\D,0}(x_i)$ looks quite like the expression for a scalar exchange Witten diagram for the bulk field dual to $\Oc$. Indeed, the form is identical, except that the bulk vertices are not integrated over all of AdS, but rather over the geodesics connecting the pairs of boundary points. This explains our nomenclature. Looking ahead to conformal partial waves for exchanged operators with spin, it is useful to think of the bulk-to-bulk propagator in \eqr{gwitt} as pulled back to the two geodesics.

Equation \eqr{3c} is a rigorous equality. We now prove it in two ways.

\subsec{Proof by direct computation}\label{iii1}

Consider the piece of \eqr{gwitt} that depends on the geodesic $\g_{12}$, which we denote $\varphi^{12}_{\D}$:
\e{31aa}{\varphi^{12}_{\D}(y(\l' )) \equiv \int_{\g_{12}}G_{b\p}(y(\l),x_1)G_{b\p}(y(\l),x_2)G_{bb}(y(\l),y(\l');\D) ~.}
In terms of $\varphi^{12}_{\D}(y(\l'))$, the formula for $\cW_{\D,0}$ becomes
\e{31a}{\cW_{\D,0}(x_i) = \int_{\g_{34}}\vphi^{12}_{\D}(y(\l'))G_{b\p}(y(\l'),x_3)G_{b\p}(y(\l'),x_4)~.}

It is useful to think of $\varphi^{12}_{\D}(y)$, for general $y$, as a cubic vertex along $\g_{12}$ between a bulk field at $y$ and two boundary fields anchored at $x_1$ and $x_2$. We may then solve for $\varphi^{12}_{\D}(y)$ as a normalizable solution of the Klein-Gordon equation with a source concentrated on $\g_{12}$. The symmetries of the problem turn out to specify this function uniquely, up to a multiplicative constant. We then pull this back to $\g_{34}$, which reduces $\cW_{\D,0}(x_i)$ to a single one-dimensional integral along $\g_{34}$. This integral can then be compared to the well-known integral representation for $G_{\D,0}(u,v)$ in \eqr{21f}, which establishes \eqr{3c}.

To make life simpler, we will use conformal symmetry to compute the geodesic Witten diagram with operators at the following positions:
\es{cpw}{ \cW_{\D,0}(u,v)&\equiv{1\over C_{12\Oc}C^{\Oc}_{~~34}}\, \langle \Oc_1(\infty)\Oc_2(0)\,P_{\D,0}\,\Oc_3(1-z)\Oc_4(1)\rangle\\& = u^{-\D_3-\D_4\over 2}G_{\D,0}(u,v)}
where, as usual, $\Oc_1(\infty) \equiv \lim_{x_1\rar\infty}x_1^{2\D_1} \Oc(x_1)$. We implement the above strategy by solving the wave equation first in global AdS, then moving to Poincar\'e coordinates and comparing with CFT. We work with the global AdS metric
\e{31b}{ds^2 = {1\over \cos^2\rho}(d\rho^2 + dt^2+\sin^2\rho d\Omega_{d-1}^2)~.}
The relation between mass and conformal dimension is
\e{31c}{m^2 = \Delta(\Delta-d) }
and so the wave equation for $\vphi^{12}_\D(y)$ away from $\g_{12}$ is
\e{31d}{ \big(\nabla^2 -\Delta(\Delta-d) \big)\vphi^{12}_\D(y)=0~,}
or
\e{31e}{ \left(\cos^2\rho \p_\rho^2+(d-1)\cot\rho\p_\rho +\cos^2\rho \p_t^2 -\Delta(\Delta-d)\right) \vphi^{12}_\D(y)=0~.}
At $\g_{12}$, there is a source. To compute \eqr{cpw}, we take $t_1 \to -\infty, t_2 \to \infty$, in which limit the geodesic becomes a line at $\rho=0$, the center of AdS. This simplifies matters because this source is rotationally symmetric. Its time-dependence is found by evaluating the product of bulk-to-boundary propagators on a fixed spatial slice,
\e{31f}{G_{b\p}(t,t_1)G_{b\p}(t,t_2) \propto e^{-\D_{12}t}~.}
Therefore, we are looking for a rotationally-symmetric, normalizable solution to the following radial equation:
\e{31g}{ \left(\cos^2\rho \p_\rho^2+(d-1)\cot\rho\p_\rho +\cos^2\rho \D_{12}^2 -\Delta(\Delta-d)\right) \vphi^{12}_\D=0~.}
The full solution, including the time-dependence and with the normalization fixed by equation \eqr{31aa}, is
\e{31h}{\vphi^{12}_\D(\rho,t) =
\beta_{\D 12}\times e^{\D_1 t_1-\D_2 t_2}
{}_2F_1\Big({\Delta+\Delta_{12}\over 2},{\Delta -\Delta_{12} \over 2};\Delta - {d-2\over 2};\cos^2 \rho\Big) \cos^{\Delta} \rho~ e^{-\Delta_{12}t } ~.}

Now we need to transform this to Poincar\'e coordinates, and pull it back to $\g_{34}$. The relation between coordinates is
\e{31l}{ e^{-2t} = u^2 +|x|^2 ~,\quad \cos^2\rho = {u^2\over u^2 +|x|^2} ~.}
Although the field $\varphi_{\D}^{12}$ is a scalar, it transforms nontrivially under the map from global to Poincar\'e AdS because the bulk-to-boundary propagators in its definition \eqr{31aa} transform as
\e{31la}{G_{b\p}(t_1,y) = |x_1|^{\D_1}G_{b\p}(y,x_1)~,\quad
G_{b\p}(t_2,y) = |x_2|^{\D_2}G_{b\p}(y,x_2) ~.}
After stripping off the power of $x_1$ needed to define the operator at infinity, the field in Poincar\'e coordinates is
\e{31lb}{\vphi^{12}_\D(u,x^i) =
\beta_{\D 12}\times
{}_2F_1\Big({\Delta+\Delta_{12}\over 2},{\Delta -\Delta_{12} \over 2};\Delta - {d-2\over 2};\cos^2 \rho\Big) \cos^{\Delta} \rho~ e^{-\Delta_{12}t } }
where on the right hand side $\rho$ and $t$ are to be viewed as functions of the Poincar\'e coordinates $u,x^i$ via \eqr{31l}.

Now we want to evaluate our Poincar\'e coordinates on the geodesic $\g_{34}$. With our choice of positions in \eqr{cpw}, $\g_{34}$ is a geodesic in an AdS$_3$ slice through AdS$_{d+1}$. 
Geodesics in Poincar\'e AdS$_3$ are semi-circles: for general boundary points $z_3,z_4$,
\e{31m}{ 2u^2+ (z-z_4)(\zb-\zb_3)+(\zb-\zb_4)(z-z_3)=0~.}
In terms of the proper length parameter $\l'$,
\es{31n}{z(\lambda')& = {z_3 +z_4 \over 2}+{z_3-z_4 \over 2} \tanh \lambda' ~,\\
\zb(\lambda')& = {\zb_3 +\zb_4 \over 2}+{\zb_3-\zb_4 \over 2} \tanh \lambda' ~,\\
u(\lambda')&= {|z_3-z_4|\over 2\cosh \lambda'}~.}
Plugging in $z_3=1-z, \zb_3=1-\zb, z_4=\zb_4=1$ into \eqr{31n} and then \eqr{31l}, we find
\es{31o}{\cos^2\rho\big|_{\g_{34}}&={1\over 2\cosh\l'}{|z|^2\over e^{-\l'}+|1-z|^2e^{\l'}}~, \\
e^{-2t}\big|_{\g_{34}}&= {e^{-\l'}+|1-z|^2e^{\l'}\over 2\cosh\l'}~.}
Therefore, the pullback of $\vphi^{12}_{\D}$ to $\g_{34}$ is
\es{31p}{\vphi^{12}_\D(u(\l'),x^i(\l')) &= \beta_{\D 12} (2\cosh\l')^{-\D_{12}-\D\over 2}(e^{-\l'}+|1-z|^2e^{\l'})^{\D_{12}-\D\over 2}|z|^{\D}\\&\times {}_2F_1\left({\Delta+\Delta_{12}\over 2},{\Delta -\Delta_{12} \over 2};\Delta - {d-2\over 2};{1\over 2\cosh\l'}{|z|^2\over e^{-\l'}+|1-z|^2e^{\l'}}\right)~.}

Finally, we need to evaluate $G_{b\p}(y(\l'),1-z)G_{b\p}(y(\l'),1)$ in \eqr{31a}. Plugging \eqr{31n} into the propagator \eqr{22d}, we get
\e{31q}{G_{b\p}(y(\l'),1-z)G_{b\p}(y(\l'),1) = {e^{\D_{34}\l'}\over |z|^{\D_3+\D_4}}~.}

We may now assemble all pieces of \eqr{31a}--\eqr{cpw}: plugging \eqr{31p} and \eqr{31q} into \eqr{31a} gives the geodesic Witten diagram. Trading $z,\zb$ for $u,v$ we find, in the parameterization \eqr{cpw},
\es{31r}{\cW_{\D,0}(u,v) = \beta_{\D 12}\,&u^{\D-\D_3-\D_4\over 2}\int_{-\infty}^{\infty} d\l' (2\cosh\l')^{-\D_{12}-\D\over 2}(e^{-\l'}+ve^{\l'})^{\D_{12}-\D\over 2}   e^{\D_{34}\l'}\\
&\times  {}_2F_1\left({\Delta+\Delta_{12}\over 2},{\Delta -\Delta_{12} \over 2};\Delta - {d-2\over 2};{1\over 2\cosh\l'}{u\over e^{-\l'}+v e^{\l'}}\right)~.}
Defining a new integration variable,
\e{31s}{\s = {e^{2\l'}\over 1+e^{2\l'}}~,}
we have
\es{31t}{\cW_{\D,0}(u,v) ={\beta_{\D 12}\over 2}\, u^{\D-\D_3-\D_4\over 2} \int_{0}^{1}& d\s\,\s^{{\D+\D_{34}-2\over 2}}(1-\s)^{{\D-\D_{34}-2\over 2}}(1-(1-v)\s)^{-\D+\D_{12}\over 2}\\&\times{}_2F_1\left({\Delta+\Delta_{12}\over 2},{\Delta -\Delta_{12} \over 2};\Delta - {d-2\over 2};{u \s(1-\s)\over 1-(1-v)\s}\right)~.}
Comparing \eqr{31t} to the integral representation \eqr{21f}, one recovers \eqr{3c} evaluated at the appropriate values of $x_1,...,x_4$. Validity of the relation \eqr{3c} for general operator positions is then assured by the identical conformal transformation properties of the two sides.

\subsec{Proof by conformal Casimir equation}\label{iii2}

In the previous section we evaluated a geodesic Witten diagram $\sW_{\D,0}$ and matched the result to a known integral expression for the corresponding conformal partial wave $W_{\D,0}$, thereby showing that $\sW_{\D,0}=W_{\D,0}$ up to a multiplicative constant. In this section we give a direct argument for the equivalence, starting from the definition of conformal partial waves.

Section \ref{iii2i} reviews the definition of conformal partial waves $W_{\D,\ell}$ as eigenfunctions of the conformal Casimir operator. We prove this definition to be satisfied by geodesic Witten diagrams $\sW_{\D,0}$ in section \ref{iii2iii}, after a small detour (section \ref{iii2ii}) to define the basic embedding space language used in the proof.

The arguments of the present section are generalized in section \ref{v5} to show that $\sW_{\D,\ell} = W_{\D,\ell}$ (again, up to a factor) for arbitrary exchanged spin $\ell$.

\subsubsection{The Casimir equation}\label{iii2i}
The generators of the $d$-dimensional conformal group $\grSO(d+1,1)$ can be taken to be the Lorentz generators $L_{AB}$ of $d+2$ dimensional Minkowski space (with $L_{AB}$ antisymmetric in $A$ and $B$ as usual). The quadratic combination $L^2\equiv{1\over2}L_{AB}L^{AB}$ is a Casimir of the algebra, i.e. it commutes with all the generators $L_{AB}$. As a result, $L^2$ takes a constant value on any irreducible representation of the conformal group, which means all states $|P^{\bf n}\cO\rangle$ in the conformal family of a primary state $|\cO\rangle$ are eigenstates of $L^2$ with the same eigenvalue. The eigenvalue depends on the dimension $\D$ and spin $\ell$ of $|\cO\rangle$, and can be shown to be \cite{Dolan:2000ut}
\e{32a}{C_2(\D,\ell) = -\D(\D-d) - \ell(\ell+d-2)~.}
The $\grSO(d+1,1)$ generators are represented on conformal fields by
\e{32b}{[L_{AB},\cO_1(x_1)] = L_{AB}^{1}\cO(x_1)~.}
where $L^1_{AB}$ is a differential operator built out of the position $x_1$ of $\cO_1$ and derivatives with respect to that position. The form of the $L^1_{AB}$ depends on the conformal quantum numbers of $\cO_1$. Equation \eqr{32b} together with conformal invariance of the vacuum imply the following identity, which holds for any state $|\alpha\rangle$:
\e{32c}{(L^1_{AB}+L^2_{AB})^2 \langle 0| \cO_1 (x_1) \cO_2 (x_2) |\alpha\rangle = \langle 0| \cO_1 (x_1) \cO_2 (x_2) L^2|\alpha\rangle~.}
Consistent with the notation for $L^2$, we have defined
\e{32d}{(L^1_{AB}+L^2_{AB})^2 \equiv \tfrac{1}2 (L^1_{AB}+L^2_{AB})(L^{1\,AB}+L^{2\,AB})~.}

As discussed in section \ref{ii1}, one obtains a conformal partial wave $W_{\D,\ell}$ by inserting into a four-point function the projection operator $P_{\D,\ell}$ onto the conformal family of a primary $\cO$ with quantum numbers $\D,\ell$:
\e{32e}{W_{\D,\ell}(x_i) =  {1\over C_{12 \cO}C_{~~34}^{\cO}}\sum_{\bf n}
\langle 0|\cO_1 (x_1)\cO_2 (x_2)|P^{\bf n} \cO\rangle \langle P^{\bf n} \cO | \cO_3 (x_3) \cO_4 (x_4) |0\rangle~.}
Applying the identity \eqr{32c} to the equation above and recalling that each state $|P^{\bf n} \cO\rangle$ is an eigenstate of $L^2$ with the same eigenvalue $C_2(\D,\ell)$, we arrive at the Casimir equation
\e{32f}{(L^1_{AB}+L^2_{AB})^2W_{\D,\ell}(x_i) = C_2(\D,\ell) W_{\D,\ell}(x_i)~.}
One can take this second-order differential equation, plus the corresponding one with $1,2\leftrightarrow 3,4$, supplemented with appropriate boundary conditions, as one's definition of $W_{\D,\ell}$ \cite{SimmonsDuffin:2012uy}. Regarding boundary conditions, it is sufficient to require that $W_{\D,\ell}$ have the correct leading behavior in the $x_2\to x_1$ and $x_4\to x_3$ limits. The correct behavior in both limits is dictated by the fact that the contribution to $W_{\D,\ell}$ of the primary $\cO$ dominates that of its descendants since those enter the OPE with higher powers of $x_{12}$ and $x_{34}$.

We will prove that geodesic Witten diagrams $\sW_{\D,0}$ are indeed proportional to conformal partial waves $W_{\D,0}$ by showing that $\sW_{\D,0}$ satisfies the Casimir equation \eqr{32f} and has the correct behavior in the $x_2\to x_1$ and $x_4\to x_3$ limits. The proof is very transparent in the embedding space formalism, which we proceed now to introduce.

\subsubsection{Embedding space}\label{iii2ii}
The embedding space formalism has been reviewed in e.g. \cite{Rychkov:lectures,Costa:2014kfa,SimmonsDuffin:2012uy}. The idea is to embed the $d$-dimensional CFT and the $d+1$ dimensional AdS on which lives the geodesic Witten diagram both into $d+2$ dimensional Minkowski space. We give this embedding space the metric
\e{32g}{ds^2 = -(dY^{-1})^2 + (dY^0)^2 + \sum_{i=1}^d (dY^i)^2~.}
The CFT will live on the projective null cone of embedding space, which is the Lorentz-invariant $d$-dimensional space defined as the set of nonzero null vectors $X$ with scalar multiples identified: $X\equiv aX$. We will use null vectors $X$ to represent points in the projective null cone with the understanding that $X$ and $aX$ signify the same point. The plane $\R^d$ can be mapped into the projective null cone via
\e{32h}{X^+(x) = a|x|^2, \quad X^-(x) = a, \quad X^i(x) = ax^i}
where we have introduced light cone coordinates $X^{\pm}=X^{-1}\pm X^0$. Of course, any nonzero choice of the parameter $a$ defines the same map.

Conformal transformations on the plane are implemented by Lorentz transformations in embedding space. As a specific example, we may consider a boost in the $0$ direction with rapidity $\lambda$. This leaves the $X^i$ coordinates unchanged, and transforms $X^{\pm}$ according to
\e{32i}{X^{+} \to e^{\lambda} X^{+}, \quad X^- \to e^{-\lambda}X^-~.}
A point $X(x)=(|x|^2,1,x^i)$ gets mapped to $(e^{\lambda}|x^2|,e^{-\lambda},x^i)$ which is projectively equivalent to $X(e^{\lambda}x^i)$. Thus boosts in the $0$ direction of embedding space induce dilatations in the plane.

Any field $\hat{\cO}$ on the null cone defines a field $\cO$ on the plane via restriction: $\cO(x) \equiv \hat{\cO}(X(x))$. Since $\hat{\cO}$ is a scalar field in embedding space, the ${\rm{SO}}(d+1,1)$ generators act on it as
\e{32j}{[L_{AB},\hat{\cO}(X)] = (X_A\p_B-X_B\p_A)\hat{\cO}(X) ~.}
The induced transformation law for $\cO$ is the correct one for a primary of dimension $\D$ if and only if $\hat{\cO}$ satisfies the homogeneity condition
\e{32k}{\hat{\cO}(aX) = a^{-\D}\hat{\cO}(X)~.}
Thus in the embedding space formalism a primary scalar field $\cO(x)$ of dimension $\Delta$ is represented by a field $\hat{\cO}(X)$ satisfying \eqr{32k}. Below, we drop the hats on embedding space fields. It should be clear from a field's argument whether it lives on the null cone (as $\cO(X)$) or on the plane (as $\cO(x)$). Capital letters will always denote points in embedding space.

Meanwhile, AdS$_{d+1}$ admits an embedding into $d+2$ dimensional Minkowski space, as the hyperboloid $Y^2 = -1$. Poincare coordinates $(u,x^i)$ can be defined on AdS via
\e{32l}{Y^+ = \frac{u^2 + |x|^2}u,\quad Y^- = \frac{1}u,\quad Y^i = \frac{x^i}u~.}
The induced metric for these coordinates is the standard one, \eqr{22a}.

The AdS hyperboloid sits inside the null cone and asymptotes toward it. As one takes $u\to 0$, the image of a point $(u,x^i)$ in AdS approaches $(Y^+,Y^-,Y^i) = u^{-1}(|x^2|,1,x^i)$ which is projectively equivalent to $X(x^i)$. In this way, the image on the projective null cone of the point $x^i\in\R^d$ marks the limit $u\to 0$ of the embedding space image of a bulk point $(u,x^i)$.

Isometries of AdS are implemented by embedding space Lorentz transformations, and so are generated by
\e{32n}{L_{AB} = Y_A\p_B - Y_B\p_A~.}
The Casimir operator $L^2 = {1\over 2}(Y_A\p_B-Y_B\p_A)(Y^A\p^B-Y^B\p^A)$ is interior to the AdS slice $Y\cdot Y = -1$. That is, for $Y$ belonging to the AdS slice, $L^2 f(Y)$ depends only on the values of $f$ on the slice. In fact, applied to scalar functions on AdS the operator $L^2$ is simply the negative of the Laplacian of AdS:
\e{32m}{L^2 f(Y) = -\nabla_Y^2 f(Y)~}
as long as $Y$ is on the AdS slice. This fact, which is not surprising given that $L^2$ is a second-order differential operator invariant under all the isometries of AdS, can be checked directly from \eqr{32n}.

\subsubsection{Geodesic Witten diagrams satisfy the Casimir equation}\label{iii2iii}
The geodesic Witten diagram ${\cal W}_{\D,0}(x_i)$ lifts to a function ${\cal W}_{\D,0}(X_i)$ on the null cone of embedding space via a lift of each of the four bulk-to-boundary propagators with the appropriate homogeneity condition
\e{32o}{
G_{b\p}(y,aX_i) = a^{-\D_i}G_{b\p}(y,X_i),\quad i=1,2,3,4~.
}
The geodesics in AdS connecting the boundary points $X_1$ to $X_2$ and $X_3$ to $X_4$ lift to curves in embedding space which can be parameterized by
\begin{align}
\begin{split}\label{32p}
Y(\lambda) &= {e^{-\lambda}X_1 + e^{\lambda}X_2\over \sqrt{-2X_1\cdot X_2}}\\
Y(\lambda') &= {e^{-\lambda'}X_3 + e^{\lambda'}X_4\over \sqrt{-2X_3\cdot X_4}}
\end{split}
\end{align}
The geodesic Witten diagram is 
\e{32q}{{\cal W}_{\Delta,0}(X_i) = \int_{\gamma_{34}}  F(X_1,X_2,Y(\lambda');\D)G_{b\p}(Y(\lambda'),X_3)G_{b\p}(Y(\lambda'),X_4)}
where we have isolated the part that depends on $X_1,X_2$:\footnote{The fact that the bulk-to-bulk propagator satisfies the Laplace equation was used to similar effect in \cite{D'Hoker:1999ni}. In particular, \cite{D'Hoker:1999ni} defines a quantity $A(y', x_1,x_2)$ that is similar to $F(X_1,X_2,Y';\D)$, except that the vertex is integrated over all of AdS instead of along a geodesic.}
\e{32r}{F(X_1,X_2,Y';\D) = \int_{\gamma_{12}}
G_{b\p}(Y(\lambda),X_1) G_{b\p}(Y(\lambda),X_2) G_{bb}(Y(\lambda),Y';\D)~.}
$F(X_1,X_2,Y';\D)$ is the lift to embedding space of $\vphi^{12}_{\D}(y)$ defined in \eqr{31aa}. The bulk arguments of the bulk-to-boundary propagators have been promoted from points $y$ in the bulk to points $Y$ in embedding space. Although the propagators are defined only on the AdS slice, there is no ambiguity because $Y(\lambda)$ and $Y'(\lambda')$ always lie in the AdS slice.

The function $F(X_1,X_2,Y';\D)$ is manifestly invariant under simultaneous ${\rm{SO}}(d+1,1)$ rotations of $X_1,X_2,Y'$, and therefore it is annihilated by $(L^{1}+L^{2}+L^{Y'})_{AB}$. This means
\e{32s}{(L^1_{AB}+L^2_{AB})F(X_1,X_2,Y';\D)=-L^{Y'}_{AB}F(X_1,X_2,Y';\D)}
and so (since of course $L^1_{AB}$ commutes with $L^{Y'}_{AB}$)
\e{32t}{(L^1_{AB} + L^2_{AB})^2F(X_1,X_2,Y';\D) = (L^{Y'})^2 F(X_1,X_2,Y';\D)~.}
Recall that $(L^{Y'})^2$ is $-\nabla^2_{Y'}$. The function $F(X_1,X_2,Y';\D)$, which depends on $Y'$ via the bulk-to-bulk propagator $G_{bb}(Y(\lambda),Y';\Delta)$, is an eigenfunction of $-\nabla^2_{Y'}$ with eigenvalue $-\Delta(\Delta-d)$. Thus we conclude that $F(X_1,X_2,Y';\D)$ is an eigenfunction of $(L^1_{AB} + L^2_{AB})^2$ with eigenvalue $C_2(\D,0)$, and therefore that
\e{32u}{(L^1_{AB} + L^2_{AB})^2\cW_{\D,0}(X_i) = C_2(\D,0)\cW_{\D,0}(X_i)~. }
Note that agreement does not hinge on what the actual eigenvalue is: it is guaranteed by the fact that the bulk-to-bulk propagator and the conformal partial wave furnish the same highest weight representation of SO($d+1,1$).

Furthermore, the behavior in the limit $x_2\to x_1$ of the bulk-to-boundary and bulk-to-bulk propagators guarantees the geodesic Witten diagram to have the power-law behavior $\cW_{\D,0}(x_i)\to (\text{constant})\times|x_{12}|^{\D-\D_1-\D_2}$ in that limit, and similarly in the $x_4\to x_3$ limit. This proves $\cW_{\D,0}$ is equal to the conformal partial wave $W_{\D,0}$ up to a constant factor.

Looking back at the proof, we can see why the bilocal function integrated between the geodesics had to be precisely the bulk-to-bulk propagator $G_{bb}(y,y';\D)$. To get \eqr{32u} we needed that function to be the appropriate eigenfunction of the Laplacian, and to get the correct limiting behavior we needed it to be the eigenfunction with normalizable boundary conditions at infinity.  It also crucial that the vertices be integrated over geodesics rather than arbitrary curves or over all of AdS.   A non-geodesic curve would introduce extra data to specify the curve, which would not be conformally invariant.  Integrating the vertices over all of AdS (which would give the full Witten diagram) allows  $y$ and $y'$ to collide, but the bulk-to-bulk propagator acted on by the wave operator picks up a source contribution when $y=y'$, hence the diagram would not be an eigenfunction of the Casimir operator in this case; indeed we know that it is a sum of eigenfunctions with different eigenvalues.

\subsec{Comments}
We close this section with a few comments.

\subsubsec{Geodesic versus ordinary Witten diagrams}

A natural question is {\it why}, intuitively, a relation like \eqr{3c} is true. Let us offer two motivational remarks.

The first is that there are two ways to integrate a bulk point while preserving conformal invariance. One is over all of AdS, which defines a Witten diagram, while the other is over a geodesic. The latter is clearly over a smaller range, which makes it seem at least plausible that it represents a conformal partial wave rather than a full correlator. Indeed, the only obvious conformally invariant objects that appear in four-point functions are the correlator itself, and the conformal partial waves.

The second is a heuristic ``derivation'' starting from the exchange Witten diagram, $\A_4^{\rm Exch}$. Consider taking the following limit of heavy external operators,
\e{34a}{\D_{1,2,3,4} \rar\infty~, \quad \D_{12}, \D_{34}~\text{fixed}~.}
As reviewed in Section \ref{ii} and computed in the next section, the full diagram equals a single trace exchange of $\Oc$, plus infinite towers of double trace exchanges of $[\Oc_1\Oc_2]_{m,0}$ and $[\Oc_3\Oc_4]_{n,0}$.  On the CFT side, the double-trace exchanges are exponentially smaller in this limit than that of the single-trace exchange, simply because the conformal partial waves decay exponentially as the internal operator dimension tends to infinity.   So the Witten diagram reduces to the single-trace block in the limit. On the bulk side, the heavy limit restricts the cubic vertices to lie on geodesics, so $\A_4^{\rm Exch}$ reduces to $\cW_{\D,0}$, the geodesic Witten diagram. This establishes equality  between $\cW_{\D,0}$ and $W_{\D,0}$ in the limit \eqr{34a}.  To complete the argument we need to use the fact that the conformal block $G_{\D,0}$ only depends on $\Delta_i$ through $\Delta_{12}$ and $\Delta_{34}$, as can be seen from the recursion relations in \cite{Dolan:2011dv}.   Furthermore, $G_{\D,0}$ and $W_{\D,0}$ only differ by a prefactor which has exponents linear in $\Delta_i$ (a form which is invariant as $\Delta_i\rightarrow \infty$); see (\ref{21db}).   Using these two facts, it follows that if $\cW_{\D,0}$ and $W_{\D,0}$ agree in the regime \eqr{34a}, then they agree for all values of $\Delta_i$ and $\Delta$.

Note that the geodesic restriction ensures that a cut down the middle of the diagram crosses only the internal line, representing the CFT primary; contrast this with the exchange Witten diagram, where integration over all of AdS ensures that the cut will sometimes cross two external lines, representing the (infinite towers of) double-trace operators.

\subsubsec{Simplification of propagators and blocks}
In even $d$, CFT$_d$ scalar conformal blocks can be resummed into hypergeometric functions. An apparently unrelated simplification occurs for AdS$_{d+1}$ scalar bulk-to-bulk propagators, which are rational functions of $S\equiv e^{-2 \s(y,y')}$ rather than hypergeometric. From \eqr{22b}, the even $d$ propagators are, at low $d$,
\es{gbb3}{d=2:&\quad {G_{bb}(y,y';\D)}=S^{\D/2}\,{1\over 1-S}\\
d=4:&\quad {G_{bb}(y,y';\D)}=S^{\D/2}\,{(3-\D)S+(\D-1)\over (\D-1)(1-S)^3}\\
d=6:&\quad {G_{bb}(y,y';\D)}=S^{\D/2}\,\frac{(5-\D) (\D-4)S^2+2 (\D-5) (\D-1)S-(\D-2) (\D-1)}{(\D-2) (\D-1) (S-1)^5}~.}
The geodesic representation of the scalar conformal blocks reveals that these simplifications have a common origin. Conversely, the lack of simplification of the propagator in odd $d$ gives a new perspective on why generic odd $d$ conformal blocks cannot be reduced to special functions.

\subsubsec{Relation to Mellin space}
It is worth noting that the spin-$\ell$ conformal block has a Mellin representation with exponential dependence on the Mellin parameter: up to normalization \cite{Fitzpatrick:2011hu},
\e{34b}{G_{\D,\ell}(s,t) = e^{\pi i({d\over 2}-\D)}\left(e^{\pi i(t+\D-d)}-1\right){\Gamma\left({\D-\ell-t\over 2}\right)\Gamma\left({2d-\D-\ell-t\over 2}\right)\over \Gamma\left({\D_1+\D_2-t\over 2}\right)\Gamma\left({\D_3+\D_4-t\over 2}\right)}P_{\D,\ell}(s,t)}
where $P_{\D,\ell}(s,t)$ is a degree-$\ell$ Mack polynomial. (In the scalar case, $\ell=0$.) It has been argued that for holographic CFTs with a gap, the Mellin amplitudes for their correlators are polynomially bounded at large $s,t$. It is interesting that despite its exponential growth at large $t$, the Mellin representation of a conformal block does have a semiclassical AdS description.

In \cite{1208.0337}, it was argued that starting with \eqr{34b}, one recovers the Mellin amplitude for the {\it full} spin-$\ell$ exchange Witten diagram by writing it as a sum over its poles and dropping all other contributions.\footnote{This is true up to polynomial contributions from contact diagrams.} Evidently, this is the Mellin transform, so to speak, of the liberation of bulk vertices from the geodesics to all of AdS.

\sec{The conformal block decomposition of scalar Witten diagrams}\label{iv}

We begin our treatment with the technically simplest case: tree-level four-point functions in AdS involving only scalar fields. All of the key steps will be visible in the decomposition of the four-point contact diagram, out of which the geometric representation of the scalar conformal block will naturally emerge. We then move on to the exchange diagram and, in the next section, to fields with spin.

\begin{figure}[t!]
   \begin{center}
 \includegraphics[width = \textwidth]{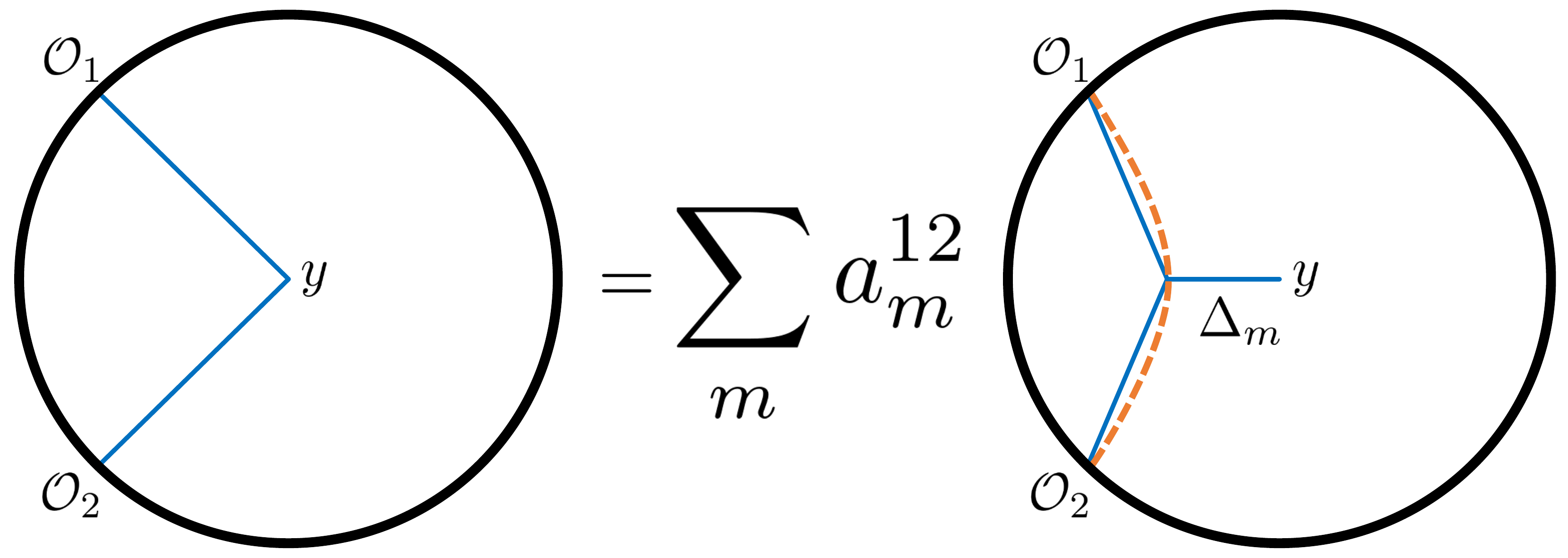}
 \caption{The identity \eqr{41a} obeyed by AdS scalar propagators. The internal line represents  bulk-to-bulk propagator for a scalar field of mass $m^2=\D_m(\D_m-d)$. $a^{12}_m$ and $\D_m$ are defined in \eqr{41b} and \eqr{41c}, respectively.}
 \label{f3}
 \end{center}
 \end{figure}

\subsec{An AdS propagator identity}\label{iv1}
The main technical tool that we will employ is an identity obeyed by AdS bulk-to-boundary propagators. Consider two scalar fields dual to gauge-invariant scalar operators $\Oc_1, \Oc_2$ of conformal dimensions $\D_{1},\D_2$, respectively. Now consider a product of their bulk-to-boundary propagators, from points $x_1$ and $x_2$ on the boundary to the same point $y$ in the bulk. Then the following identity holds:
\e{41a}{G_{b\p}(y, x_1)G_{b\p}(y,x_2) = \sum_{m=0}^{\infty}a^{12}_m\,\varphi^{12}_{\D_m}(y)}
where $\varphi^{12}_{\D_m}(y)$
is the field solution defined in \eqr{31aa}. The bulk-to-bulk propagator $G_{bb}(y(\l),y;\D_m)$, running from the geodesic to the original bulk point $y$, is for a scalar field with mass $m_m^2=\D_m(\D_m-d)$, where
\e{41b}{\D_m = \D_1+\D_2+2m~.}
The $a^{12}_m$ are coefficient functions of $\D_1, \D_2$ and $d$:
\e{41c}{a^{12}_m = {1\over \b_{\D_m 12}}{(-1)^m\over m!}{(\D_1)_m(\D_2)_m\over \left(\D_1+\D_2+m-{d\over 2}\right)_m}~.}
This identity is depicted in Figure \ref{f3}.

In words, the original bilinear is equal to an infinite sum of three-point vertices integrated over the geodesic $\g_{12}$, for fields of varying masses $m_m^2 = \D_m(\D_m-d)$. To prove this, we work in global AdS with $t_1\rar -\infty, t_2\rar+\infty$, whereupon $\g_{12}$ becomes a worldline at $\rho=0$. We already solved for $\varphi^{12}_{\D_m}(y)$ in \eqr{31h}. Plugging that solution into \eqr{41a}, we must solve
\e{41d}{(\cos\rho)^{\D_1+\D_2} = \sum_{m=0}^{\infty} a^{12}_m \b_{\D_m 12}(\cos\rho)^{\D_m}{}_2F_1\left({\D_m+\D_{12}\over 2},{\D_m-\D_{12}\over 2};\D_m-{d-2\over 2};\cos^2\rho\right)~.
}
Expanding as a power series in $\cos^2\rho$, the unique solution is given by $\D_m$ in \eqr{41b} and $a^{12}_m$ in \eqr{41c}.

The identity \eqr{41a} is suggestive of a bulk operator product expansion, where the propagation of two boundary fields to the same bulk point is replaced by an infinite sum over field solutions. Note that the dimensions $\D_m$ are those of the scalar double-trace operators $[\Oc_1\Oc_2]_{m,0}$ at leading order in $1/N$. As we now show, this fact ensures that the decomposition of a given Witten diagram involving $G_{b\p}(y,x_1)G_{b\p}(y,x_2)$ includes the exchange of $[\Oc_1\Oc_2]_{m,0}$, consistent with the generalized free field content of the dual CFT.

\begin{figure}[t!]
   \begin{center}
~~~  \includegraphics[width = .44\textwidth]{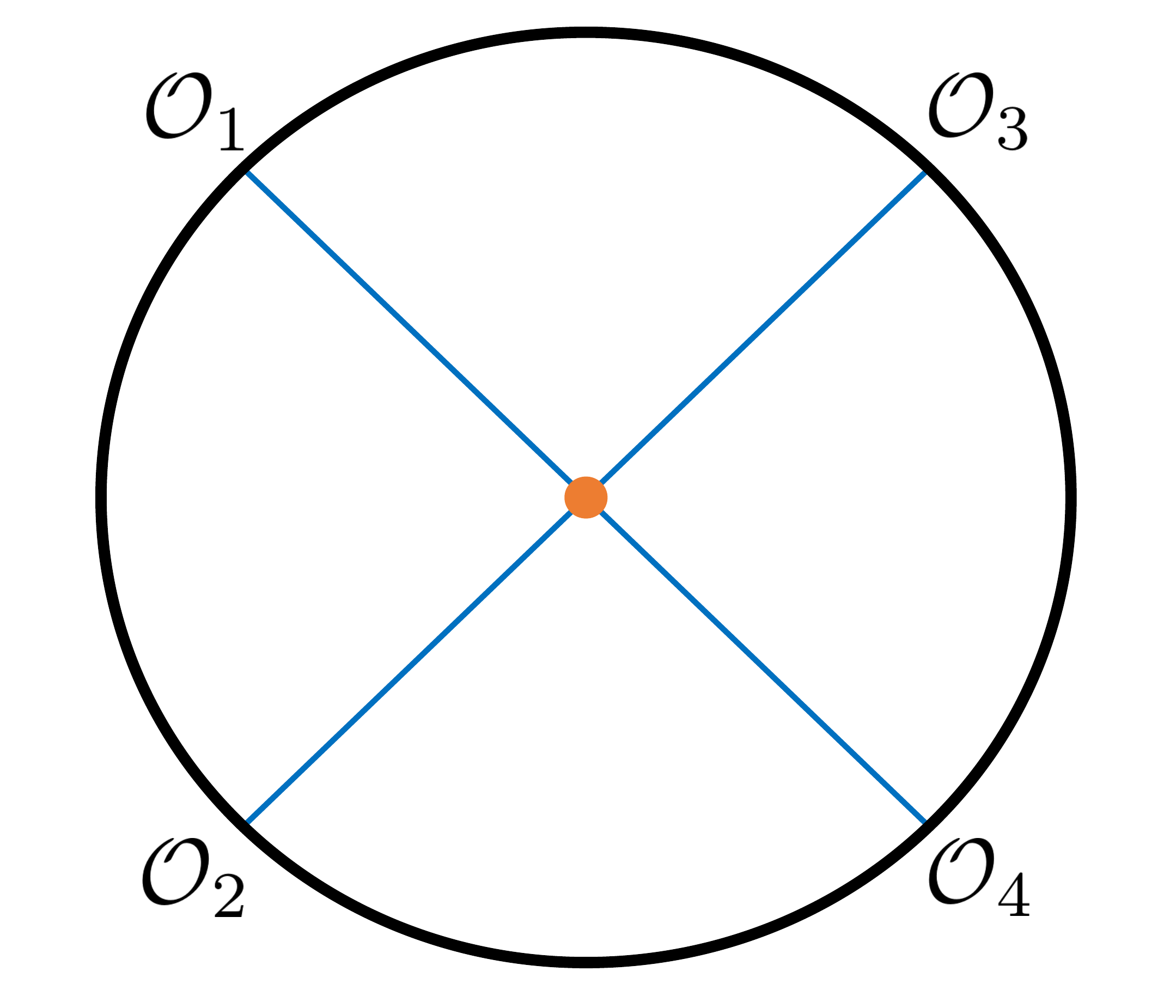}

  \includegraphics[width = .625\textwidth]{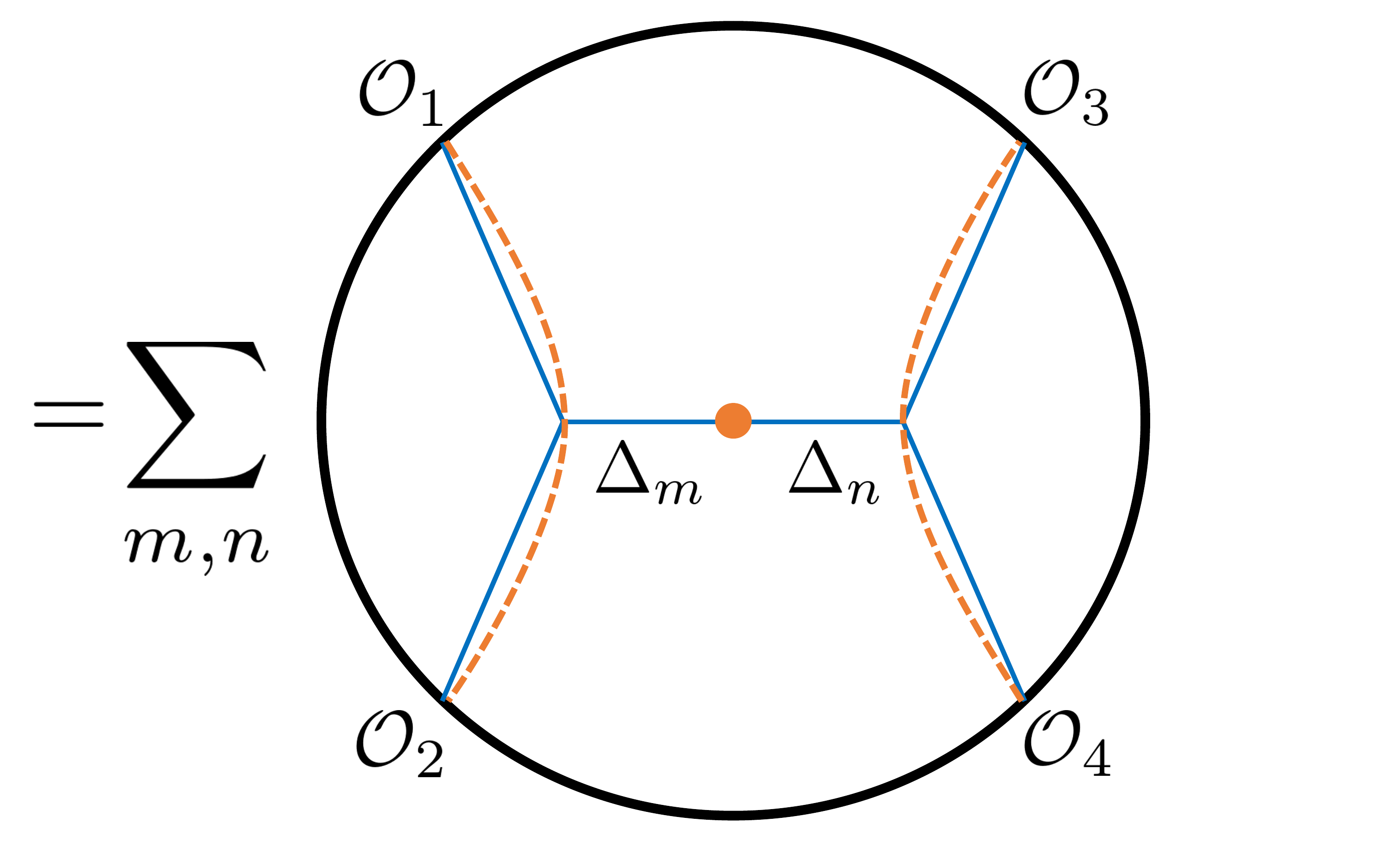}%

  \includegraphics[width = .98\textwidth]{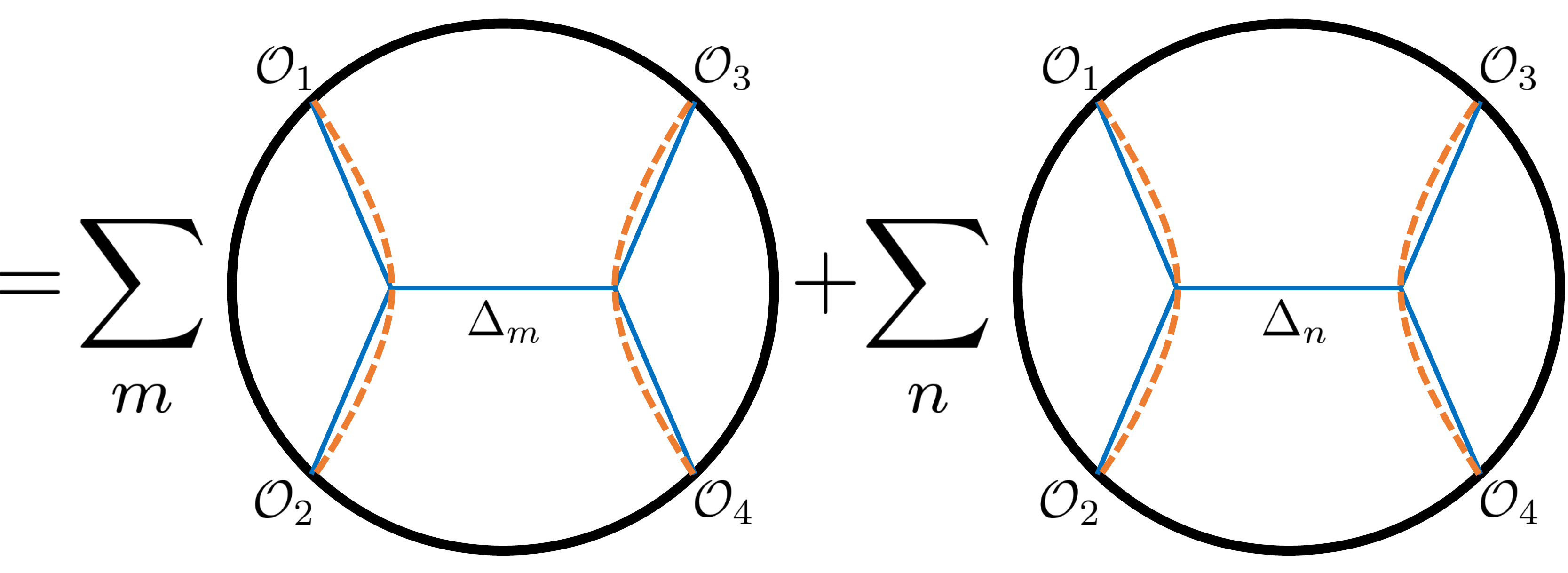}%
 \caption{The decomposition of a four-point scalar contact diagram into conformal partial waves disguised as geodesic Witten diagrams. Passage to the second line uses \eqr{41f}, and passage to the last line uses \eqr{41g}. The last line captures the infinite set of CFT exchanges of the double-trace operators $[\Oc_1\Oc_2]_{m,0}$ and $[\Oc_3\Oc_4]_{n,0}$. We have suppressed OPE coefficients; the exact result is in equation \eqr{41i}. }
\label{f4}
  \end{center}
 \end{figure}

\subsec{Four-point contact diagram}
We want to compute the four-point scalar contact diagram \eqr{22g}, for all operator dimensions $\D_i$ generic. We reproduce the integral here:
\e{41e}{D_{\D_1\D_2\D_3\D_4}(x_i) = \int_y G_{b\p}(y,x_1)\,G_{b\p}(y,x_2)\,G_{b\p}(y,x_3)\,G_{b\p}(y,x_4)~.}
A helpful pictorial representation of the following calculation is given in Figure \ref{f4}.

Using our geodesic toolkit, the evaluation of this diagram is essentially trivial. First, we use the identity \eqr{41a} on the pairs $(12)$ and $(34)$. This yields
\es{41f}{D_{\D_1\D_2\D_3\D_4}(x_i) =  \sum_{m,n}a^{12}_ma^{34}_n&\int_{\g_{12}}\int_{\g_{34}}G_{b\p}(y(\l),x_1)\,G_{b\p}(y(\l),x_2)\\
\times &\int_y G_{bb}(y(\l),y;\D_m)\,G_{bb}(y,y(\l');\D_n)\\
\times &\,G_{b\p}(y(\l'),x_3)\,G_{b\p}(y(\l'),x_4)~.}
Next, we use
\e{prop}{G_{bb}(y,y';\D) = \Big\langle y\Big|{1\over \nabla^2-m^2}\Big|y'\Big\rangle}
to represent the product of bulk-to-bulk propagators integrated over the common bulk point $y$ as
\es{41g}{\int_y G_{bb}(y(\l),y;\D_m)\,G_{bb}(y,y(\l');\D_n)&= \,{G_{bb}(y(\l),y(\l');\D_m)-G_{bb}(y(\l),y(\l');\D_n)\over m_m^2-m_n^2}}
where we used completeness, $\int_y |y\rangle \langle y|=1$. The integrated product is thus replaced by a difference of unintegrated propagators from $\g_{12}$ to $\g_{34}$. This leaves us with
\es{41h}{&D_{\D_1\D_2\D_3\D_4}(x_i) =  \sum_{m,n}{a^{12}_ma^{34}_n\over m_m^2-m_n^2}\times\\\Bigg(&\int_{\g_{12}}\int_{\g_{34}}G_{b\p}(y(\l),x_1)G_{b\p}(y(\l),x_2)\times G_{bb}(y(\l),y(\l');\D_m)\times \,G_{b\p}(y(\l'),x_3)G_{b\p}(y(\l'),x_4)\\-&\int_{\g_{12}}\int_{\g_{34}}G_{b\p}(y(\l),x_1)G_{b\p}(y(\l),x_2)\times G_{bb}(y(\l),y(\l');\D_n)\times \,G_{b\p}(y(\l'),x_3)G_{b\p}(y(\l'),x_4)\Bigg)~.}
But from \eqr{gwitt}, we now recognize the last two lines as conformal partial waves\text{!} Thus, we have
\e{41i}{D_{\D_1\D_2\D_3\D_4}(x_i) = \sum_{m,n}{a^{12}_ma^{34}_n\over m_m^2-m_n^2}\left(\cW_{\D_m,0}(x_i) - \cW_{\D_n,0}(x_i)\right)~.}
This is the final result. In the CFT notation of section \ref{ii}, we write this as a pair of single sums over double-trace conformal partial waves,
\e{41j}{D_{\D_1\D_2\D_3\D_4}(x_i) = \sum_{m} P^{(12)}_1(m,0)\, W_{\D_m,0}(x_i) + \sum_{n} P^{(34)}_1(n,0)\,W_{\D_n,0}(x_i)}
with squared OPE coefficients
\es{41k}{P_1^{(12)}(m,0) &= \left(\b_{\D_m 12}\,a^{12}_m\right) \left(\b_{\D_m 34}\sum_n{a^{34}_n\over m_m^2-m_n^2}\right)\\
P_1^{(34)}(n,0) &=\left(\b_{\D_n 34}\, a^{34}_n \right)\left( \b_{\D_n 12}\sum_m{a^{12}_m\over m_n^2-m_m^2}\right)}
where $m^2=\D(\D-d)$ as always. The structure of the answer is manifestly consistent with CFT expectations: only double-trace operators $[\Oc_1\Oc_2]_{m,0}$ and $[\Oc_3\Oc_4]_{n,0}$ are exchanged.

We will analyze this result more closely after computing the exchange diagram.

\subsec{Four-point exchange diagram}

\begin{figure}[t!]
   \begin{center}
 \includegraphics[width = \textwidth]{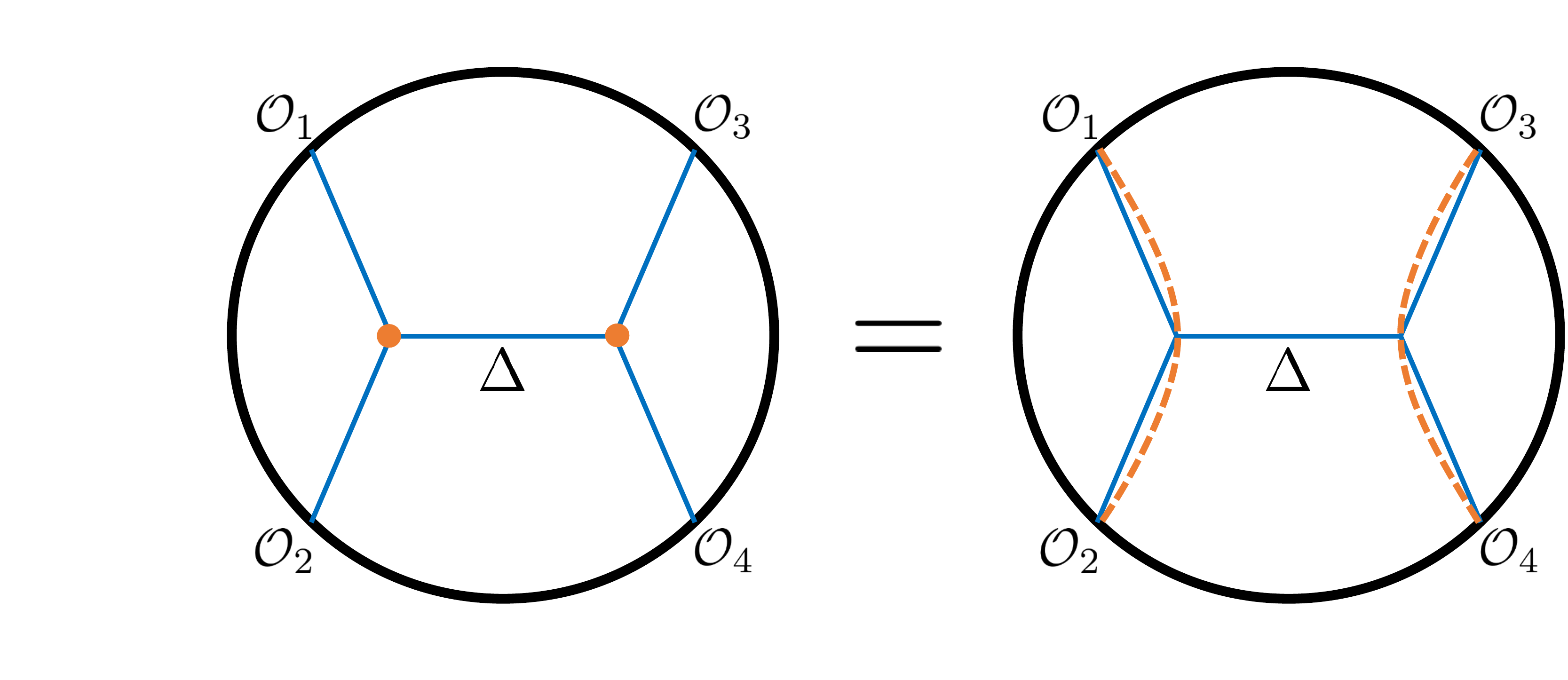}
 \includegraphics[width = \textwidth]{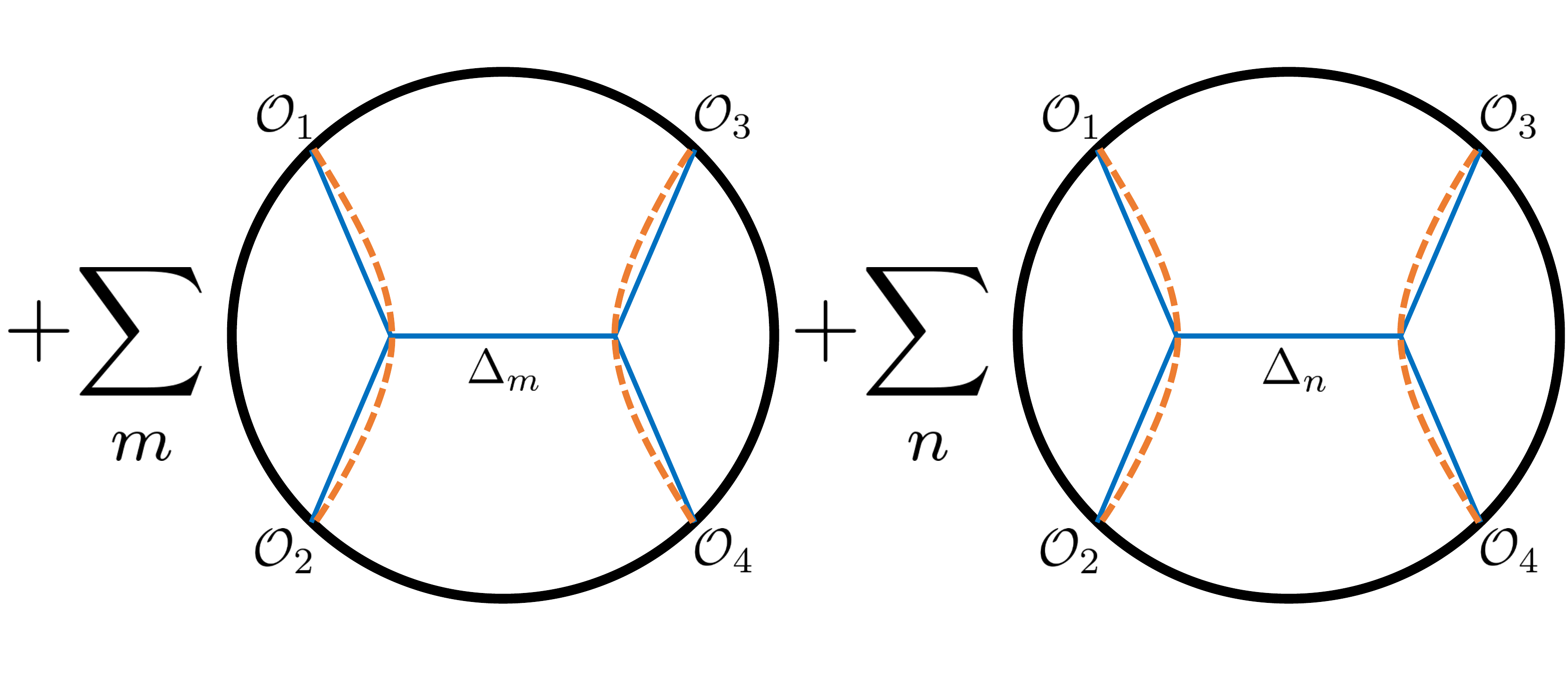}%
 \caption{The decomposition of a four-point scalar exchange diagram (upper left) into conformal partial waves, for an exchanged scalar $\phi$ of mass $m^2=\D(\D-d)$. We have skipped the intermediate steps, which are nearly identical to those of the contact diagram. The term in the upper right captures the single-trace exchange of the scalar operator dual to $\phi$. The second line captures the infinite set of CFT exchanges of the double-trace operators $[\Oc_1\Oc_2]_{m,0}$ and $[\Oc_3\Oc_4]_{n,0}$. We have suppressed OPE coefficients; the exact result is in equations \eqr{43d}--\eqr{43e}.}
\label{f5}
\end{center}
 \end{figure}

Turning to the scalar exchange diagram, we reap the real benefits of this approach: unlike an approach based on brute force integration, this case is no harder than the contact diagram. A pictorial representation of the final result is given in Figure \ref{f5}.

We take all external dimensions $\D_i$, and the internal dimension $\D$, to be generic. The diagram is computed as
\e{43a}{{\cal A}_4^{\rm Exch}(x_i) = \int_{y} \int_{y'}G_{b\p}(y,x_1)G_{b\p}(y,x_2)\times G_{bb}(y,y';\D)\times G_{b\p}(y',x_3)G_{b\p}(y',x_4)~.}

Expanding in the $s$-channel (12)-(34), the algorithm is the same as the contact case. First, use \eqr{41a} twice to get
\es{43b}{{\cal A}_4^{\rm Exch}(x_i)= \sum_{m,n}a^{12}_ma^{34}_n&\int_{\g_{12}}\int_{\g_{34}}G_{b\p}(y(\l),x_1)G_{b\p}(y(\l),x_2)\\
\times &\int_{y}\int_{y'} G_{bb}(y(\l),y;\D_m)G_{bb}(y,y';\D)G_{bb}(y',y(\l');\D_n)\\
\times &\,G_{b\p}(y(\l'),x_3)G_{b\p}(y(\l'),x_4)~.}
This is of the same form as the contact diagram, only we have three bulk-to-bulk propagators and two integrations. We again use \eqr{prop} to turn the second line into a sum over terms with a single bulk-to-bulk propagator:
\es{43c}{&\int_{y}\int_{y'} G_{bb}(y(\l),y;\D_m)G_{bb}(y,y';\D)G_{bb}(y',y(\l');\D_n)\\
=~& {G_{bb}(y(\l),y(\l');\D_m)\over (m_m^2-m_{\D}^2)(m_m^2-m_n^2)}+{G_{bb}(y(\l),y(\l');\D)\over (m_{\D}^2-m_m^2)(m_{\D}^2-m_n^2)}+{G_{bb}(y(\l),y(\l');\D_n)\over (m_n^2-m_m^2)(m_n^2-m_{\D}^2)}~.}
 Recognizing the remaining integrals as conformal partial waves, we reach our final result:
\e{43d}{{\cal A}_4^{\rm Exch}(x_i) = C_{12\D}C^{\D}_{~~34}W_{\D,0}(x_i) + \sum_{m} P^{(12)}_1(m,0)\, W_{\D_m,0}(x_i) + \sum_{n} P^{(34)}_1(n,0)\,W_{\D_n,0}(x_i)}
where
\es{43e}{C_{12\D}C^{\D}_{~~34} &=\left(\beta_{\D 12}\sum_m {a^{12}_m\over m_{\D}^2-m_m^2}\right)\left( \beta_{\D 34}\sum_n {a^{34}_n\over m_{\D}^2-m_n^2}\right)\\
P_1^{(12)}(m,0) &=\left( \beta_{\D_m 12}\,{a^{12}_m\over m_m^2 - m_{\D}^2}\right)\left(\beta_{\D_m 34}\sum_n {a^{34}_n\over m_m^2-m_n^2}\right)\\
P_1^{(34)}(n,0) &=\left(\beta_{\D_n 34}\,{a^{34}_n\over m_n^2-m_{\D}^2}\right)\left( \beta_{\D_n 12}\sum_m{a^{12}_m\over m_n^2 - m_m^2}\right) ~.}
Its structure is precisely as required by AdS/CFT: in addition to the double-trace exchanges of $[\Oc_1\Oc_2]_{m,0}$ and $[\Oc_3\Oc_4]_{n,0}$, there is a single-trace exchange of the operator dual to the exchanged field in the bulk of dimension $\D$.

Comparing \eqr{43e} to \eqr{41k}, we can immediately read off a new identity relating the double-trace OPE coefficients of the contact and exchange diagrams:
\e{43f}{{P_1^{(12)}(m,0)\big|_{\rm Contact}\over P_1^{(12)}(m,0)\big|_{\rm Exch}} = m_m^2-m_{\D}^2}
and likewise for $P_1^{(34)}(n,0)$. This is quite simple. One can quickly check this against the $d=4$ example in Appendix B of \cite{ElShowk:2011ag}.

\subsec{Further analysis}

\subsubsec{OPE factorization}

Notice that the squared OPE coefficients in \eqr{43e} and \eqr{41k} factorize naturally into terms associated with the (12) and (34) channels. To emphasize this, it is useful to define\footnote{We observe a likeness between $\a^{34}_s$ and calculations in \cite{1007.2412} of $\ell=0$ double-trace anomalous dimensions due to heavy operator exchange; see Section 4.3 therein. It is not immediately clear to us whether there is a deeper statement to be made.}
\e{44a}{\a^{34}_s \equiv \sum_n{a^{34}_n\over m_s^2-m_n^2}}
for some mass squared $m_s^2=\D_s(\D_s-d)$, and similarly for $\a^{12}_s$. This allows us to write the Witten diagrams in a tidy form as
\e{44b}{D_{\D_1\D_2\D_3\D_4}(x_i) = \sum_m a^{12}_m \a^{34}_m\, \cW_{\D_m,0}(x_i) +\sum_n \a^{12}_n a^{34}_n\, \cW_{\D_n,0}(x_i)}
and
\e{44c}{{\cal A}_4^{\rm Exch}(x_i) = \a^{12}_{\D}\a^{34}_{\D}\,\cW_{\D,0}(x_i) + \sum_m {a^{12}_m\a^{34}_m\over m_m^2-m_{\D}^2}\, \,\cW_{\D_m,0}(x_i)+ \sum_n\,{ \a^{12}_n a^{34}_n\over m_n^2-m_{\D}^2}\, \cW_{\D_n,0}(x_i)~.}
For compactness in the above equations we have used $\cW_{\D,0}$ in place of $W_{\D,0}$. Recall that $\cW_{\D,0}$ is a rescaling of the standard conformal partial wave, $\cW_{\D,0}(x_i) = \b_{\D 12}\b_{\D 34}W_{\D,0}(x_i)$. The coefficient relating $\cW_{\D,0}$ to $W_{\D,0}$ clearly factorizes.

Writing the OPE coefficients in terms of the coefficients $a^{12}_m, a^{34}_n$ and masses $m_m, m_n, m_{\D}$ makes their origin transparent. But the sum defining $\a^{34}_s$ can actually be performed, yielding
\es{44d}{\a^{34}_s &={\Gamma(\D_3+\D_4)\over \Gamma(\D_3)\Gamma(\D_4)} \left(F(\D_s,\D_3,\D_4)+F(d-\D_s,\D_3,\D_4)\right)}
where
\es{}{F(\D_s,\D_3,\D_4) &\equiv{1\over \left(\D_s-{d\over 2}\right)(\D_s-\D_3-\D_4)}\\&\quad\times{}_4F_3\Farg{{\D_3+\D_4\over 2},{\D_3+\D_4+1\over 2},{\D_3+\D_4-\D_s\over 2},\D_3+\D_4-{d\over 2}}{{\D_3+\D_4\over 2}-{d\over 4},{\D_3+\D_4\over 2}-{d-2\over 4},{\D_3+\D_4-\D_s+2\over 2}}{-1}~.}

\subsubsec{Recovering logarithmic singularities}
Recall from Section \ref{ii} that when the external operator dimensions obey $\D_1+\D_2-\D_3-\D_4 \in 2\mathbb{Z}$, logarithms appear in tree-level Witten diagrams due to anomalous dimensions of double-trace operators. In brute force calculation of the AdS integrals, these logarithms are extracted by isolating the relevant integration range. In Mellin space, they appear as double poles in the Mellin amplitude.

In the present approach, these logarithms fall out trivially as algebraic conditions. Considering the scalar four-point contact diagram written in the form \eqr{41i}, for instance, we see that terms for which $m_m^2=m_n^2$ give rise to derivatives of conformal blocks, and hence to logarithms. This is equivalent to the condition $\D_m=\D_n$ or $\D_m = d-\D_n$. Since $d\in\mathbb{Z}$, both of these are equivalent to $\D_1+\D_2-\D_3-\D_4 \in 2\mathbb{Z}$, which is precisely the integrality condition stated above. Identical structure is visible in \eqr{43c}: logarithms will appear when any of $m_m^2, m_n^2, m_{\D}^2$ coincide.

As an explicit example, let us consider $D_{\D\D\D\D}(x_i)$. Then \eqr{41j} can be split into $m\neq n$ and $m=n$ terms, the latter of which yield logarithms:
\es{44e}{D_{\D\D\D\D}(x_i) = \sum_{n=0}^{\infty}2 a^{\D\D}_n\left(\sum_{m\neq  n}{a^{\D\D}_m\over m_n^2-m_m^2}\right) \cW_{2\D+2n,0}(x_i)+ \left({(a^{\D\D}_n)^2\over \p_nm_n^2}\right)\p_n\cW_{2\D+2n,0}(x_i) ~.}
This takes the form of the $\ell=0$ terms in \eqr{22k}, with
\e{44f}{ P_1(n,0) = 2\b_{(2\D+2n)\,\D\D}^2 a^{\D\D}_n\left(\sum_{m\neq  n}{a^{\D\D}_m\over m_n^2-m_m^2}\right)+{(a^{\D\D}_n)^2\over \p_nm_n^2}\p_n\left(\b_{(2\D+2n)\,\D\D}^2\right)}
and
\e{44fa}{\half P_0(n,0) \g_1(n,0) = {(a^{\D\D}_n)^2\over \p_nm_n^2}\b_{(2\D+2n)\,\D\D}^2~.}

As an aside, we note the conjecture of \cite{Heemskerk:2009pn}, proven in \cite{Fitzpatrick:2011dm}, that
\e{44g}{P_1(n,\ell) = \half \p_n(P_0(n,\ell)\g_1(n,\ell))~.}
We have checked in several examples that this is obeyed by \eqr{44f}--\eqr{44fa}. It would be interesting to prove it using generalized hypergeometric identities.

\subsec{Taking stock}
We close this section with some perspective. Whereas traditional methods of computing Witten diagrams are technically involved and require explicit bulk integration \cite{D'Hoker:1998mz} and/or solution of differential equations \cite{D'Hoker:1999ni}, the present method skips these steps with a minimum of technical complexity. It is remarkable that for neither the contact nor exchange diagrams have we performed any integration: the integrals have instead been absorbed into sums over, and definitions of, conformal partial waves.

For the contact diagram/$D$-function, we have presented an efficient algorithm for its decomposition into spin-0 conformal blocks in position space. Specific cases of such decompositions have appeared in previous works \cite{ElShowk:2011ag,Heemskerk:2009pn}, although no systematic treatment had been given. Moreover, perhaps the main virtue of our approach is that exchange diagrams are no more difficult to evaluate than contact diagrams.

$D$-functions also appear elsewhere in CFT, including in weak coupling perturbation theory. For example, the four-point function of the {\bf 20'} operator in planar $\N=4$ SYM at weak coupling is given, at order $\l$, by \cite{hep-th/9811155}
\e{44l}{\langle \Oc_{\bf 20'}(x_1) \Oc_{\bf 20'}(x_2) \Oc_{\bf 20'}(x_3) \Oc_{\bf 20'}(x_4)\rangle\big|_{\l} \propto \overline D_{1111}(z,\zb)}
where $\overline D_{1111}(z, \zb)$ was defined in \eqr{22h}. The ubiquity of $D$-functions at weak coupling may be related to constraints of crossing symmetry in the neighborhood of free fixed points \cite{1506.04659}.

\sec{Spinning exchanges and conformal blocks}\label{v}

The OPE of two scalar primary operators yields not just other scalar primaries but also primaries transforming in symmetric traceless tensor representations of the Lorentz group.  We refer to such a rank-$\ell$ tensor as a spin-$\ell$ operator.   Thus, for the full conformal block decomposition of a correlator of scalar primaries we need to include blocks describing spin-$\ell$ exchange. The expression for such blocks as geodesic Witten diagrams turns out to be the natural extension of the scalar exchange case.  The exchanged operator is now described by a massive spin-$\ell$ field in the bulk, which couples via its pullback to the geodesics connecting the external operator insertion points. This was drawn in Figure \ref{f1}.

In this section we do the following.  We give a fairly complete account of the spin-$1$ case, showing how to decompose a Witten diagram involving the exchange of a massive vector field, and establishing that the geodesic diagrams reproduce known results for spin-1 conformal blocks.   We also give an explicit treatment of the spin-2 geodesic diagram, again checking that we reproduce known results for the spin-2 conformal blocks.   More generally, we use the conformal Casimir equation to prove that our construction yields the correct blocks for arbitrary $\ell$.

\subsection{Known results}

Conformal blocks with external scalars and internal spin-$\ell$ operators were studied in the early work of Ferrara et. al.  \cite{Ferrara:1971vh}.   They obtained expressions for these blocks as double integrals.  It is easy to verify that their form for the scalar exchange block precisely coincides with our geodesic Witten diagram expression (\ref{gwitt}).  We thus recognize the double integrals as integrals over pairs of geodesics.   Based on this, we expect agreement for general $\ell$, although we have not so far succeeded in showing this due to the somewhat complicated form for the general spin-$\ell$ bulk-to-bulk propagator \cite{Costa:2014kfa,Bekaert:2014cea}. Some more discussion is in section \ref{Fercomp}. We will instead use other arguments to establish the validity of our results.

Dolan and Osborn \cite{Dolan:2003hv} studied these blocks using the conformal Casimir equation.   Closed-form expressions in terms of hypergeometric functions were obtained in dimensions $d=2, 4, 6$.    For example, in $d=2$ we have
\es{twod}{
G_{\D,\ell}(z,\zb)&= |z|^{\Delta-\ell }\times \\
&\quad \Big[z^\ell {}_2F_1\left({\D-\Delta_{12}+\ell\over 2},{\D+\Delta_{34}+\ell\over 2},\D+\ell;z\right)\\& \quad\times {}_2F_1\left({\D-\Delta_{12}-\ell\over 2},{\D+\Delta_{34}-\ell\over 2},\D-\ell;\zb\right) + (z\leftrightarrow \zb) \Big]}
and in $d=4$ we have
\es{fourd}{G_{\D,\ell}(z,\zb)&= |z|^{\D-\ell }{1\over z-\zb} \times \\ &\quad\Big[ z^{\ell+1} {}_2F_1\left({\D-\Delta_{12}+\ell\over 2}, {\D+\Delta_{34}+\ell\over 2};\D+\ell;z\right)\\ & \quad\quad\times {}_2F_1\left({\D-\Delta_{12}-\ell\over 2}-1, {\D+\Delta_{34}-\ell\over 2}-1;\D-\ell-2;\zb\right)-(z\leftrightarrow \zb)\Big]}
The $d=6$ result is also available, taking the same general form, but it is more complicated.  Note that the $d=2$ result is actually a sum of two irreducible blocks, chosen so as to be even under parity.  The irreducible $d=2$ blocks factorize holomorphically, since the global conformal algebra splits up as  sl(2,$\bR$) $\oplus$ sl(2,$\bR$).  An intriguing fact is that the $d=4$ block  is expressed as a sum of two terms, each of which ``almost" factorizes holomorphically.    Results in arbitrary dimension are available in series form.

Since  the results of Dolan and Osborn are obtained as solutions of the conformal Casimir equation, and we will show that our geodesic Witten diagrams are solutions of the same equation with the same boundary conditions, this will constitute exact agreement.  Note, though, that the geodesic approach produces the solution in an integral representation. It is not obvious by inspection that these results agree with those in \cite{Dolan:2003hv}, but we will verify this in various cases to assuage any doubts that our general arguments are valid.  As noted above, in principle a more direct comparison is to the formulas of Ferrara et. al. \cite{Ferrara:1971vh}.

\subsection{Geodesic Witten diagrams with spin-$\ell$ exchange: generalities}

Consider a CFT$_d$ primary operator which carries scaling dimension $\D$ and transforms in the rank-$\ell$ symmetric traceless tensor representation of the (Euclidean) Lorentz group.   The AdS$_{d+1}$ bulk dual to such an operator is a symmetric traceless tensor field $h_{\mu_1 \ldots \mu_{\ell}}$ obeying the field equations
\es{spin_l_eqs}{& \nabla^2 h_{\mu_1 \ldots \mu_{\ell}} -  [\D(\D-d)-\ell]h_{\mu_1 \ldots \mu_{\ell}} =0~,\\
& \nabla^{\mu_1} h_{\mu_1 \ldots \mu_{\ell}} =0~.}
Our proposal is that the conformal partial wave $W_{\D,\ell}(x_i)$ is given by the same expression as in (\ref{gwitt}) except that now the bulk-to-bulk propagator is that of the spin-$\ell$ field pulled back to the geodesics. The latter defines the spin-$\ell$ version of the geodesic Witten diagram, $\cW_{\D,\ell}(x_i)$: its precise definition is
\es{gwitt2}{&{\cal W}_{\D,\ell}(x_i)\equiv\\& \int_{\g_{12}}\int_{\g_{34}}G_{b\p}(y(\l), x_1)G_{b\p}(y(\l),x_2)\times G_{bb}(y(\l),y(\l');\D,\ell)\times G_{b\p}(y(\l'),x_3)G_{b\p}(y(\l'),x_4)}
and $G_{bb}(y(\l),y(\l');\D,\ell)$ is the pulled-back spin-$\ell$ propagator,
\e{pbprop}{
G_{bb}(y(\l),y(\l');\D,\ell) \equiv [G_{bb}(y,y';\D)]_{\mu_1 \ldots \mu_{\ell}, \nu_1 \ldots \nu_{\ell}}{d y^{\mu_1} \over d\lambda} \ldots {d y^{\mu_\ell} \over d\lambda}{d y'^{\nu_1} \over d\lambda'} \ldots {d y'^{\nu_\ell} \over d\lambda'}\Big|_{y=y(\l),\, y'=y(\l')}~.}

To explicitly evaluate this we will use the same technique as in section \ref{iii1}. Namely, the integration over one geodesic can be expressed as a normalizable spin-$\ell$ solution of the equations (\ref{spin_l_eqs}) with a geodesic source.  Inserting this result, we obtain an expression for the geodesic Witten diagram as an integral over the remaining geodesic. If we call the above normalizable solution $h_{\nu_1 \ldots \nu_{\ell}}$, then the analog of (\ref{31a}) is
\be\label{genspin}
\cW_{\D,\ell}(x_i) = \int_{\gamma_{34}}  h_{\nu_1 \ldots \nu_{\ell}}(y(\lambda')){d y'^{\nu_1} \over d\lambda'} \ldots {d y'^{\nu_\ell} \over d\lambda'} G_{b\p}(y(\l'),x_3)G_{b\p}(y(\l'),x_4)~.
\ee

As in section \ref{iii1}, we will specifically compute
\es{cpw2}{\cW_{\D,\ell}(z,\zb)&\equiv {1\over C_{12\Oc}C^{\Oc}_{~~34}}\,\langle \Oc_1(\infty)\Oc_2(0)\,P_{\D,\ell}\,\Oc_3(1-z)\Oc_4(1)\rangle \\&= |z|^{-\D_3-\D_4}G_{\D,\ell}(z,\zb)}
now written in terms of $(z,\zb)$ instead of $(u,v)$ to facilitate easier comparison with \eqr{twod} and \eqr{fourd}. We recall that this reduces $\g_{12}$ to a straight line at the origin of global AdS. The form of $\gamma_{34}$ is given in (\ref{31n}), from which the pullback is computed using
\bea
 \cos^2 \rho\big|_{\g_{34}}& =& { 1\over 2\cosh\lambda'} { |z|^2\over  e^{-\lambda'}+|1-z|^2 e^{\lambda'} }~,\cr
e^{2t}\big|_{\g_{34}}&=&{2\cosh \lambda' \over e^{-\lambda'}+|1-z|^2e^{\lambda'} }~,\cr
e^{2i\phi}\big|_{\g_{34}}&=& {(1-z) e^{\lambda'}+e^{-\lambda'} \over (1-\zb) e^{\lambda'}+e^{-\lambda'} }~.
\eea
We also recall
\e{qqq}{G_{b\p}(y(\l'),1-z)G_{b\p}(y(\l'),1) = {e^{\D_{34}\l'}\over |z|^{\D_3+\D_4}}~.}
Carrying out this procedure for all dimensions $d$ at once presents no particular complications.  However, it does not seem easy to find the solution $ h_{\nu_1 \ldots \nu_{\ell}}$  for all $\ell$ at once.  For this reason, below we just consider the two simplest cases of $\ell=1, 2$, which suffice for illustrating the general procedure.

\subsec{Evaluation of geodesic Witten diagram: spin-1}\label{v3}

In the global AdS$_{d+1}$ metric
\be
ds^2 = {1\over \cos^2 \rho}(d\rho^2+ dt^2+\sin^2 \rho d\Omega_{d-1}^2)
\ee
we seek a normalizable solution of
\be
 \nabla^2 A_{\mu} -  [\D(\D-d)-1]A_{\mu} =0~,\quad \nabla^{\mu} A_{\mu} =0
\ee
which is spherically symmetric and has time dependence $e^{-\Delta_{12}t}$.   A suitable ansatz is
\be
A_\mu dx^\mu =   A_t(\rho,t) dt + A_\rho(\rho,t)d\rho~.
\ee
Assuming the time dependence $e^{-\Delta_{12}t}$, the divergence free condition implies
\be
\p_\rho\left(\tan^{d-1}\rho  A_\rho\right)-\Delta_{12}\tan^{d-1}\rho A_t=0
\ee
and the components of the wave equation are
\es{}{&{\cos^{d-1}\rho \over \sin^{d-1}\rho}\p_\rho \left({\sin^{d-1}\rho\over \cos^{d-3}\rho}\p_\rho A_t \right) +\left(\Delta_{12}^2 \cos^2\rho  -(\D-1)(\D-d+1)\right)A_t \\
&\quad-2\Delta_{12}\cos \rho \sin\rho A_\rho =0  \\
& {\cos^{d-1}\rho \over \sin^{d-1}\rho}\p_\rho \left({\sin^{d-1}\rho\over \cos^{d-3}\rho}\p_\rho A_\rho  \right)+\left(\Delta_{12}^2 \cos^2\rho  -{d-1\over \sin^2\rho}-(\D-1)(\D-d+1)\right)A_\rho \\
&\quad +2\Delta_{12} \cos\rho\sin\rho A_t =0~.}
The normalizable solution is
\bea\label{vecsol}
A_\rho &=& \Delta_{12} \sin \rho (\cos\rho)^{\D} {_2{F_1}}\left({\D+\Delta_{12}+1\over 2},{\D-\Delta_{12}+1\over 2},\D-{d-2\over 2};\cos^2\rho  \right)e^{-\Delta_{12}t}\cr
A_t &=& {1\over \Delta_{12}\tan^{d-1}\rho} \p_\rho(\tan^{d-1} \rho  A_\rho )
\eea
where we have inserted a factor of $\Delta_{12}$ in $A_\rho$ to ensure a smooth $\Delta_{12}\rightarrow 0$ limit.  In particular, setting $\Delta_{12}=0$  we have $A_\rho=0$ and
\be
A_t= (\cos \rho)^{\D-1} {_2{F_1}}\left({\D+1\over 2},{\D-1\over 2},\D-{d-2\over 2};\cos^2\rho  \right)~.
\ee
It is now straightforward to plug into (\ref{genspin}) to obtain an integral expression for the conformal block.   Because the general formula is rather lengthy we will only write it out explicitly in the case $\Delta_{12}=0$.   In this case we find (not paying attention to overall normalization factors)
\es{}{\cW_{\D,1}(z,\zb) &=|z|^{\D-\Delta_3-\Delta_4-1}(1-|1-z|^2)\\
&\quad \int_0^1\! d\s \s^{{\D+\Delta_{34}-1\over 2}}(1-\s)^{{\D-\Delta_{34}-1\over 2}} \big(1-(1-|1-z|^2)\s\big)^{-{\D+1\over 2}}
\\ &\quad \quad\times {_2{F_1}}\Bigg({\D+1\over 2},{\D-1\over 2},\D-{d-2\over 2}; {|z|^2\s(1-\s)\over 1-(1-|1-z|^2)\s}\Bigg)~.}
Setting $d=2, 4$, it is straightforward to verify that the series expansion of this integral reproduces the known $d=2, 4$ results in (\ref{twod}),(\ref{fourd}) for $\Delta_{12}=0$.  We have also verified agreement for $\Delta_{12}\neq 0$.

\subsec{Evaluation of geodesic Witten diagram: spin-2}

In this section we set $\Delta_{12}=0$ to simplify formulas a bit.   We need to solve
\be
 \nabla^2 h_{\mu\nu} -  [\D(\D-d)-2]h_{\mu\nu} =0~,\quad \nabla^{\mu} h_{\mu\nu} =0~,\quad h^\mu_\mu=0~.
\ee
$h_{\mu\nu}$ should be static and spherically symmetric, which implies the general ansatz
\be
 h_{\mu\nu}dx^\mu dx^\nu = f_{\rho\rho}(\rho)g_{\rho\rho}d\rho^2+ f_{tt}(\rho)g_{tt}dt^2
+{1\over d-1} f_{\phi\phi}(\rho)\tan^2\rho \,d\Omega^2_{d-1}~.
\ee
We first impose the divergence free and tracelessness conditions.  We have
\be h^\mu_\mu = f_{\rho\rho}+f_{tt}+f_{\phi\phi}~.
\ee
We use this to eliminate $f_{\phi\phi}$,
\be
f_{\phi\phi}=-f_{\rho\rho}-f_{tt}~.
\ee
Moving to the divergence, only the component $\nabla^\mu h_{\mu\rho}$ is not automatically zero.  We find
\be
 \nabla^\mu h_{\mu\rho} = f_{\rho\rho}' + { d+1\over \cos \rho\sin\rho} f_{\rho\rho} -{\cos\rho\over \sin\rho} f_{\rho\rho} +{\cos\rho\over \sin\rho}f_{tt}=0
 \ee
which we solve as
\be f_{tt} = -\tan\rho f_{\rho\rho}' +\left(1-{d+1\over \cos^2\rho}\right)f_{\rho\rho}~.
\ee
We then work out the $\rho\rho$ component of the field equation,
\be
 \nabla^2 h_{\rho\rho}-[\D(\D-d)-2]h_{\rho\rho} = f_{\rho\rho}''+\left({d+3\over \cos\rho\sin\rho}-2\cot\rho \right)f_{\rho\rho}'-{(\D+2)(\D-d-2)\over \cos^2\rho}f_{\rho\rho}~.
 \ee
Setting this to zero, the normalizable solution is
\be
 f_{\rho\rho} = (\cos \rho)^{\D+2}  {_2{F_1}}\left({\D\over 2},{\D+2\over 2},\D-{d-2\over 2};\cos^2\rho\right)~.
\ee
This completely specifies the solution, and we now have all we need to plug into (\ref{genspin}).   We refrain from writing out the somewhat lengthy formulas.   The series expansion of the result matches up with (\ref{twod}) and (\ref{fourd}) as expected.

\subsec{General $\ell$: proof via conformal Casimir equation}\label{v5}

As in the case of scalar exchange, the most efficient way to verify that a geodesic Witten diagram yields a conformal partial wave is to check that it is an eigenfunction of the conformal Casimir operator with the correct eigenvalue and asymptotics.

We start from the general expression \eqr{gwitt2}. A rank-$n$ tensor on AdS is related to a tensor on the embedding space via
\be
T_{\mu_1 \ldots \mu_n} = {\p Y^{M_1} \over \p y^{\mu_1}}\ldots {\p Y^{M_n} \over \p y^{\mu_n}}T_{M_1 \ldots M_n}~.
\ee
In particular, this holds for the bulk-to-bulk propagator of the spin-$\ell$ field, and so we can write
\be
G_{bb}(y,y';\D,\ell) = [G_{bb}(Y,Y';\D)]_{M_1 \ldots M_{\ell}, N_1 \ldots N_{\ell}}{d Y^{M_1} \over d\lambda} \ldots {d Y^{M_\ell} \over d\lambda}{d Y'^{N_1} \over d\lambda'} \ldots {d Y'^{N_\ell} \over d\lambda'}~.
\ee
Now, $[G_{bb}(Y,Y';\D)]_{M_1 \ldots M_{\ell}, N_1 \ldots N_{\ell}}$ only depends on $Y$ and $Y'$.  Since $Y^M {dY^M\over d\lambda} = {1\over 2} {d\over d\lambda} (Y\cdot Y) =0$, when pulled back to the geodesics the only contributing structure is
\be\label{propstruc}
[G_{bb}(Y,Y';\D)]_{M_1 \ldots M_{\ell}, N_1 \ldots N_{\ell}} = f(Y\cdot Y') Y'_{M_1} \ldots Y'_{M_\ell}Y_{N_1} \ldots Y_{N_\ell}~.
\ee
We also recall a few other useful  facts.  Lifted to the embedding space, the geodesic connecting boundary points $X_1$ and $X_2$ is
\be
Y(\lambda) = {e^\lambda X_1 + e^{-\lambda} X_2 \over \sqrt{-2 X_1 \cdot X_2} }~.
\ee
The bulk-to-boundary propagator lifted to the embedding space is
\be\label{bbform}
G_{b\partial}(X_i,Y)\propto  (X_i\cdot Y)^{-\Delta_i}~.
\ee

We follow the same strategy as in the case of scalar exchange.  We start by isolating the part of the diagram that contains all the dependence on $X_{1,2}$,
\es{}{&F_{M_1 \ldots M_\ell}(X_1,X_2,Y';\D)=\\ &\int_{\gamma_{12}} G_{b\partial}(X_1,Y(\lambda))G_{b\partial}(X_2,Y(\lambda))[G_{bb}(Y(\lambda),Y';\D)]_{M_1 \ldots M_{\ell}, N_1 \ldots N_{\ell}}{d Y^{M_1} \over d\lambda} \ldots {d Y^{M_\ell} \over d\lambda}~.}
Here, $Y(\lambda)$ lives on $\gamma_{12}$, but $Y'$ is left arbitrary. This is the spin-$\ell$ generalization of $\vphi^{12}_{\D}(y)$ defined in \eqr{31aa}, lifted to embedding space. We now argue that this is annihilated by the SO(d+1,1) generators $L^1_{AB} +L^2_{AB} + L^{Y'}_{AB}$.  This generator is the sum of three generators in the scalar representation, plus a ``spin" term acting on the free indices $N_1 \ldots N_\ell$.   This operator annihilates any expression of the form $g(X_1 \cdot X_2, X_1 \cdot Y', X_2 \cdot Y') X_{N_1}\ldots X_{N_\ell}$, where each $X$ stands for either $X_1$ or $X_2$.  To show this, we just note the SO(d+1,1) invariance of the dot products, along with the fact that $X_N$ is the normal vector to the $X^2=0$ surface and so is also SO(d+1,1) invariant.
From (\ref{propstruc})-(\ref{bbform}) we see that $F_{N_1 \ldots N_\ell}(X_1,X_2,Y';\D)$ is of this form, and so is annihilated by $L^1_{AB} +L^2_{AB} + L^{Y'}_{AB}$.  We can therefore write
\bea\label{cas}
(L^1_{AB} +L^2_{AB})^2 F_{N_1 \ldots N_\ell}(X_1,X_2,Y';\D) &=& (L^{Y'}_{AB})^2F_{N_1 \ldots N_\ell}(X_1,X_2,Y';\D)\cr &=&C_2(\D,\ell) F_{N_1 \ldots N_\ell}(X_1,X_2,Y';\D)
\eea
where we used that $(L^{Y'}_{AB})^2$ is acting on the spin-$\ell$ bulk-to-bulk propagator, which is an eigenfunction of the conformal Casimir operator\footnote{Note that the conformal Casimir is equal to the spin-$\ell$ Laplacian up to a constant shift: $(L_{AB}^{Y'})^2 = \nabla^2_{\ell}+\ell(\ell+d-1)$ \cite{Pilch:1984xx}.} with eigenvalue \eqr{32c}. The relation \eqr{cas} holds for all $Y'$, and hence holds upon integrating $Y'$ over $\gamma_{34}$ with any weight.   Hence we arrive at the conclusion
\be
(L^1_{AB} +L^2_{AB})^2{\cal W}_{\D,\ell}(x_i) = C_2(\D,\ell){\cal W}_{\D,\ell}(x_i)
\ee
which is the same eigenvalue equation obeyed by the spin-$\ell$ conformal partial wave, $W_{\D,\ell}(x_i)$. The short distance behavior as dictated by the OPE is easily seen to match in the two cases, establishing that we have the same eigenfunction.   We conclude that the spin-$\ell$ geodesic Witten diagram is, up to normalization, equal to the spin-$\ell$ conformal partial wave.

\subsection{Comparison to double integral expression of Ferrara et. al.}
\label{Fercomp}

It is illuminating to compare our expression (\ref{gwitt2}) to equation (50) in \cite{Ferrara:1973vz}, which gives the general result (in $d=4$) for the scalar conformal partial wave with spin-$\ell$ exchange, written as a double integral.   We will rewrite the result in \cite{Ferrara:1973vz} in a form permitting easy comparison to our formulas.   First, it will be useful to rewrite the scalar bulk-to-bulk propagator (\ref{22b}) by applying a quadratic transformation to the hypergeometric function,
\be\label{bbmod}
G_{bb}(y,y';\Delta)=\xi^\Delta {}_2F_1\left({\Delta\over 2},{\Delta+1\over 2},\Delta+1-{d\over 2};\xi^2\right)~.
\ee
Next, recall that in embedding space the geodesics are given by (\ref{32p}), from which we compute the quantity $\xi$ with one point on each geodesic
\be
\xi^{-1} = -Y(\lambda) \cdot Y(\lambda')  = {1\over 2} { e^{\lambda+\lambda'}x_{13}^2 + e^{\lambda-\lambda'}x_{14}^2 +e^{-\lambda+\lambda'}x_{23}^2 + e^{-\lambda-\lambda'}x_{24}^2 \over x_{12} x_{34}  }~.
\ee
We also define a modified version as
\be
\xi_-^{-1} = -{dY(\lambda)\over d\lambda}\cdot {dY(\lambda')\over d\lambda'}={1\over 2} { e^{\lambda+\lambda'}x_{13}^2 - e^{\lambda-\lambda'}x_{14}^2 -e^{-\lambda+\lambda'}x_{23}^2 + e^{-\lambda-\lambda'}x_{24}^2 \over x_{12} x_{34}  }~.
\ee
Comparing to  \cite{Ferrara:1973vz}, we have $\xi^{-1}=\lambda_+$ and $\xi_-^{-1}=\lambda_-$.

With these definitions in hand, it is not hard to show that the result of \cite{Ferrara:1973vz} takes the form
\es{}{W_{\Delta,\ell}(x_i)= \int_{\g_{12}}\int_{\g_{34}}&G_{b\p}(y(\l), x_1)G_{b\p}(y(\l),x_2)\\
&  \times C'_{\ell}(2\xi_{-}^{-1})G_{bb}(y(\l),y(\l');\D)\times G_{b\p}(y(\l'),x_3)G_{b\p}(y(\l'),x_4)~.}
Here $G_{bb}(y(\l),y(\l');\D)$ is the scalar bulk-to-bulk propagator (\ref{bbmod}), and $C'_{\ell}(x)$ is a Gegenbauer polynomial.   This obviously looks very similar to our expression (\ref{gwitt2}), and indeed agrees with it for $\ell=0$.   The two results must be equal (up to normalization) since they are both expressions for the same conformal partial wave.  If we assume that equality holds for the integrand, then we find the interesting result that the pullback of the spin-$\ell$ propagator, as written in (\ref{pbprop}), is equal to $C'_{\ell}(2\xi_{-}^{-1})G_{bb}(y(\l),y(\l');\D)$.   The general spin-$\ell$ propagator is very complicated (see \cite{Costa:2014kfa,Bekaert:2014cea}), but apparently has a simple relation to the scalar propagator when pulled back to geodesics.    It would be interesting to verify this.

\subsection{Decomposition of spin-1 Witten diagram into conformal blocks}

\begin{figure}[t!]
   \begin{center}
 \includegraphics[width = .95\textwidth]{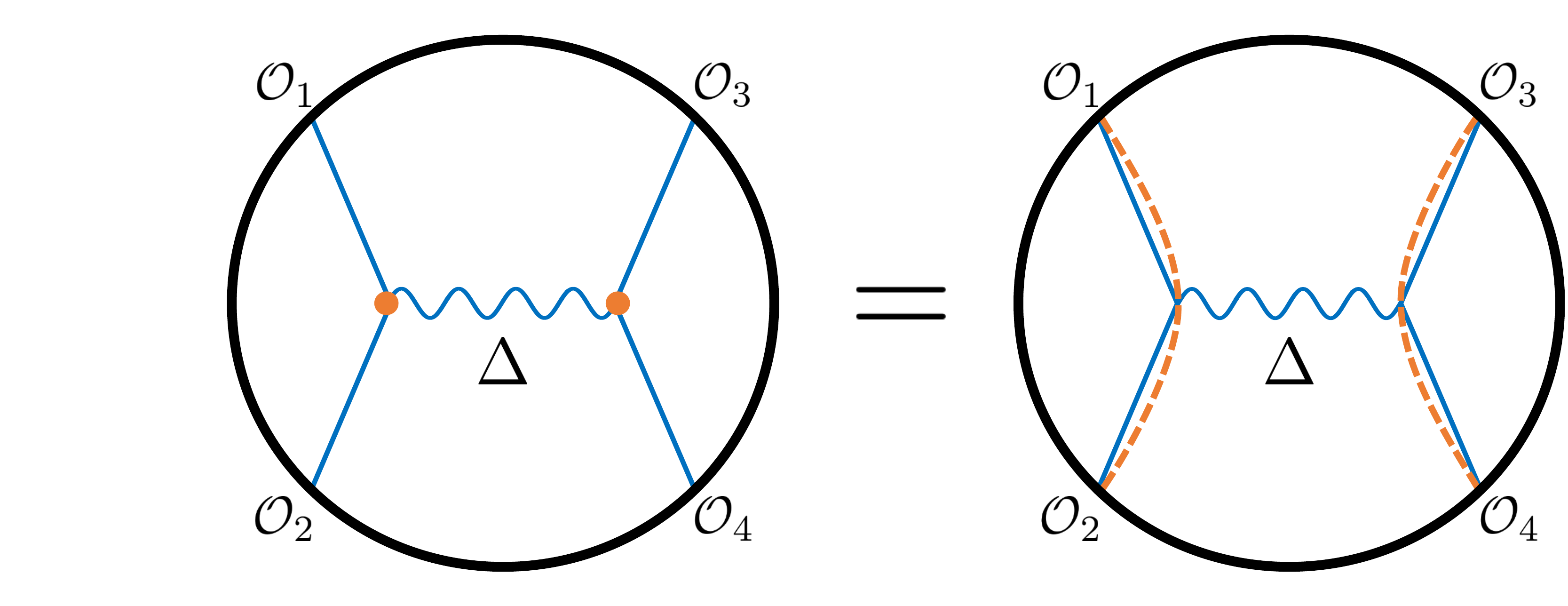}
 \includegraphics[width = .95\textwidth]{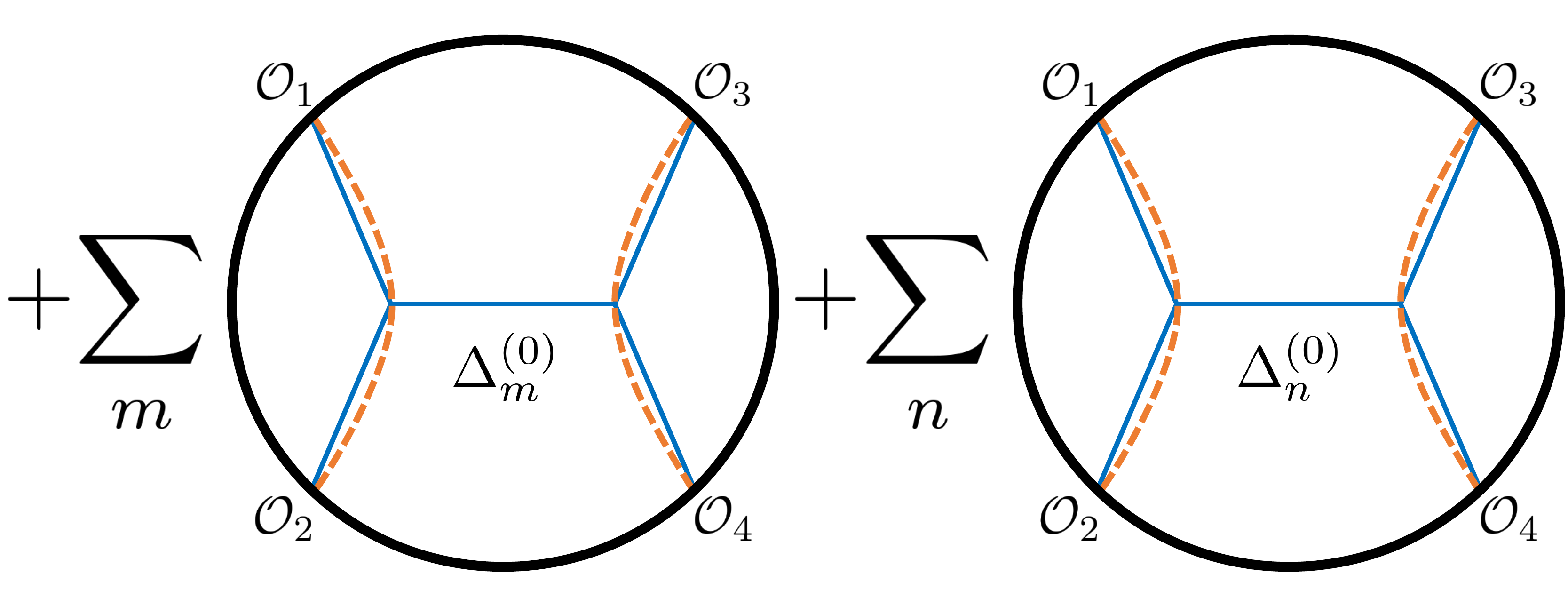}
  \includegraphics[width = .95\textwidth]{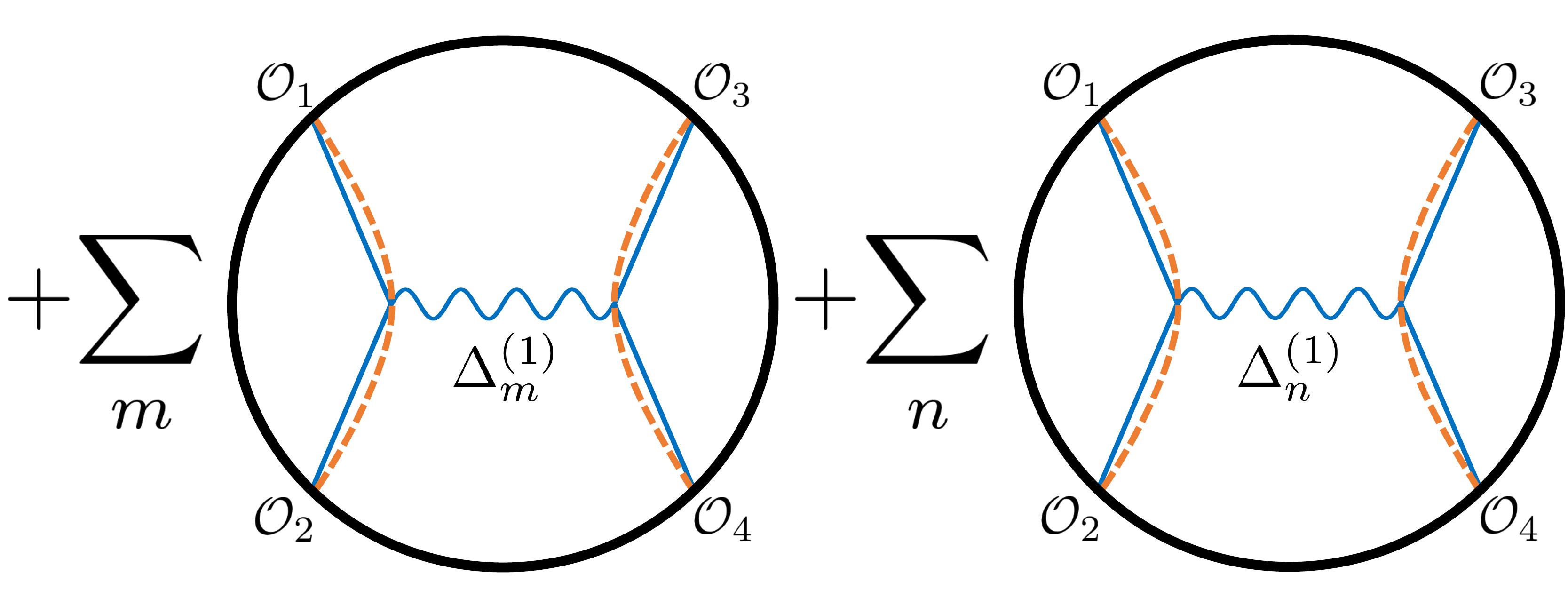}
 \caption{The decomposition of a four-point vector exchange diagram (upper left) into conformal partial waves. The term in the upper right captures the single-trace exchange of the dual vector operator. The second line captures the CFT exchanges of the $\ell=0$ double-trace operators $[\Oc_1\Oc_2]_{m,0}$ and $[\Oc_3\Oc_4]_{n,0}$. Likewise, the final line captures the CFT exchanges of the $\ell=1$ double-trace operators $[\Oc_1\Oc_2]_{m,1}$ and $[\Oc_3\Oc_4]_{n,1}$.}
\label{f6}
\end{center}
 \end{figure}

In the case of scalar exchange diagrams, we previously showed how to decompose a Witten diagram into a sum of geodesic Witten diagrams, the latter being identified with conformal partial waves of both single- and double-trace exchanges.   We now wish to extend this to the case of higher spin exchange; we focus here on the case of spin-1 exchange for simplicity.  A picture of the final result is given in Figure \ref{f6}.

As discussed in section \ref{ii}, given two scalar operators in a generalized free field theory, we can form scalar double trace primaries with schematic form $[\Oc_1\Oc_2]_{m,0}\sim \Oc_1 \p^{2m}\Oc_2$ and dimension $\D^{(12)}(m,0) = \Delta_1 + \Delta_2 +2m+O(1/N^2)$, and vector primaries $[\Oc_1\Oc_2]_{m,1} \sim \Oc_1 \p^{2m}\p_{\mu}\Oc_2$ with dimension $\D^{(12)}(m,1) = \Delta_1 + \Delta_2 +1+2m+O(1/N^2)$. The analysis of \cite{Heemskerk:2009pn}, and later \cite{Fitzpatrick:2011dm, Costa:2014kfa,Bekaert:2014cea} demonstrated that these conformal blocks, and their cousins $[\Oc_3\Oc_4]_{n,0}$ and $[\Oc_3\Oc_4]_{n,1}$, should appear in the decomposition of the vector exchange Witten diagram, together with the exchange of a single-trace vector operator.  The computations below will confirm this expectation.

The basic approach is the same as in the scalar case, although the details are more complicated.   Before diving in, let us note the main new features.   In the scalar case a basic step was to write, in \eqr{41a}, the product of two bulk-to-boundary propagators  $G_{b\partial}(y,x_1)G_{b\partial}(y,x_2)$ as a sum over solutions $\vphi^{12}_{\D}(y)$ of the scalar wave equation sourced on the $\gamma_{12}$ geodesic.   Here, we will similarly need a decomposition of $G_{b\partial}(y,x_1)\nabla_\mu G_{b\partial}(y,x_2)$, where $\nabla_{\mu}$ is a covariant derivative with respect to bulk coordinates $y$.  It turns out that this can be expressed as a sum over  massive spin-1 solutions and derivatives of massive scalar solutions.  This translates into  the statement that the spin-1 exchange Witten diagram decomposes as a sum of spin-1 and spin-0 conformal blocks, as noted above.

Now to the computation.   We consider a theory of massive  scalars coupled to a massive vector field via couplings  $\phi_i \nabla_\mu \phi_j A^\mu$.  The Witten diagram with vector exchange is then
\be\label{witt1}
\A_4^{\text{Vec}}(x_i) = \int_{y}\int_{y'} G_{b\partial}(y,x_1)\nabla_\mu  G_{b\partial}(y,x_2) \times G_{bb}^{\mu\nu}(y,y';\D) \times G_{b\partial}(y',x_3)\nabla_\nu  G_{b\partial}(y',x_4)~.
\ee

Our first task is to establish the expansion
\be\label{spin1eq}
 G_{b\partial}(y,x_1)\nabla_\mu  G_{b\partial}(y,x_2)  = \sum_m \left( c_m A_{m,\mu}(y) + b_m \nabla_\mu \vphi_m(y) \right)
 \ee
where $ A_{m,\mu}(y) $ and $\vphi_m(y)$ denote the solutions to the massive spin-1 and spin-0 equations sourced on $\gamma_{12}$, found earlier in sections \ref{v3} and \ref{iv1}, respectively.\footnote{$\vphi_m$ is just $\vphi^{12}_m$, whose superscript we suppress for clarity, and likewise for $\vphi_n$ and $\vphi^{34}_n$.} $m$ labels the masses of the bulk fields, to be determined shortly. We will not attempt to compute the coefficients $c_m$ and $b_m$, which is straightforward but involved, contenting ourselves to determining the spectrum of conformal dimensions appearing in the expansion, and showing how the expansion coefficients can be obtained if desired.

Following the scalar case, we work in global AdS and send $t_1 \rightarrow -\infty$, $t_2 \rightarrow \infty$.  Dropping normalizations, as we shall do throughout this section, we have
\es{}{G_{b\partial}(y,x_1)\nabla_\rho G_{b\partial}(y,x_2) &= \sin\rho\,(\cos\rho)^{\Delta_1+\Delta_2-1}e^{-\Delta_{12}t} \\
G_{b\partial}(y,x_1)\nabla_t G_{b\partial}(y,x_2) &= (\cos\rho)^{\Delta_1+\Delta_2}e^{-\Delta_{12}t}~.}
Letting $\Delta_m^{(\ell)}$ denote the dimension of the corresponding spin, we have, from \eqr{vecsol} and \eqr{31h},
\es{}{\vphi_m & = {}_2F_1\Big({\Delta^{(0)}_m+\Delta_{12}\over 2},{\Delta_m^{(0)} -\Delta_{12} \over 2};\Delta_m^{(0)} - {d-2\over 2};\cos^2 \rho\Big)( \cos \rho)^{\Delta_m^{(0)}}~ e^{-\Delta_{12}t }\\
A_{m,\rho}&= \Delta_{12} \sin \rho\, (\cos\rho)^{\Delta_m^{(1)}} {_2{F_1}}\left({\Delta_m^{(1)}+\Delta_{12}+1\over 2},{\Delta_m^{(1)}-\Delta_{12}+1\over 2},\Delta_m^{(1)}-{d-2\over 2};\cos^2\rho  \right)e^{-\Delta_{12}t}\\
A_{m,t} &= {1\over \Delta_{12}\tan^{d-1}\rho} \p_\rho(\tan^{d-1} \rho A_{m,\rho} )}
The various terms have the following powers $(\cos^2 \rho)^k$ in an expansion in powers of $\cos^2 \rho$,
\bea
G_{b\partial}(y,x_1)\nabla_\rho G_{b\partial}(y,x_2): && \quad k={\Delta_1+\Delta_2-1\over 2} + q\quad
\cr A_{m,\rho}: && \quad k={\Delta^{(1)}_m\over 2} + q \cr
\nabla_\rho \vphi_m:&&\quad k= {\Delta^{(0)}_m-1\over 2}+q \cr
G_{b\partial}(y,x_1)\nabla_t G_{b\partial}(y,x_2): && \quad k={\Delta_1+\Delta_2\over 2} \cr
A_{m,t}: && \quad k={\Delta^{(1)}_m-1\over 2} + q \cr
\nabla_t \vphi_m:&&\quad k= {\Delta^{(0)}_m\over 2}+q
\eea
where $q=0, 1, 2, \ldots$.  Comparing, we see that we have the right number of free coefficients for  (\ref{spin1eq}) to hold, provided we have the following spectrum of dimensions appearing
\bea\label{539}
\Delta_m^{(0)}& = & \Delta_1+\Delta_2 + 2m \cr
 \Delta_m^{(1)}& = & \Delta_1+\Delta_2+1 + 2m
\eea
with $m=0, 1, 2, \ldots$.   The formulas above can be used to work out the explicit coefficients $c_m$ and $b_m$.    We noted at the beginning of this subsection that this spectrum of dimensions coincides with the expected spectrum of double-trace scalar and vector operators appearing in the OPE, at leading order in large $N$.

We may now rewrite (\ref{witt1}) as\footnote{Following the precedent of Section \ref{iv}, all quantities with an $m$ subscript refer to the double-trace operators appearing in the $\Oc_1\Oc_2$ OPE, and those with an $n$ subscript refer to the double-trace operators appearing in the $\Oc_3\Oc_4$ OPE. }
\be\label{witt2}
\A_4^{\text{Vec}}(x_i) =\sum_{m,n} \int_{y}\int_{y'}   \left( c_m A_{m,\mu}(y) + b_m \nabla_\mu \vphi_m(y) \right) G_{bb}^{\mu\nu}(y,y';\D)   \left( c_n A_{n,\nu}(y') + b_n \nabla_\nu \vphi_n(y') \right)~.
\ee
We expand this out in an obvious fashion as
\be\label{4terms}
\A_4^{\text{Vec}}(x_i) = \A_{AA}(x_i)+ \A_{A\phi}(x_i)+ \A_{\phi A}(x_i)+ \A_{\phi\phi}(x_i)~.
\ee
The next step is to relate each term to geodesic Witten diagrams, which we now do in turn.

\subsubsection{ $\A_{AA}$}

We have
\be
\A_{AA} = \sum_{m,n} c_mc_n \int_{y}\int_{y'} A_{m,\mu}(y)  G_{bb}^{\mu\nu}(y,y';\D)A_{n,\nu}(y')~.
\ee
The solution  $A_{m,\mu}(y)$ can be expressed as
\bea\label{Aint}
A^\mu_{m}(y)&=& \int_{\gamma_{12}} G_{b\partial}(y(\l),x_1)\nabla_\nu G_{b\partial}(y(\l),x_2)G_{bb}^{\mu\nu}(y(\lambda),y;\Delta^{(1)}_m)\cr
&=&-\Delta_2 \int_{\gamma_{12}} G_{b\partial}(y(\l),x_1)G_{b\partial}(y(\l),x_2){dy_\nu(\lambda)\over d\lambda} G_{bb}^{\mu\nu}(y(\lambda),y;\Delta^{(1)}_m)~.
\eea
The second equality follows from the relation $\nabla_\mu G_{b\partial}(x,y(\lambda)) = -\Delta {dy_\mu(\lambda)\over d\lambda} G_{b\partial}(x,y(\lambda))$, which is easily verified for a straight line geodesic at the center of global AdS, and hence is true in general. Using this we obtain (dropping the normalization, as usual)
\bea\label{WAA}
\A_{AA}&=& \sum_{m,n}  c_mc_n \int_{y}\int_{y'}  \int_{\gamma_{12}}  \int_{\gamma_{34}} \Big[G_{b\partial}(y(\lambda),x_1) G_{b\partial}(y(\lambda),x_2){dy_\mu(\lambda)\over d\lambda} \Big]\cr
&& \quad\quad\quad\quad \times\Big[ G_{bb}^{\mu\nu}(y(\lambda),y;\Delta^{(1)}_m)G_{bb,\nu\alpha}(y,y';\Delta) G_{bb}^{\alpha \beta}(y',y(\lambda');\Delta^{(1)}_n)\Big]\cr
&&\quad\quad\quad\quad \times\Big[ G_{b\partial}(y(\lambda),x_3) G_{b\partial}(y(\lambda),x_4){dy'_\beta(\lambda')\over d\lambda'}\Big]~.
\eea
The bulk-to-bulk propagator for the vector field obeys
\be
\big(\nabla^2 -m^2 \big) G_{bb}^{\mu\nu}(y,y';\Delta)= \delta^{\mu\nu}(y-y')
\ee
where $\delta^{\mu\nu}(y-y') $ denotes a linear combination of $g^{\mu\nu} \delta(y-y')$ and $\nabla^\mu \nabla^\nu \delta(y-y')$.  Using this, and the fact that the propagator is divergence free at non-coincident points, we can verify the composition law
\be\label{comp}
\int_{y'} G_{bb}^{\mu\nu}(y,y';\Delta) G_{bb,\nu\alpha}(y',y'';{\Delta'}) =  {1\over m^2-(m')^{2}} \Big( {G_{bb}}^{\mu}_{~\alpha}(y,y'';{\Delta}) - {G_{bb}}^{\mu}_{~\alpha}(y,y'';{\Delta'})  \Big)~.
\ee
We use this relation twice within  (\ref{WAA}) to obtain a sum of three terms, each with a single vector bulk-to-bulk propagator.  Note also that these propagators appear pulled back to the geodesics.  Each term is thus a geodesic Witten diagram with an exchanged vector, that is, a spin-1 conformal partial wave.  The spectrum of spin-1 operators that appears is
\be
\Delta~,\quad \Delta_1 + \Delta_2 +1 +2m~,\quad  \Delta_3 + \Delta_4 +1 +2n~,\quad m,n=0, 1, 2, \ldots
\ee
So the contribution of $\A_{AA}$ is a sum of spin-1 conformal blocks with internal dimensions corresponding to the original exchanged field, along with the expected spin-1 double trace operators built out of the external scalars.

\subsubsection{$\A_{A \phi}$ and $\A_{\phi A}$}

We start with
\be
\A_{\phi A}= \sum_{m,n} c_mb_n \int_{y}\int_{y'}\nabla_\mu \vphi_m(y) G^{\mu\nu}_{bb}(y,y';\Delta) A_{n,\nu}(y')~.
\ee
Next we integrate by parts in $y$, use $\nabla_\mu G_{bb}^{\mu\nu}(y,y';\D) \propto \nabla^\nu \delta(y-y') $, and integrate by parts again, to get
\be
\A_{\phi A} =  \sum_{m,n} b_mc_n \int_{y'}\nabla_\mu \vphi_m(y') A^\mu_n(y')~.
\ee
Now we write $A^\mu_{n}(y')$ as an integral over $\gamma_{34}$ as in (\ref{Aint}) and then again remove the bulk-to-bulk propagator by integrating by parts.  This yields
\be
\A_{\phi A} =  \sum_{m,n} b_mc_n \int_{\gamma_{34}} G_{b\partial}(y(\lambda'),x_3)G_{b\partial}(y(\lambda'),x_4){d y^\mu(\lambda')\over d\lambda'}\nabla_\mu \vphi_m(y(\lambda'))~.
\ee
Writing $\vphi_m$ as an integral sourced on $\gamma_{12}$ we obtain
\es{}{&\A_{\phi A} =  \sum_{m,n} b_mc_n\times\\
& \int_{\gamma_{12}} \int_{\gamma_{34}} G_{b\partial}(y(\lambda),x_1)G_{b\partial}(y(\lambda),x_2){d\over d\lambda'}G_{bb}(y(\lambda),y(\lambda');\Delta^{(0)}_m)
G_{b\partial}(y(\lambda'),x_3)G_{b\partial}(y(\lambda'),x_4)~.}
Integrating by parts and using ${d\over d\lambda'} \big( G_{b\partial}(y(\lambda'),x_3)G_{b\partial}(y(\lambda'),x_4)\big) \propto G_{b\partial}(y(\lambda'),x_3)G_{b\partial}(y(\lambda'),x_4)$
we see that $\A_{\phi A}$ decomposes into a sum of spin-0 exchange geodesic Witten diagrams.    That is, $\A_{\phi A}$ contributes a sum of spin-0 blocks with conformal dimensions
\be
\Delta_1 + \Delta_2 + 2m~,\quad m=0, 1, 2, \ldots
\ee

By the same token $\A_{A \phi}$ yields a sum of spin-0 blocks with conformal dimensions %
\e{}{\Delta_3 + \Delta_4 + 2n~,\quad n=0, 1, 2, \ldots}

\subsubsection{$\A_{\phi\phi}$}

We have
\be
\A_{\phi\phi}= \sum_{m,n} b_mb_n \int_{y}\int_{y'} \nabla_\mu \vphi_m(y)G^{\mu\nu}_{bb}(y,y';\D)\nabla_\nu \vphi_n(y')~.
\ee
Integration by parts reduces this to
\be
\A_{\phi\phi}= \sum_{m,n} b_mb_n \int_{y} \nabla^\mu \vphi_m(y) \nabla_\mu \vphi_n(y)~.
\ee
Now rewrite the scalar solutions as integrals over the respective geodesic sources,
\bea\label{Waa}
\A_{\phi\phi}&=& \sum_{m,n} b_mb_n\int_{y'} \int_{\gamma_{12}}\int_{\gamma_{34}}G_{b\partial}(y(\lambda),x_1) G_{b\partial}(y(\lambda),x_2) G_{b\partial}(y(\lambda),x_3) G_{b\partial}(y(\lambda),x_4) \cr
&& \quad\quad\quad\quad\quad\quad\quad\quad\quad\quad\times \nabla^{\mu'}G_{bb}(y(\lambda),y';\Delta^{(0)}_m) \nabla_{\mu'}G_{bb}(y',y(\lambda');\Delta^{(0)}_n)~.
\eea
The composition law analogous to (\ref{comp}) is easily worked out to be
\es{}{\int_{y'} \nabla^{\mu'}G_{bb}(y(\lambda),y';\Delta^{(0)}_m) \nabla_{\mu'}G_{bb}(y',y(\lambda');\Delta^{(0)}_n) &= c_{mn} G_{bb}(y(\lambda),y(\lambda');\Delta^{(0)}_m) \\&+ d_{mn} G_{bb}(y(\lambda),y(\lambda');\Delta^{(0)}_n)}
with some coefficients $c_{mn}$ and $d_{mn}$ that we do not bother to display here. Inserting this in (\ref{Waa}) we see that $\A_{\phi\phi}$ decomposes into a sum of scalar blocks with conformal dimensions
\be
\Delta_1+\Delta_2 +2m~,\quad \Delta_3+\Delta_4 +2n~, \quad m,n=0,1,2,\ldots
\ee

\subsubsection{Summary}

We have shown that the Witten diagram involving the exchange of a spin-1 field of dimension $\D$ decomposes into a sum of spin-1 and spin-0 conformal blocks.    The full spectrum of conformal blocks appearing in the decomposition is
\bea
{\rm scalar:}&& \Delta_1+\Delta_2 +2n~,\quad \Delta_3+\Delta_4 +2n \cr
{\rm vector:}&& \Delta~,\quad \Delta_1+\Delta_2+1 +2n~,\quad \Delta_3+\Delta_4 +1+2n
\eea
where $n=0,1,2,\ldots$. This matches the spectrum expected from $1/N$ counting, including single- and double-trace operator contributions.   With some patience, the formulas above can be used to extract the coefficient of each conformal block, but we have not carried this out in full detail here.

While we have not explored this in any detail, it seems likely that the above method can be directly generalized to the case of arbitrary spin-$\ell$ exchange. The split \eqr{4terms} will still be natural, and a higher spin version of \eqr{comp} should hold.

\sec{Discussion and future work}\label{vi}

\begin{figure}[t!]
   \begin{center}
    \hspace*{\fill}%
 \includegraphics[width = 0.46\textwidth]{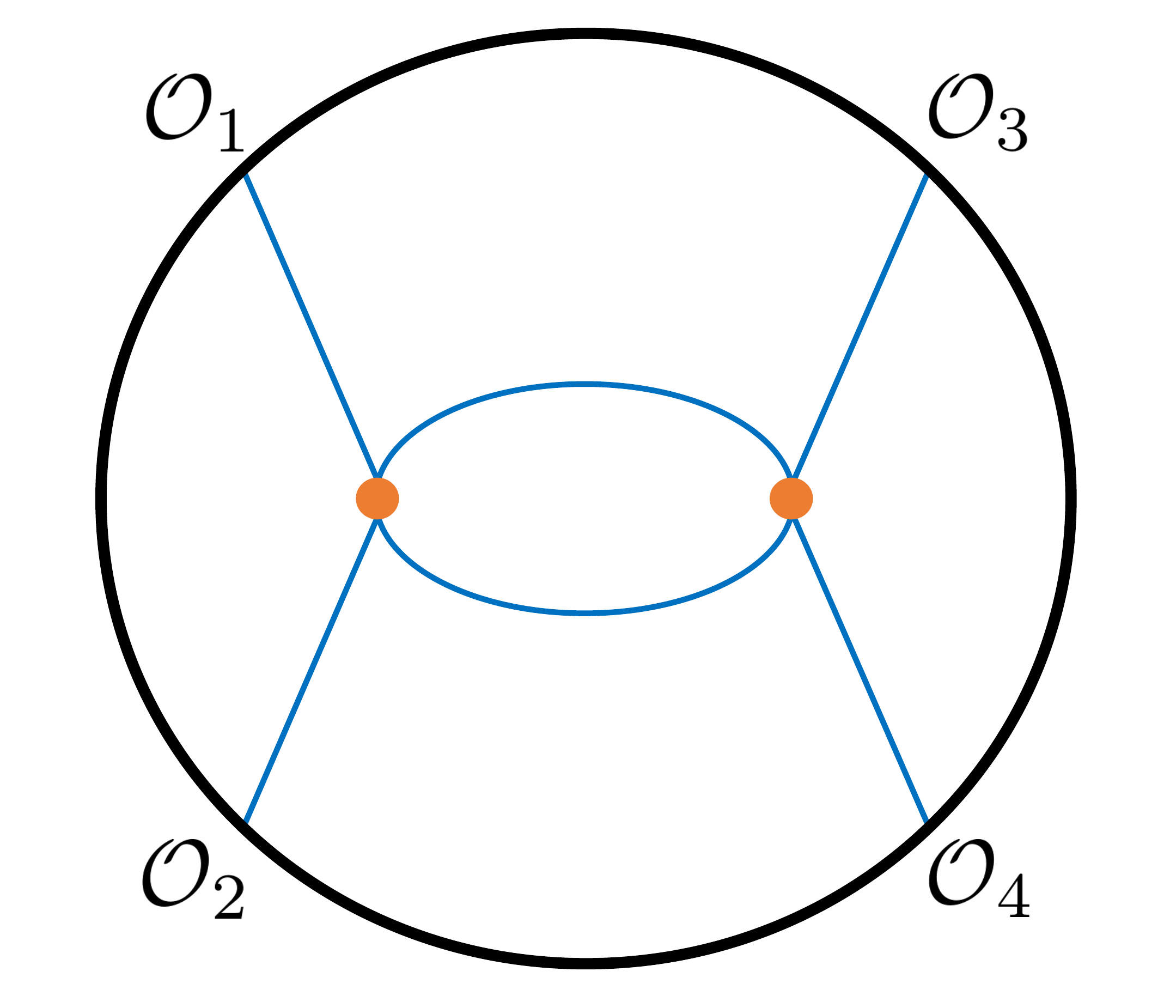}\hfill
  \includegraphics[width = 0.46\textwidth]{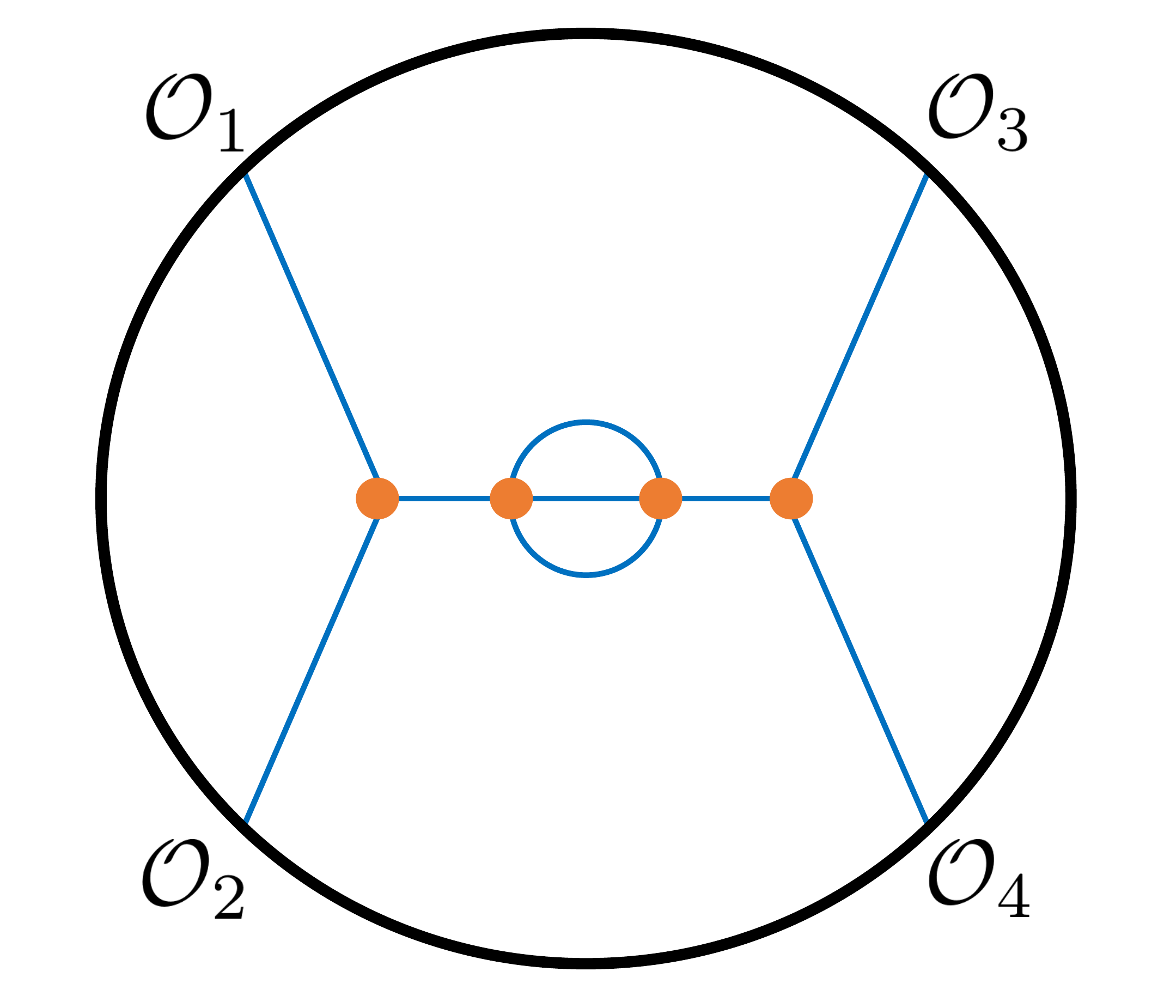}%
    \hspace*{\fill}%
 \caption{Some examples of loop diagrams that can be written as infinite sums over tree-level diagrams, and hence decomposed into conformal blocks using our methods.}
 \label{f7}
 \end{center}
 \end{figure}

In this paper, we have shed new light on the underlying structure of tree-level scattering amplitudes in AdS. Four-point scalar amplitudes naturally organize themselves into geodesic Witten diagrams; recognizing these as CFT conformal partial waves signals the end of the computation, and reveals a transparency between bulk and boundary with little technical effort required. We are optimistic that this reformulation extends, in some manner, to computations of generic holographic correlation functions in AdS/CFT. To that end, we close with some concrete observations and proposals, as well as a handful of future directions.
\vs
\noindent
{\bf $\bullet$~Adding loops}
\vs
It is clearly of interest to try to generalize our techniques to loop level. We first note that there is a special class of loop diagrams that we can compute already using these methods: namely, those that can be written as an infinite sum of tree-level exchange diagrams \cite{Penedones:2010ue}. For the same reason, this is the only class of loop diagrams whose Mellin amplitudes are known \cite{Penedones:2010ue}. These diagrams only involve bulk-to-bulk propagators that all start and end at the same points; see Figure \ref{f7} for examples. Careful study of the resulting sums would be useful.

More generally, though, we do not yet know how to decompose generic diagrams into geodesic objects. This would seem to require a ``geodesic identity'' analogous to \eqr{41a} that applies to a pair of bulk-to-bulk propagators, rather than bulk-to-boundary propagators. It would be very interesting to find these, if they exist. Such identities would also help to decompose an exchange Witten diagram in the crossed channel.

\vs
\noindent
{\bf $\bullet$~Adding legs}
\vs

\begin{figure}[t!]
   \begin{center}
 \includegraphics[width = 0.47\textwidth]{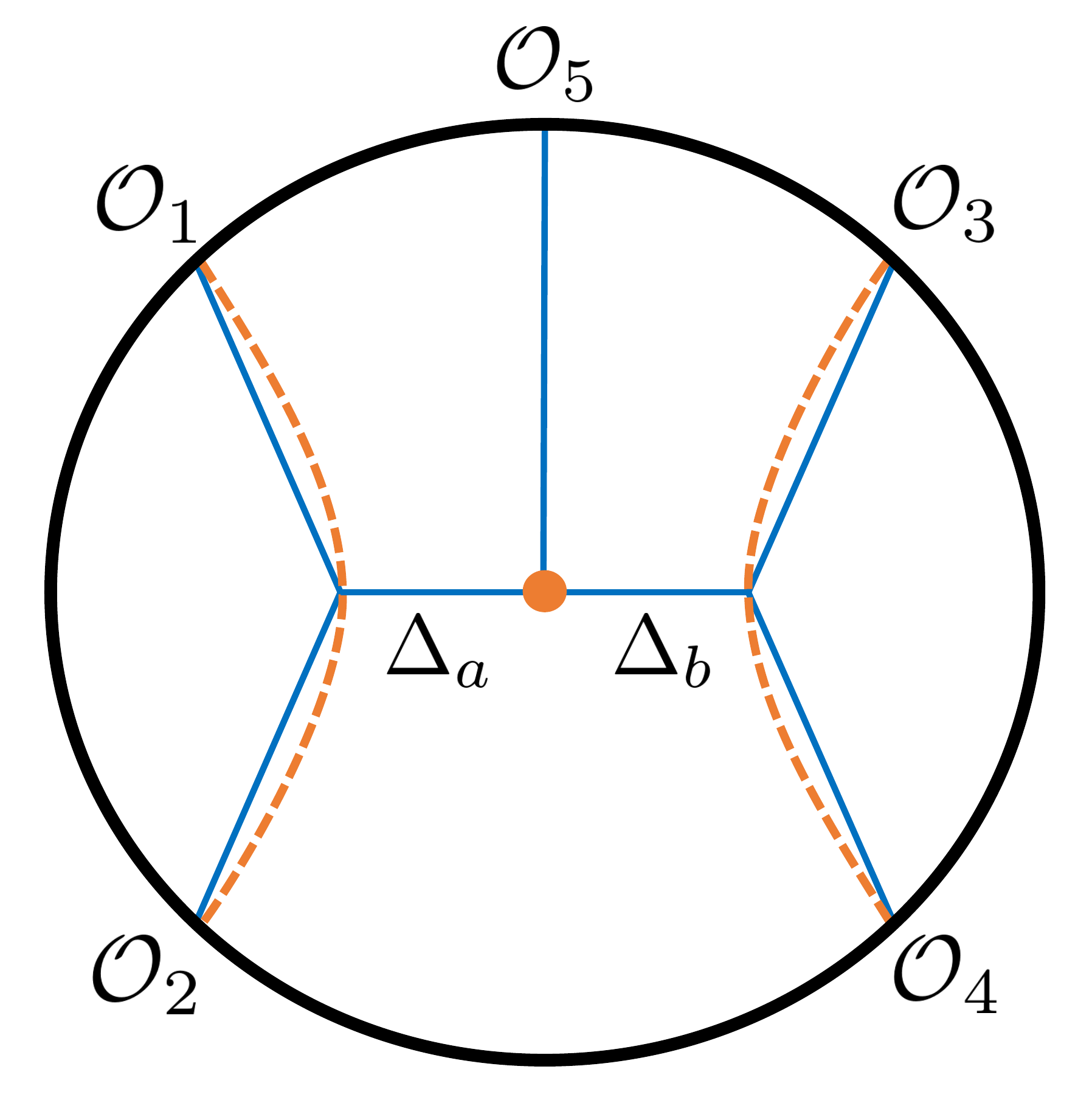}
 \caption{This is the basic constituent that emerges in applying our technology to the decomposition of a five-point tree-level Witten diagram. However, it is not equal to the five-point conformal partial wave, as discussed in the text.}
 \label{f8}
 \end{center}
 \end{figure}

Consider for example a five-point correlator of scalar operators $\langle \Oc_1(x_1)\ldots \Oc_5(x_5) \rangle$.    We can define associated conformal partial waves by inserting projection operators as
\be\label{fiveblock}
W_{\Delta,\ell;\Delta',\ell'}(x_i)= \langle \Oc_1(x_1)\Oc_2(x_2) P_{\Delta,\ell} \Oc_5(x_5)P_{\Delta',\ell'}\Oc_3(x_3)\Oc_4(x_4) \rangle~.
\ee
Using the OPE on $\Oc_1 \Oc_2$ and $\Oc_3 \Oc_4$, reduces this to three-point functions.   The question is, can we represent $ W_{\Delta,\ell;\Delta',\ell'}(x_i)$ as a geodesic Witten diagram?

Suppose we try to dismantle a tree level five-point Witten diagram. For definiteness, we take $\ell'=\ell=0$.  All tree level five-point diagrams will lead to the same structures upon using our geodesic identities: namely, they can be written as sums over geodesic-type diagrams, each as in Figure \ref{f8}, which we label $\widehat \cW_{\D_a,0;\D_b,0}(x_i)$. This is easiest to see starting from a $\phi^5$ contact diagram, and using \eqr{41a} on the pairs of propagators (12) and (34). In that case, $\D_a \in \lbrace\D_m\rbrace$ and $\D_b\in\lbrace \D_n\rbrace$. As an equation, Figure \ref{f8} reads
\es{5b}{\widehat \cW_{\D_a,0;\D_b,0}(x_i)&=\int_{\g_{12}}\int_{\g_{34}}
G_{b\p}(y(\l),x_1)G_{b\p}(y(\l),x_2)\\
&\times \int_{y_5} G_{bb}(y(\l),y_5;\D_a)\,G_{b\partial}(y_5,x_5) \,G_{bb}(y_5,y(\l');\D_b)\\
&\times \,G_{b\p}(y(\l'),x_3)G_{b\p}(y(\l'),x_4)~.}
Note that the vertex at $y_5$, indicated by the orange dot in the figure, must be integrated over all of AdS.  Could these diagrams be computing $W_{\Delta_a,0;\Delta_b,0}(x_i)$ as defined above?  The answer is no, as a simple argument shows.  Suppose we set $\Delta_5=0$ in \eqr{5b}, which requires $\D_a=\D_b\equiv \D$.  From (\ref{fiveblock})  it is clear that we must recover the four-point conformal partial wave with the exchanged primary $(\Delta,0)$.   So we should ask whether (\ref{5b}) reduces to the expression for the four-point geodesic Witten diagram, $\cW_{\D,0}$.   Using $G_{b\partial}(y_5,x_5)|_{\Delta_5=0}\propto 1$,  the integral over $y_5$ becomes
\be\int_{y_5} G_{bb}(y(\l),y_5;\D)G_{bb}(y_5,y(\l');\Delta)\propto {\partial \over \partial m_\Delta^2} G_{bb}(y(\l),y(\l');\Delta)~.
\ee
Therefore, the $\Delta_5=0$ limit of \eqr{5b} does not give back the four-point partial wave, but rather its derivative with respect to $m_\Delta^2$, which is a different object.

We conclude that although we can decompose a five-point Witten diagram into a sum of diagrams of the type in Figure \ref{f8}, this is not the conformal block decomposition.  This raises two questions:  what is the meaning of this decomposition in CFT terms, and (our original question) what diagram computes the five-point partial wave?

\vs
\noindent
{\bf $\bullet$~External operators with spin}
\vs

Another obvious direction in which to generalize is to consider correlation functions of operators carrying spin. As far as the conformal blocks go, partial information is available.  In particular, \cite{Costa:2011dw} obtained expressions for such blocks as differential operators acting on blocks with external scalars, but this approach is limited to the case in which the exchanged operator is a symmetric traceless tensor, since only such operators appear in the OPE of two scalar operators. The same approach was taken in \cite{1505.03750}. Explicit examples of mixed symmetry exchange blocks were given in \cite{1411.7351}.

Our formulation in terms of geodesic Witten diagrams suggests an obvious proposal for the AdS computation of an arbitrary conformal partial wave: take our usual expression (\ref{geowit}), now with the bulk-to-boundary and bulk-to-bulk propagators corresponding to the fields dual to the respective operators. Of course, there are many indices here which have to be contracted, and there will be inequivalent ways of doing so.  But this is to be expected, as in the general case there are multiple conformal blocks for a given set of operators, corresponding to the multiplicity of ways in which one spinning primary can appear in the OPE of two other spinning primaries. It will be interesting to see whether this proposal turns out to be valid.  As motivation, we note that it would be quite useful for bootstrap purposes to know all the conformal blocks that arise in the four-point function of stress tensors.

A related pursuit would be to decompose all four-point scalar contact diagrams, including any number of derivatives at the vertices. This would involve a generalization of \eqr{spin1eq} to include more derivatives.

\vs
\noindent \bul {\bf Virasoro blocks and AdS$_3$/CFT$_2$}
\vs

Our calculations give a new perspective on how to construct the dual of a generic Virasoro conformal block: starting with the geodesic Witten diagram, we dress it with gravitons. Because Virasoro blocks depend on $c$, a computation in semiclassical AdS gravity would utilize a perturbative $1/c\sim G_N$ expansion. In \cite{Hijano:2015qja}, we put the geodesic approach to use in constructing the holographic dual of the heavy-light Virasoro blocks of \cite{Fitzpatrick:2015zha}, where one geodesic essentially backreacts on AdS to generate a conical defect or black hole geometry. It would be worthwhile to pursue a $1/c$ expansion around the geodesic Witten diagrams more generally.

A closely related question is how to decompose an AdS$_3$ Witten diagram into Virasoro, rather than global, blocks. For a tree-level diagram involving light external operators like those considered here, there is no difference, because the large $c$ Virasoro block with light external operators reduces to the global block \cite{zamo}. It will be interesting to see whether loop diagrams in AdS$_3$ are easier to analyze using Virasoro symmetry.

\vs
\noindent \bul {\bf Assorted comments}
\vs
The geodesic approach to conformal blocks should be useful in deriving various CFT results, not only mixed symmetry exchange conformal blocks. For example, the conformal blocks in the limits of large $\tau$, $\ell$ or $d$ \cite{1212.3616, 1212.4103, 1305.4604, 1502.01437, 1504.00772, 1305.0004, Vos:2014pqa}, and subleading corrections to these, should be derivable using properties of AdS propagators. One can also ask whether there are similar structures present in bulk spacetimes besides AdS. For instance, an analog of the geodesic Witten diagram in a thermal spacetime would suggest a useful ingredient for parameterizing holographic thermal correlators. Perhaps the existence of a dS/CFT correspondence suggests similar structures in de Sitter space as well.

It is natural to wonder whether there are analogous techniques to those presented here that are relevant for holographic correlators of nonlocal operators like Wilson loops or surface operators, perhaps involving bulk minimal surfaces.

Let us close by noting a basic fact of our construction: even though a conformal block is not a semiclassical object per se, we have given it a representation in terms of classical fields propagating in a smooth spacetime geometry. In a bulk theory of quantum gravity putatively dual to a finite $N$ CFT, we do not yet know how to compute amplitudes. Whatever the prescription, there is, evidently, a way to write the answer using geodesic Witten diagrams. It would be interesting to understand how this structure emerges.

\section*{Acknowledgments}

We thank Eric D'Hoker, Liam Fitzpatrick, Tom Hartman, Daniel Jafferis, Juan Maldacena, Joao Penedones and Sasha Zhiboedov for helpful discussions. EP wishes to thank the KITP and Strings 2015 for hospitality during this project.   P.K. is supported in part by NSF grant PHY-1313986.  This research was supported in part by the National Science Foundation under Grant No. NSF PHY11-25915. E.P. is supported by the Department of Energy under Grant No. DE-FG02-91ER40671.

\bibliographystyle{ssg}
\bibliography{biblio}

\begingroup\raggedright\begin{thebibliography}{10}

\bibitem{Ferrara:1971vh}
S.~Ferrara, A.~F. Grillo, and R.~Gatto, ``{Manifestly conformal covariant
  operator-product expansion},'' {\em Lett. Nuovo Cim.} {\bf 2S2} (1971)
  1363--1369. [Lett. Nuovo Cim.2,1363(1971)].

\bibitem{Ferrara:1973vz}
S.~Ferrara, A.~F. Grillo, G.~Parisi, and R.~Gatto, ``{Covariant expansion of
  the conformal four-point function},'' {\em Nucl. Phys.} {\bf B49} (1972)
  77--98.

\bibitem{Ferrara:1974ny}
S.~Ferrara, R.~Gatto, and A.~F. Grillo, ``{Properties of Partial Wave
  Amplitudes in Conformal Invariant Field Theories},'' {\em Nuovo Cim.} {\bf
  A26} (1975) 226.

\bibitem{Dolan:2000ut}
F.~A. Dolan and H.~Osborn, ``{Conformal four point functions and the operator
  product expansion},'' {\em Nucl. Phys.} {\bf B599} (2001) 459--496,
  \href{http://xxx.lanl.gov/abs/hep-th/0011040}{{\tt hep-th/0011040}}.

\bibitem{Dolan:2003hv}
F.~A. Dolan and H.~Osborn, ``{Conformal partial waves and the operator product
  expansion},'' {\em Nucl. Phys.} {\bf B678} (2004) 491--507,
  \href{http://xxx.lanl.gov/abs/hep-th/0309180}{{\tt hep-th/0309180}}.

\bibitem{Dolan:2011dv}
F.~A. Dolan and H.~Osborn, ``{Conformal Partial Waves: Further Mathematical
  Results},'' \href{http://xxx.lanl.gov/abs/1108.6194}{{\tt 1108.6194}}.

\bibitem{Costa:2011dw}
M.~S. Costa, J.~Penedones, D.~Poland, and S.~Rychkov, ``{Spinning Conformal
  Blocks},'' {\em JHEP} {\bf 11} (2011) 154,
  \href{http://xxx.lanl.gov/abs/1109.6321}{{\tt 1109.6321}}.

\bibitem{Rattazzi:2008pe}
R.~Rattazzi, V.~S. Rychkov, E.~Tonni, and A.~Vichi, ``{Bounding scalar operator
  dimensions in 4D CFT},'' {\em JHEP} {\bf 12} (2008) 031,
  \href{http://xxx.lanl.gov/abs/0807.0004}{{\tt 0807.0004}}.

\bibitem{ElShowk:2012ht}
S.~El-Showk, M.~F. Paulos, D.~Poland, S.~Rychkov, D.~Simmons-Duffin, and
  A.~Vichi, ``{Solving the 3D Ising Model with the Conformal Bootstrap},'' {\em
  Phys. Rev.} {\bf D86} (2012) 025022,
  \href{http://xxx.lanl.gov/abs/1203.6064}{{\tt 1203.6064}}.

\bibitem{hep-th/9711200}
J.~M. Maldacena, ``{The Large N limit of superconformal field theories and
  supergravity},'' {\em Int. J. Theor. Phys.} {\bf 38} (1999) 1113--1133,
  \href{http://xxx.lanl.gov/abs/hep-th/9711200}{{\tt hep-th/9711200}}. [Adv.
  Theor. Math. Phys.2,231(1998)].

\bibitem{hep-th/9802109}
S.~S. Gubser, I.~R. Klebanov, and A.~M. Polyakov, ``{Gauge theory correlators
  from noncritical string theory},'' {\em Phys. Lett.} {\bf B428} (1998)
  105--114, \href{http://xxx.lanl.gov/abs/hep-th/9802109}{{\tt
  hep-th/9802109}}.

\bibitem{Witten:1998qj}
E.~Witten, ``{Anti-de Sitter space and holography},'' {\em Adv. Theor. Math.
  Phys.} {\bf 2} (1998) 253--291,
  \href{http://xxx.lanl.gov/abs/hep-th/9802150}{{\tt hep-th/9802150}}.

\bibitem{Liu:1998ty}
H.~Liu and A.~A. Tseytlin, ``{On four point functions in the CFT / AdS
  correspondence},'' {\em Phys. Rev.} {\bf D59} (1999) 086002,
  \href{http://xxx.lanl.gov/abs/hep-th/9807097}{{\tt hep-th/9807097}}.

\bibitem{Liu:1998th}
H.~Liu, ``{Scattering in anti-de Sitter space and operator product
  expansion},'' {\em Phys. Rev.} {\bf D60} (1999) 106005,
  \href{http://xxx.lanl.gov/abs/hep-th/9811152}{{\tt hep-th/9811152}}.

\bibitem{Freedman:1998bj}
D.~Z. Freedman, S.~D. Mathur, A.~Matusis, and L.~Rastelli, ``{Comments on 4
  point functions in the CFT / AdS correspondence},'' {\em Phys. Lett.} {\bf
  B452} (1999) 61--68, \href{http://xxx.lanl.gov/abs/hep-th/9808006}{{\tt
  hep-th/9808006}}.

\bibitem{D'Hoker:1998mz}
E.~D'Hoker and D.~Z. Freedman, ``{General scalar exchange in AdS(d+1)},'' {\em
  Nucl. Phys.} {\bf B550} (1999) 261--288,
  \href{http://xxx.lanl.gov/abs/hep-th/9811257}{{\tt hep-th/9811257}}.

\bibitem{D'Hoker:1999pj}
E.~D'Hoker, D.~Z. Freedman, S.~D. Mathur, A.~Matusis, and L.~Rastelli,
  ``{Graviton exchange and complete four point functions in the AdS / CFT
  correspondence},'' {\em Nucl. Phys.} {\bf B562} (1999) 353--394,
  \href{http://xxx.lanl.gov/abs/hep-th/9903196}{{\tt hep-th/9903196}}.

\bibitem{D'Hoker:1999jp}
E.~D'Hoker, S.~D. Mathur, A.~Matusis, and L.~Rastelli, ``{The Operator product
  expansion of N=4 SYM and the 4 point functions of supergravity},'' {\em Nucl.
  Phys.} {\bf B589} (2000) 38--74,
  \href{http://xxx.lanl.gov/abs/hep-th/9911222}{{\tt hep-th/9911222}}.

\bibitem{Hoffmann:2000tr}
L.~Hoffmann, A.~C. Petkou, and W.~Ruhl, ``{A Note on the analyticity of AdS
  scalar exchange graphs in the crossed channel},'' {\em Phys. Lett.} {\bf
  B478} (2000) 320--326, \href{http://xxx.lanl.gov/abs/hep-th/0002025}{{\tt
  hep-th/0002025}}.

\bibitem{Hoffmann:2000mx}
L.~Hoffmann, A.~C. Petkou, and W.~Ruhl, ``{Aspects of the conformal operator
  product expansion in AdS / CFT correspondence},'' {\em Adv. Theor. Math.
  Phys.} {\bf 4} (2002) 571--615,
  \href{http://xxx.lanl.gov/abs/hep-th/0002154}{{\tt hep-th/0002154}}.

\bibitem{Penedones:2010ue}
J.~Penedones, ``{Writing CFT correlation functions as AdS scattering
  amplitudes},'' {\em JHEP} {\bf 03} (2011) 025,
  \href{http://xxx.lanl.gov/abs/1011.1485}{{\tt 1011.1485}}.

\bibitem{Paulos:2011ie}
M.~F. Paulos, ``{Towards Feynman rules for Mellin amplitudes},'' {\em JHEP}
  {\bf 10} (2011) 074, \href{http://xxx.lanl.gov/abs/1107.1504}{{\tt
  1107.1504}}.

\bibitem{Fitzpatrick:2011ia}
A.~L. Fitzpatrick, J.~Kaplan, J.~Penedones, S.~Raju, and B.~C. van Rees, ``{A
  Natural Language for AdS/CFT Correlators},'' {\em JHEP} {\bf 11} (2011) 095,
  \href{http://xxx.lanl.gov/abs/1107.1499}{{\tt 1107.1499}}.

\bibitem{Fitzpatrick:2011hu}
A.~L. Fitzpatrick and J.~Kaplan, ``{Analyticity and the Holographic
  S-Matrix},'' {\em JHEP} {\bf 10} (2012) 127,
  \href{http://xxx.lanl.gov/abs/1111.6972}{{\tt 1111.6972}}.

\bibitem{Costa:2012cb}
M.~S. Costa, V.~Goncalves, and J.~Penedones, ``{Conformal Regge theory},'' {\em
  JHEP} {\bf 12} (2012) 091, \href{http://xxx.lanl.gov/abs/1209.4355}{{\tt
  1209.4355}}.

\bibitem{Fitzpatrick:2011dm}
A.~L. Fitzpatrick and J.~Kaplan, ``{Unitarity and the Holographic S-Matrix},''
  {\em JHEP} {\bf 10} (2012) 032, \href{http://xxx.lanl.gov/abs/1112.4845}{{\tt
  1112.4845}}.

\bibitem{1410.4185}
V.~Gonçalves, J.~Penedones, and E.~Trevisani, ``{Factorization of Mellin
  amplitudes},'' \href{http://xxx.lanl.gov/abs/1410.4185}{{\tt 1410.4185}}.

\bibitem{Hartman:2013mia}
T.~Hartman, ``{Entanglement Entropy at Large Central Charge},''
  \href{http://xxx.lanl.gov/abs/1303.6955}{{\tt 1303.6955}}.

\bibitem{Asplund:2014coa}
C.~T. Asplund, A.~Bernamonti, F.~Galli, and T.~Hartman, ``{Holographic
  Entanglement Entropy from 2d CFT: Heavy States and Local Quenches},'' {\em
  JHEP} {\bf 02} (2015) 171, \href{http://xxx.lanl.gov/abs/1410.1392}{{\tt
  1410.1392}}.

\bibitem{Fitzpatrick:2015zha}
A.~L. Fitzpatrick, J.~Kaplan, and M.~T. Walters, ``{Virasoro Conformal Blocks
  and Thermality from Classical Background Fields},''
  \href{http://xxx.lanl.gov/abs/1501.05315}{{\tt 1501.05315}}.

\bibitem{Fitzpatrick:2014vua}
A.~L. Fitzpatrick, J.~Kaplan, and M.~T. Walters, ``{Universality of
  Long-Distance AdS Physics from the CFT Bootstrap},'' {\em JHEP} {\bf 08}
  (2014) 145, \href{http://xxx.lanl.gov/abs/1403.6829}{{\tt 1403.6829}}.

\bibitem{Hijano:2015rla}
E.~Hijano, P.~Kraus, and R.~Snively, ``{Worldline approach to semi-classical
  conformal blocks},'' {\em JHEP} {\bf 07} (2015) 131,
  \href{http://xxx.lanl.gov/abs/1501.02260}{{\tt 1501.02260}}.

\bibitem{Alkalaev:2015wia}
K.~B. Alkalaev and V.~A. Belavin, ``{Classical conformal blocks via AdS/CFT
  correspondence},'' \href{http://xxx.lanl.gov/abs/1504.05943}{{\tt
  1504.05943}}.

\bibitem{Hijano:2015qja}
E.~Hijano, P.~Kraus, E.~Perlmutter, and R.~Snively, ``{Semiclassical Virasoro
  Blocks from AdS$_3$ Gravity},''
  \href{http://xxx.lanl.gov/abs/1508.04987}{{\tt 1508.04987}}.

\bibitem{Rychkov:lectures}
V.~Rychkov, ``{EPFL Lectures on Conformal Field Theory in D$\geq$ 3 Dimensions,
  https://sites.google.com/site/slavarychkov/home},''.

\bibitem{D'Hoker:2002aw}
E.~D'Hoker and D.~Z. Freedman, ``{Supersymmetric gauge theories and the AdS /
  CFT correspondence},'' in {\em {Strings, Branes and Extra Dimensions: TASI
  2001: Proceedings}}, pp.~3--158, 2002.
\newblock \href{http://xxx.lanl.gov/abs/hep-th/0201253}{{\tt hep-th/0201253}}.

\bibitem{Zamo:lectures}
A.~Zamolodchikov, ``{ Conformal symmetry in two-dimensional space: recursion
  represen- ation of conformal block},'' in {\em { Theoretical and Mathematical
  Physics, Vol. 73, Issue 1, pp 1088-1093}}, 1987.

\bibitem{zamo}
A.~Zamolodchikov, ``{Conformal symmetry in two-dimensions: An explicit
  recurrence formula for the conformal partial wave amplitude},'' {\em
  Commun.Math.Phys.} {\bf 96} (1984) 419--422.

\bibitem{1307.6856}
F.~Kos, D.~Poland, and D.~Simmons-Duffin, ``{Bootstrapping the $O(N)$ vector
  models},'' {\em JHEP} {\bf 1406} (2014) 091,
  \href{http://xxx.lanl.gov/abs/1307.6856}{{\tt 1307.6856}}.

\bibitem{1208.0337}
A.~L. Fitzpatrick and J.~Kaplan, ``{AdS Field Theory from Conformal Field
  Theory},'' {\em JHEP} {\bf 02} (2013) 054,
  \href{http://xxx.lanl.gov/abs/1208.0337}{{\tt 1208.0337}}.

\bibitem{Joaotalk}
J.~Penedones, ``{Talk at Strings 2015},''.

\bibitem{1503.01409}
J.~Maldacena, S.~H. Shenker, and D.~Stanford, ``{A bound on chaos},''
  \href{http://xxx.lanl.gov/abs/1503.01409}{{\tt 1503.01409}}.

\bibitem{Heemskerk:2009pn}
I.~Heemskerk, J.~Penedones, J.~Polchinski, and J.~Sully, ``{Holography from
  Conformal Field Theory},'' {\em JHEP} {\bf 10} (2009) 079,
  \href{http://xxx.lanl.gov/abs/0907.0151}{{\tt 0907.0151}}.

\bibitem{Heemskerk:2010ty}
I.~Heemskerk and J.~Sully, ``{More Holography from Conformal Field Theory},''
  {\em JHEP} {\bf 09} (2010) 099, \href{http://xxx.lanl.gov/abs/1006.0976}{{\tt
  1006.0976}}.

\bibitem{ElShowk:2011ag}
S.~El-Showk and K.~Papadodimas, ``{Emergent Spacetime and Holographic CFTs},''
  {\em JHEP} {\bf 10} (2012) 106, \href{http://xxx.lanl.gov/abs/1101.4163}{{\tt
  1101.4163}}.

\bibitem{Hartman:2014oaa}
T.~Hartman, C.~A. Keller, and B.~Stoica, ``{Universal Spectrum of 2d Conformal
  Field Theory in the Large c Limit},'' {\em JHEP} {\bf 09} (2014) 118,
  \href{http://xxx.lanl.gov/abs/1405.5137}{{\tt 1405.5137}}.

\bibitem{Arutyunov:2002fh}
G.~Arutyunov, F.~A. Dolan, H.~Osborn, and E.~Sokatchev, ``{Correlation
  functions and massive Kaluza-Klein modes in the AdS / CFT correspondence},''
  {\em Nucl. Phys.} {\bf B665} (2003) 273--324,
  \href{http://xxx.lanl.gov/abs/hep-th/0212116}{{\tt hep-th/0212116}}.

\bibitem{Muck:1998rr}
W.~Mueck and K.~S. Viswanathan, ``{Conformal field theory correlators from
  classical scalar field theory on AdS(d+1)},'' {\em Phys. Rev.} {\bf D58}
  (1998) 041901, \href{http://xxx.lanl.gov/abs/hep-th/9804035}{{\tt
  hep-th/9804035}}.

\bibitem{D'Hoker:1998gd}
E.~D'Hoker and D.~Z. Freedman, ``{Gauge boson exchange in AdS(d+1)},'' {\em
  Nucl. Phys.} {\bf B544} (1999) 612--632,
  \href{http://xxx.lanl.gov/abs/hep-th/9809179}{{\tt hep-th/9809179}}.

\bibitem{D'Hoker:1999jc}
E.~D'Hoker, D.~Z. Freedman, S.~D. Mathur, A.~Matusis, and L.~Rastelli,
  ``{Graviton and gauge boson propagators in AdS(d+1)},'' {\em Nucl. Phys.}
  {\bf B562} (1999) 330--352,
  \href{http://xxx.lanl.gov/abs/hep-th/9902042}{{\tt hep-th/9902042}}.

\bibitem{D'Hoker:1999ni}
E.~D'Hoker, D.~Z. Freedman, and L.~Rastelli, ``{AdS / CFT four point functions:
  How to succeed at z integrals without really trying},'' {\em Nucl. Phys.}
  {\bf B562} (1999) 395--411,
  \href{http://xxx.lanl.gov/abs/hep-th/9905049}{{\tt hep-th/9905049}}.

\bibitem{Banks:1998nr}
T.~Banks and M.~B. Green, ``{Nonperturbative effects in AdS in five-dimensions
  x S**5 string theory and d = 4 SUSY Yang-Mills},'' {\em JHEP} {\bf 05} (1998)
  002, \href{http://xxx.lanl.gov/abs/hep-th/9804170}{{\tt hep-th/9804170}}.

\bibitem{Brodie:1998ke}
J.~H. Brodie and M.~Gutperle, ``{String corrections to four point functions in
  the AdS / CFT correspondence},'' {\em Phys. Lett.} {\bf B445} (1999)
  296--306, \href{http://xxx.lanl.gov/abs/hep-th/9809067}{{\tt
  hep-th/9809067}}.

\bibitem{Chalmers:1998wu}
G.~Chalmers and K.~Schalm, ``{The Large N(c) limit of four point functions in
  N=4 superYang-Mills theory from Anti-de Sitter supergravity},'' {\em Nucl.
  Phys.} {\bf B554} (1999) 215--236,
  \href{http://xxx.lanl.gov/abs/hep-th/9810051}{{\tt hep-th/9810051}}.

\bibitem{Intriligator:1998ig}
K.~A. Intriligator, ``{Bonus symmetries of N=4 superYang-Mills correlation
  functions via AdS duality},'' {\em Nucl. Phys.} {\bf B551} (1999) 575--600,
  \href{http://xxx.lanl.gov/abs/hep-th/9811047}{{\tt hep-th/9811047}}.

\bibitem{hep-th/0002154}
L.~Hoffmann, A.~C. Petkou, and W.~Ruhl, ``{Aspects of the conformal operator
  product expansion in AdS / CFT correspondence},'' {\em Adv. Theor. Math.
  Phys.} {\bf 4} (2002) 571--615,
  \href{http://xxx.lanl.gov/abs/hep-th/0002154}{{\tt hep-th/0002154}}.

\bibitem{hep-th/0002170}
G.~Arutyunov and S.~Frolov, ``{Four point functions of lowest weight CPOs in
  N=4 SYM(4) in supergravity approximation},'' {\em Phys. Rev.} {\bf D62}
  (2000) 064016, \href{http://xxx.lanl.gov/abs/hep-th/0002170}{{\tt
  hep-th/0002170}}.

\bibitem{hep-th/0003218}
E.~D'Hoker, J.~Erdmenger, D.~Z. Freedman, and M.~Perez-Victoria, ``{Near
  extremal correlators and vanishing supergravity couplings in AdS / CFT},''
  {\em Nucl. Phys.} {\bf B589} (2000) 3--37,
  \href{http://xxx.lanl.gov/abs/hep-th/0003218}{{\tt hep-th/0003218}}.

\bibitem{hep-th/0005182}
G.~Arutyunov, S.~Frolov, and A.~C. Petkou, ``{Operator product expansion of the
  lowest weight CPOs in N=4 SYM(4) at strong coupling},'' {\em Nucl. Phys.}
  {\bf B586} (2000) 547--588,
  \href{http://xxx.lanl.gov/abs/hep-th/0005182}{{\tt hep-th/0005182}}.
  [Erratum: Nucl. Phys.B609,539(2001)].

\bibitem{hep-th/0009106}
B.~Eden, A.~C. Petkou, C.~Schubert, and E.~Sokatchev, ``{Partial
  nonrenormalization of the stress tensor four point function in N=4 SYM and
  AdS / CFT},'' {\em Nucl. Phys.} {\bf B607} (2001) 191--212,
  \href{http://xxx.lanl.gov/abs/hep-th/0009106}{{\tt hep-th/0009106}}.

\bibitem{hep-th/0601148}
F.~A. Dolan, M.~Nirschl, and H.~Osborn, ``{Conjectures for large N
  superconformal N=4 chiral primary four point functions},'' {\em Nucl. Phys.}
  {\bf B749} (2006) 109--152,
  \href{http://xxx.lanl.gov/abs/hep-th/0601148}{{\tt hep-th/0601148}}.

\bibitem{hep-th/0611123}
L.~Cornalba, M.~S. Costa, J.~Penedones, and R.~Schiappa, ``{Eikonal
  Approximation in AdS/CFT: Conformal Partial Waves and Finite N Four-Point
  Functions},'' {\em Nucl. Phys.} {\bf B767} (2007) 327--351,
  \href{http://xxx.lanl.gov/abs/hep-th/0611123}{{\tt hep-th/0611123}}.

\bibitem{1011.0780}
S.~Raju, ``{BCFW for Witten Diagrams},'' {\em Phys. Rev. Lett.} {\bf 106}
  (2011) 091601, \href{http://xxx.lanl.gov/abs/1011.0780}{{\tt 1011.0780}}.

\bibitem{1201.6449}
S.~Raju, ``{New Recursion Relations and a Flat Space Limit for AdS/CFT
  Correlators},'' {\em Phys. Rev.} {\bf D85} (2012) 126009,
  \href{http://xxx.lanl.gov/abs/1201.6449}{{\tt 1201.6449}}.

\bibitem{Costa:2014kfa}
M.~S. Costa, V.~Gonçalves, and J.~Penedones, ``{Spinning AdS Propagators},''
  {\em JHEP} {\bf 09} (2014) 064, \href{http://xxx.lanl.gov/abs/1404.5625}{{\tt
  1404.5625}}.

\bibitem{Bekaert:2014cea}
X.~Bekaert, J.~Erdmenger, D.~Ponomarev, and C.~Sleight, ``{Towards holographic
  higher-spin interactions: Four-point functions and higher-spin exchange},''
  {\em JHEP} {\bf 03} (2015) 170, \href{http://xxx.lanl.gov/abs/1412.0016}{{\tt
  1412.0016}}.

\bibitem{0907.2407}
G.~Mack, ``{D-independent representation of Conformal Field Theories in D
  dimensions via transformation to auxiliary Dual Resonance Models. Scalar
  amplitudes},'' \href{http://xxx.lanl.gov/abs/0907.2407}{{\tt 0907.2407}}.

\bibitem{Nandan:2011wc}
D.~Nandan, A.~Volovich, and C.~Wen, ``{On Feynman Rules for Mellin Amplitudes
  in AdS/CFT},'' {\em JHEP} {\bf 05} (2012) 129,
  \href{http://xxx.lanl.gov/abs/1112.0305}{{\tt 1112.0305}}.

\bibitem{1410.4717}
L.~F. Alday, A.~Bissi, and T.~Lukowski, ``{Lessons from crossing symmetry at
  large N},'' {\em JHEP} {\bf 06} (2015) 074,
  \href{http://xxx.lanl.gov/abs/1410.4717}{{\tt 1410.4717}}.

\bibitem{1411.1675}
V.~Gonçalves, ``{Four point function of $\mathcal{N}=4$ stress-tensor
  multiplet at strong coupling},'' {\em JHEP} {\bf 04} (2015) 150,
  \href{http://xxx.lanl.gov/abs/1411.1675}{{\tt 1411.1675}}.

\bibitem{1506.04659}
L.~F. Alday and A.~Zhiboedov, ``{Conformal Bootstrap With Slightly Broken
  Higher Spin Symmetry},'' \href{http://xxx.lanl.gov/abs/1506.04659}{{\tt
  1506.04659}}.

\bibitem{Joaocalc}
J.~Penedones, ``{Unpublished},''.

\bibitem{Dobrev:1975ru}
V.~K. Dobrev, V.~B. Petkova, S.~G. Petrova, and I.~T. Todorov, ``{Dynamical
  Derivation of Vacuum Operator Product Expansion in Euclidean Conformal
  Quantum Field Theory},'' {\em Phys. Rev.} {\bf D13} (1976) 887.

\bibitem{SimmonsDuffin:2012uy}
D.~Simmons-Duffin, ``{Projectors, Shadows, and Conformal Blocks},'' {\em JHEP}
  {\bf 04} (2014) 146, \href{http://xxx.lanl.gov/abs/1204.3894}{{\tt
  1204.3894}}.

\bibitem{1007.2412}
A.~L. Fitzpatrick, E.~Katz, D.~Poland, and D.~Simmons-Duffin, ``{Effective
  Conformal Theory and the Flat-Space Limit of AdS},'' {\em JHEP} {\bf 07}
  (2011) 023, \href{http://xxx.lanl.gov/abs/1007.2412}{{\tt 1007.2412}}.

\bibitem{hep-th/9811155}
F.~Gonzalez-Rey, I.~Y. Park, and K.~Schalm, ``{A Note on four point functions
  of conformal operators in N=4 superYang-Mills},'' {\em Phys. Lett.} {\bf
  B448} (1999) 37--40, \href{http://xxx.lanl.gov/abs/hep-th/9811155}{{\tt
  hep-th/9811155}}.

\bibitem{Pilch:1984xx}
K.~Pilch and A.~N. Schellekens, ``{Formulae for the Eigenvalues of the
  Laplacian on Tensor Harmonics on Symmetric Coset Spaces},'' {\em J. Math.
  Phys.} {\bf 25} (1984) 3455.

\bibitem{1505.03750}
A.~C. Echeverri, E.~Elkhidir, D.~Karateev, and M.~Serone, ``{Deconstructing
  Conformal Blocks in 4D CFT},'' \href{http://xxx.lanl.gov/abs/1505.03750}{{\tt
  1505.03750}}.

\bibitem{1411.7351}
M.~S. Costa and T.~Hansen, ``{Conformal correlators of mixed-symmetry
  tensors},'' {\em JHEP} {\bf 02} (2015) 151,
  \href{http://xxx.lanl.gov/abs/1411.7351}{{\tt 1411.7351}}.

\bibitem{1212.3616}
A.~L. Fitzpatrick, J.~Kaplan, D.~Poland, and D.~Simmons-Duffin, ``{The Analytic
  Bootstrap and AdS Superhorizon Locality},'' {\em JHEP} {\bf 1312} (2013) 004,
  \href{http://xxx.lanl.gov/abs/1212.3616}{{\tt 1212.3616}}.

\bibitem{1212.4103}
Z.~Komargodski and A.~Zhiboedov, ``{Convexity and Liberation at Large Spin},''
  {\em JHEP} {\bf 1311} (2013) 140,
  \href{http://xxx.lanl.gov/abs/1212.4103}{{\tt 1212.4103}}.

\bibitem{1305.4604}
L.~F. Alday and A.~Bissi, ``{Higher-spin correlators},'' {\em JHEP} {\bf 1310}
  (2013) 202, \href{http://xxx.lanl.gov/abs/1305.4604}{{\tt 1305.4604}}.

\bibitem{1502.01437}
A.~Kaviraj, K.~Sen, and A.~Sinha, ``{Analytic bootstrap at large spin},''
  \href{http://xxx.lanl.gov/abs/1502.01437}{{\tt 1502.01437}}.

\bibitem{1504.00772}
A.~Kaviraj, K.~Sen, and A.~Sinha, ``{Universal anomalous dimensions at large
  spin and large twist},'' {\em JHEP} {\bf 07} (2015) 026,
  \href{http://xxx.lanl.gov/abs/1504.00772}{{\tt 1504.00772}}.

\bibitem{1305.0004}
A.~L. Fitzpatrick, J.~Kaplan, and D.~Poland, ``{Conformal Blocks in the Large
  $D$ Limit},'' {\em JHEP} {\bf 1308} (2013) 107,
  \href{http://xxx.lanl.gov/abs/1305.0004}{{\tt 1305.0004}}.

\bibitem{Vos:2014pqa}
G.~Vos, ``{Generalized Additivity in Unitary Conformal Field Theories},''
  \href{http://xxx.lanl.gov/abs/1411.7941}{{\tt 1411.7941}}.

\end{thebibliography}\endgroup

\end{document}